\documentclass{aa}  
\usepackage[varg]{txfonts}
\usepackage{multirow}
\usepackage{graphicx}
\usepackage{color}
\usepackage{soul}
\usepackage{adjustbox}
\usepackage{supertabular}
\usepackage{longtable,lscape}
\usepackage{rotating}   
\usepackage[T1]{fontenc}
\usepackage{rotating}
\usepackage{booktabs}  
\usepackage{siunitx}  
\usepackage[colorlinks, citecolor={blue}]{hyperref}

\begin{document} 

\title{A MaNGA view of isolated galaxy mergers in the star-forming \\ Main Sequence} 
\titlerunning{A MaNGA view of isolated galaxy mergers in the star-forming \\ Main Sequence}

\author{
P.~V\'asquez-Bustos\inst{1,2}
\and
M.~Argudo-Fernández\inst{1,2,3}
\and
M.~Boquien\inst{4}
\and
N.~Castillo-Baeza\inst{1}
\and
A.~Castillo-Rencoret\inst{1}
\and
D.~Ariza-Quintana\inst{2}
}

\institute{
Instituto de F\'isica, Pontificia Universidad Cat\'olica de Valpara\'iso, Casilla 4059, Valpara\'iso, Chile.
\and
Departamento de Física Teórica y del Cosmos, Edificio Mecenas, Campus Fuentenueva, Universidad de Granada, 18071 Granada, Spain. 
\and 
Instituto Carlos I de Física Teórica y Computacional, Facultad de Ciencias, Universidad de Granada, 18071 Granada, Spain.
\and
Universit\'e C\text{\^o}te d’Azur, Observatoire de la C\text{\^o}te d’Azur, CNRS, Laboratoire Lagrange, 06000 Nice, France.
}

   \date{Received July 11, 2024; accepted February 13, 2025}

 
  \abstract
   {Many open questions in the complex process of galaxy evolution during interactions remain, as each stage presents a different timing of star formation.}
   {We aim to better understand the processes triggered in galaxies by interactions, considering low-density environments, making in-situ interaction between the members the main process driving their evolution.}
   {In this work we carry out an analysis of star-formation and nuclear activity in the different stages during a galaxy merger identified in isolated systems (isolated galaxies, isolated pairs, and isolated triplets) using integral field spectroscopy from the SDSS-IV/MaNGA project. We classify galaxies into close pairs, pre-mergers, mergers, and post-mergers (including galaxies with post-starburst spectroscopic features), for a total sample of 137 galaxies. We constrained their star formation history from spectro-photometric SED fitting with CIGALE, and used spatially resolved WHAN diagrams, with other MaNGA data products to explore if there is any connection of their physical properties with their merging stage.}
   {In general, galaxies show characteristic properties intrinsically related to each stage of the merger process. Galaxies in the merger and post-merger stages  present higher star formation activity (measured by their integrated sSFR). In the merger stage, the fraction of strong AGN spaxels is comparable to the fraction of spaxels with pure star-formation emission, with no difference between AGN activity in close pairs and strongly interacting galaxies with the same stellar mass.}
   {Our results support the scenario where galaxy interactions triggers star-formation and nuclear activity on galaxies. Nonetheless, AGN has a minor role in quenching galaxies following a merger, as AGN feedback might not have had sufficient time to inhibit star formation. In addition, we found that the quenching process in post-mergers galaxies with post-starburst emission is happening outside-in, being an observational proof of the effect of interactions on the quenching process. The transforming processes after a recent major galaxy interaction may happen slowly on isolated environments, where the system evolves in a common dark matter halo without any perturbation of external galaxies.}


   \maketitle
%

\section{Introduction}
\label{Sec:intro}

Galaxy interactions and mergers play a fundamental role in the evolution of galaxies, intrinsically influencing their shape, growth, and physical properties. The gravitational interactions between galaxies, usually forming structures such as tidal tails \citep{1972ApJ...178..623T,1996ApJ...464..641M}, and the friction between the gas and dust have major effects on the galaxies involved. Due to their nature, these processes are chaotic, making them difficult to physically understand. The result of a galaxy merger depends on a wide variety of parameters such as relative size/composition or galaxy morphology, and kinematic parameters such as collision angle and relative velocity, thus necessitating studies of galaxy interactions in both, controlled environments and simulations \citep{1999ASPC..160..351A, 2005A&A...437...69B, 2007A&A...468...61D,2010MNRAS.404..590L,2011MNRAS.415.3750M}, but also using photometric and spectroscopic observational data 
\citep{2003A&A...405...31B,2004AJ....128..163L,2007ApJ...660L..51L,2008ApJ...681..232L,2016PASJ...68...96M,2019MNRAS.487.2491E,2019A&A...631A..51P,2022MNRAS.517L..92E,2022ApJ...937...97C,2023A&A...669A..23G}, with all together providing a global picture to investigate these processes.

During decades many different studies have shown that these evolutionary mechanisms are responsible for changes in the internal physical properties of galaxies, mainly the enhancement of star formation, strongly implying a connection between galaxy interactions and the birth of stars \citep{1997MNRAS.292..835R}. Star formation during galaxy interactions may happen in particular regions through abrupt bursts in the regions where galaxy interactions occur \citep{1972ApJ...178..623T,1987AJ.....93.1011K,2000ApJ...530..660B, 2003ApJ...582..668B, 2003MNRAS.346.1189L,2006ApJ...652..270B, 2007ApJ...671.1538B, 2009ApJ...704..324R, 2009ApJ...697.1971J, 2019A&A...631A..51P, 2022MNRAS.516.4922R}, and the presence of star forming regions in gas clouds in tidal tails or in the bridge of material between galaxies \citep{1999ApJ...523L.133N,2006MNRAS.372..839N,2008AJ....135..548D,2009AJ....137.4561B,2010AJ....140.2124B,2011A&A...533A..19B,2013LNP...861..327D,2014A&A...566A..97J,2021MNRAS.503.2866D,2021ApJ...923L..21P}. 

These violent interactions trigger and accelerate the processes of star formation in galaxies. Galaxies that have suffered a recent interaction transit from a state of starburst or enhanced star formation \citep[e.g. ][]{1985MNRAS.214...87J,2015ApJ...807L..16K,2021MNRAS.501.1046M,2021MNRAS.502.4815T,2022ApJ...940...31L} to a quenching state, occurring faster than in normal (non interacting) galaxies \citep{2007ApJ...665..265F,2010ApJ...721..193P,2012ApJ...753..167B,2016A&A...587A..72B,2019A&A...631A..51P}. It has been also observed that, in addition to an increment of star formation, galaxy interactions may trigger the nuclear activity of galaxies, which might be also connected to star formation quenching by active galactic nuclei (AGN) feedback processes \citep[e.g.,][]{1996ARA&A..34..749S,2007MNRAS.375.1017A,2010MNRAS.401.1043D,2018MNRAS.476.2308W,2019MNRAS.487.2491E}.

In particular, AGN is one of the main drivers for the transition from star-forming disk to passive spheroidal galaxies \citep{2003MNRAS.346.1055K}. In an AGN, the supermassive black hole (SMBH) growing through mass accretion at the centre of massive galaxies releases large amounts of energy into their surroundings, as well as into the galaxy itself \citet{1998AJ....115.2285M,2000ApJ...539L...9F,2004MNRAS.351..169M,2013ARA&A..51..511K}. How the energy accretion or AGN feedback (the posteriori re-deposition of energy and momentum into the interstellar medium through outflows and radiation) generate changes in galaxies, such as in the stellar populations in the bulge and outgrowth \citep{2007MNRAS.382.1415S,2008ApJ...675.1025S,2015MNRAS.453L..83M,2018RMxAA..54..217S,2020MNRAS.492.3073L} or significant increment in the SFR of galaxies \citep{2011ApJ...739...57K,2012A&A...540A.109S,2016MNRAS.458L..34E,2020ApJ...901...66W}, are some of the currently unknown topics. There is some evidence that major mergers are the main driver of high luminosity AGNs \citep{2011MNRAS.418.2043E,2012MNRAS.419..687R,2015MNRAS.452..774K}. Moreover, AGNs in interacting systems, through AGN feedback, may have an important role in regulating star formation in these interacting systems, but this argument is still under debate due to the recent incorporation of integral field unit (IFU) spectroscopy. Some studies reveal that there is a correlation between these phenomena \citep{2011MNRAS.418.2043E,2013MNRAS.430.3128E,2014AJ....148..137L,2014MNRAS.441.1297S,2018PASJ...70S..37G}, while others contradict it \citep{2011ApJ...726...57C,2017MNRAS.466..812V,2019ApJ...882..141M,2021ApJ...909..124S}. 

Because of the aforementioned, many open questions in the complex process of galaxy evolution during interactions still remain, as each stage of the process presents a different timing of star formation \citep[e.g.,][]{2018A&A...614A..66S,2008ApJS..175..356H,2015MNRAS.449...49R}. In addition, the bibliography distinguishes between two main types of interactions depending of the presence of cold and molecular gas: ``wet" mergers, which are mergers of galaxies where a galaxy contains at least a substantial amount of gas to form stars; and ``dry" mergers, which are mergers of galaxies with low gas content and no associated with starburst or star-formation activity \citep{2006ApJ...636L..81N,2010ApJ...718.1158L,2011hsa6.conf..173E}. But also depending on the mass ratio between the galaxies: minor merger, when the stellar mass of the companion is typically less than 1/10; and major merger, when the companion galaxy has similar stellar mass \citep{2012A&A...539A..45L,2014A&A...567A..68D}. Therefore the different stage of the interactions may present different properties in the galaxies depending of the type of merger \citep{2010ApJ...715..202H,2014A&A...572A..25M,2017MNRAS.464.3882W,2022ApJ...937...97C}. The analysis of galaxy interactions might be much better understood if considering low-density environments, making in-situ interaction between the members the main process driving their evolution, with a minimal contamination of other external environment processes that may happen, for instance, within a galaxy cluster \citep{2016ApJ...822L..33H}. For this reason, the SDSS-based catalogue of Isolated Galaxies (SIG), Isolated Pairs (SIP), and Isolated Triplets (SIT) presented in \citet{2015A&A...578A.110A} are ideal to investigate these processes. Moreover, as it was shown in \citet{2023A&A...670A..63V}, we can easily identify the presence of galaxy interactions in isolated systems with high values of the tidal strength parameter ($Q_{trip} > -2$ for the SIT). In this regard, in this work we carry out an analysis of star-formation and nuclear activity in the different stages during a galaxy merger identified in isolated systems (isolated galaxies, isolated pairs, and isolated triplets). To better understand the physical processes occurring in the galaxies, we used IFU spectroscopy from the MaNGA \citep[MApping Nearby Galaxies at APO;][]{2015ApJ...798....7B,2015AJ....149...77D} survey. \\\\
This study is organised as follows. In Sect.~\ref{Sec:merger-data} we explain how we select galaxies at different stages of the merger process within the SIG, SIP, and SIT catalogues that have available MaNGA data. The methods we used to constrain the star-formation histories (SFH) from spectro-photometric spectral energy distribution (SED) fitting, as well as emission line diagnostic diagrams to explore nuclear activity, are presented in Sect.~\ref{Sec:merger-met}. Our results and their corresponding discussion are presented in sections \ref{Sec:merger-res} and \ref{Sec:merger-dis}, respectively. A brief summary and the conclusions of this analysis can be found in Sect.~\ref{Sec:merger-con}. Throughout the study, a cosmology with $\Omega_{\Lambda_{0}} = 0.7$, $\Omega_{\rm m_{0}} = 0.3$, and $H_0=70$\,km\,s$^{-1}$\,Mpc$^{-1}$ is assumed.


\section{Data and the sample} 
\label{Sec:merger-data}

\subsection{MaNGA data}

Integral field spectroscopy (IFS) provides a more complete understanding of the emissions in different regions and components of galaxies. In particular, optical IFS is helping us to understand the local versus global universal relations in galaxies \citep{2016MNRAS.463.2513B}, and for instance, how low ionization emission-line regions are not only found in the nuclear regions of galaxies \citep{2016MNRAS.461.3111B}. 

For this study we used spectroscopic data from the MaNGA survey which provides IFU data products for 10.000 galaxies in the local universe ($z\,<\,0.15$) as part of the SDSS-IV legacy survey \citep{2017AJ....154...28B,2022ApJS..259...35A}. MaNGA uses hexagonal fibre bundles, each containing between 19 to 127 spectroscopic fibres, ensuring the delivery of a minimum spatially resolved spectrum coverage of 1.5 effective radius per galaxy in the primary sample, and 2.5 effective radius for the secondary sample. It also covers a spectral range from $3600$ Å to $10300$ Å, with a resolution of $R~=~2000$. MaNGA data products include sky-subtracted spectrophotometrically calibrated spectra and rectified three-dimensional data cubes \citep{2016AJ....152...83L,2021AJ....161...52L} and high level data products, including stellar kinematics (velocity and velocity dispersion), emission-line properties (kinematics, fluxes, and equivalent widths), and spectral indices \citep[][e.g., D4000 and the Lick indices,]{1983ApJ...273..105B,1984ApJ...287..586B,1985ApJS...57..711F,1997A&A...325.1025P,2005MNRAS.362...41G}. See \citet{2019AJ....158..231W} and \citet{2019AJ....158..160B} for more information about the MaNGA data products and its pipeline. The MaNGA dataset can be accessed via Marvin\footnote{\url{https://sdss-marvin.readthedocs.io/en/latest/}} tools \citep{2019AJ....158...74C}.

For this study we used $H_{\alpha}$ and $H_{\beta}$ emission line maps, with the $D_n\,4000$ map and the spectrum in each spaxel, to constraint the star formation history, as explained in Sec.~\ref{sec:SED}. We also used the [NII] emission map with the equivalent widths $W_{H_{\alpha}}$ and $W_{NII}$ to create diagnostic diagrams of the nuclear activity, as explained in Sec.~\ref{sec:WHAN}. 

\subsection{Interacting galaxies in isolated systems}

As introduced in Sect.~\ref{Sec:intro}, for this study we used the SIG, SIP, and SIT catalogues compiled by \citet{2015A&A...578A.110A} based on the SDSS survey data. The catalogues contain 3702 isolated galaxies, 1240 isolated pairs, and 315 isolated triplets, respectively, in the local Universe ($z$~$\leq$~$0.080$). The systems are isolated with no nearest neighbours within $\rm \Delta v\, \leq\, 500\, km \,s^{-1}$ line-of-sight velocity difference with a spatial projected radius of the field of 1\,Mpc. Galaxies in the SIG, and central galaxies in the SIP and SIT \citep[named as A galaxies in][]{2015A&A...578A.110A}, are brighter than an r-band apparent magnitude m$_r$\,=\,15.7. This allows the search for satellite galaxies up to 2 magnitude fainter within the spectroscopic completeness limit of the SDSS (m$_r$\,=\,17.7). Therefore satellite galaxies are fainter, and also generally less massive, than central galaxies. Galaxies in the SIP and SIT are physically bound at a projected distance up to $\rm d\, \leq\, 450\, kpc$ within $\rm \Delta v\, \leq\, 160\, km \,s^{-1}$ line-of-sight velocity difference with respect to the central (the brightest) galaxy. \citet{2015A&A...578A.110A} also provides a quantification of the local and the large-scale environment for galaxies in the SIG, SIP, and SIT using local number-densities and tidal strength parameters. We refer to \citet{2015A&A...578A.110A} for more information about the compilation of the catalogues. In addition to the aforementioned catalogues, we used a non-public sample of 12 isolated mergers that were removed when compiling the SIG \citep[hereafter the SIM, for analogy with the samples from][]{2015A&A...578A.110A}. We used the Marvin tools to access to the MaNGA dataset, finding that 54 SIT galaxies, 156 SIP galaxies, 348 SIG galaxies, and 6 SIM galaxies have available data products. Note that, in the case of SIP and SIT samples, there is not always MaNGA observations for all the galaxies belonging to the same system. Since we are not interested on studying interactions within the same system, which is beyond the scope of this work and it would be limited to a few individual systems, we use these samples to select galaxies at different interaction stages to explore the impact of the interaction on the galaxies with available MaNGA data.

To consider the level of interaction between all members in the SIP and SIT, independently of the available MaNGA data, we use the tidal strength parameter Q. As presented in \citet{2023A&A...670A..63V}, isolated triplets with $Q_{\rm trip}~>~-2$ show compact configurations due to interactions happening in the systems. In particular, triplets with $Q_{\rm trip}~>~-0.45$ present on-going mergers and strong interactions \citep{2023A&A...670A..63V}. Taking into account that the sample of merging/strong interacting galaxies in the SIT is small (37 triplets, without considering the availability of IFU data), we followed the same selection criteria to identify interacting galaxies in the SIP, using in this case the $Q_A$ parameter, as defined in \citet{2015A&A...578A.110A}. Given that the SIG, SIP, and SIT were selected homogeneously following the same isolation criteria, we also look for visually disturbed isolated galaxies in the SIG. We performed a visual inspection of the SDSS three-colour images of the galaxies in the SIT, SIP, SIG, and SIM, to identify and classify the level of interaction. To identify possible features, fainter than SDSS images, we also used three-colour images from Pan-STARRS 1 \citep{2016arXiv161205560C} for a better visualization of the structure of the galaxies. We also used spectroscopic MaNGA data to identify possible post-starburst galaxies, since these galaxies might have probably suffered a recent past interaction. The details of the identification of the galaxies in different interaction stage are presented in Sect.~\ref{Sec:merger-cat}. 

\section{Methodology}
\label{Sec:merger-met}

\subsection{Merging stage}
\label{Sec:merger-cat}

\begin{figure}
    \centering
\includegraphics[width=\columnwidth, trim={0.5cm 0.4cm 1.5cm 1.5cm},clip]{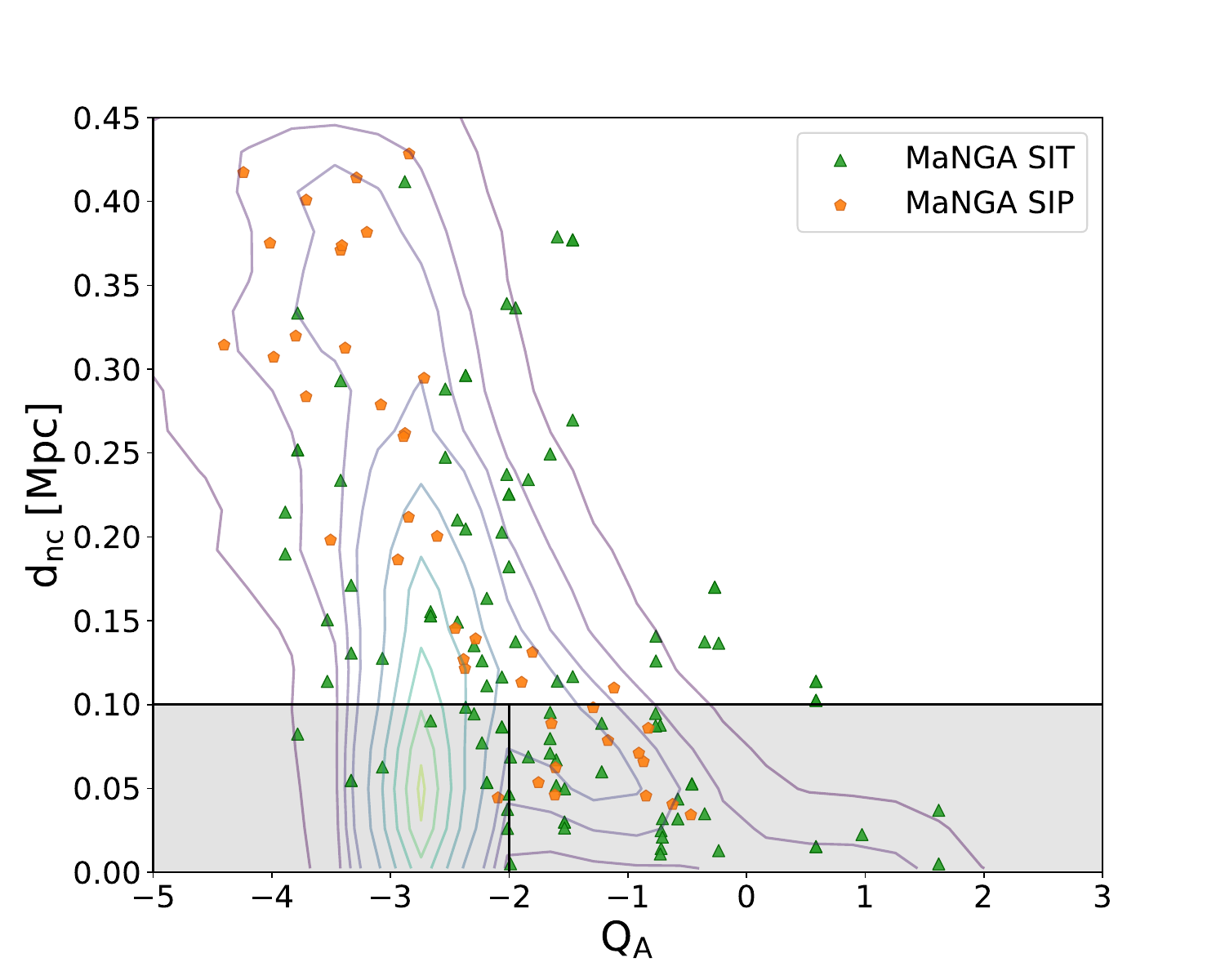}
    \caption{Projected distance to the nearest companion $d_{nc}$, in Mpc, with respect to the tidal strength of the central galaxy on the system $Q_A$. Contour lines correspond to all the galaxies in the SIP and SIT, of them, green triangles and orange hexagons indicate galaxies in the SIT and SIP, respectively, with MaNGA data. The horizontal black solid line delimit the gray shaded area with $d_{nc}$\,$\leq$\,100\,kpc that we use to delimit close pairs. Within this area, the vertical black solid line at $Q_A\,=\,-2$ is used to identify interacting galaxies, following \cite{2023A&A...670A..63V}.}
    \label{fig:QA}
\end{figure}

To better understand the complex processes happening during galaxy mergers, it is necessary to divide the galaxies in different stages of the interaction, starting from close pairs, through the merging process, and to the post-merger scenario. For instance, star formation can reach its maximum during a pre-merger or merging period or rapidly decay in a short time-scale after a major merger event. We classify the galaxies in the following categories according to their merging stage as detailed below:  

\begin{itemize}
    \item  Close Pair (CP): Close pairs of galaxies in the SIT and the SIP with projected distance $d~\leq~100$\,kpc (see Fig.~\ref{fig:QA}). For these galaxies no signs of interactions between the members, such as tidal arms and tails or deformations, should be detectable in a visual inspection. Therefore this sample can be used as a control sample for the other categories, where signs of interaction are visually present. An example of a galaxy in this category is shown in Fig.~\ref{fig:cp_maps}.
    
    \item Pre-Merger (PrM): Low to mild level of interaction is visible appreciable between physically bound galaxies in the SIP and SIT, with slight deformations of the spiral arms (in the case of late-type galaxies), with projected distance $d~\leq~100$\,kpc and local tidal strength $Q_A~<~-2$ (within the left side of the shaded area in Fig.~\ref{fig:QA}). Under this category, visible tidal tails and bridges of materials between galaxies can be observed, however the two galaxies can be independently identified, therefore the process of merging of the nuclei has not started yet. Note that the galaxies in this category may have undergone an interaction process that only deformed them a long time ago. An example of a galaxy in this category is shown in Fig.~\ref{fig:prm_maps}.
    
    \item Merger (M): This category covers the stage where galaxy nuclei merge. Under this category, large deformations in most of the visible regions of galaxies, accretion of matter from one galaxy to another, as well as tidal arms and bridges can be visually observed in the optical SDSS images. For these cases it was found that both galaxies can be present in a single MaNGA field of view (FoV). An example of a galaxy in this category is shown in Fig.~\ref{fig:m_maps}.
    
    \item Post-Merger (PsM): This category considers the final stages of an interaction. This sample is composed of galaxies in the SIG and the SIM with visible signs of interaction but with no presence of an observable close companion galaxy. We also include some galaxies with post-starburst spectroscopic signatures. The criteria followed to select these galaxies is explained in Sec.~\ref{sec:psm}. An example of a galaxy in the PsM category is shown in Fig.~\ref{fig:psm_maps}. 
\end{itemize}

Note that each point in Fig.~\ref{fig:QA} does not always correspond to a pair of galaxies that belong to the same system, it may correspond to a single galaxy in the case of SIP and SIT systems with MaNGA products for only one galaxy. 
The sample of interacting galaxies at different merging stages is composed of a total of 137 galaxies. The number of galaxies for each category is shown in the Table~\ref{tab:Ngal}, while the number of central and satellites for each merger classification is presented in Table~\ref{tab:Ngal_sc}. 

\begin{table}
    \centering
    \setlength{\arrayrulewidth}{0.1mm}
    \begin{tabular}{ccc}
    \hline\hline
      Merging stage  & $\rm N_{galaxies}$ & $\rm p_{merger}$\\ 
    \hline
    CP  & 40 & 0.19\,$\pm$\,0.19 \\
    PrM  & 21 & 0.49\,$\pm$\,0.36 \\
    M & 6 & 0.78\,$\pm$\,0.12 \\
    PsM (PSB)  & 63 (7) &  0.35\,$\pm$\,0.30 (0.30\,$\pm$\,0.29) \\
    \hline
     \end{tabular}
    \caption[Number of galaxies in each merging stage.]{Number of galaxies classified in each merging stage: close pair (CP), pre-merger (PrM), merger (M), and post-merger (PsM), where of them, 7 are classified as post-starburst (PSB). The total number of galaxies considered in this work is 137 galaxies. The last column presents the mean value (and standard deviation) of the probability that a galaxy presents merger morphology, parameterised with the $\rm p_{merger}$ value according to \citet{2018MNRAS.476.3661D}.}
    \label{tab:Ngal}
\end{table}

\begin{table}
    \centering
    \setlength{\arrayrulewidth}{0.1mm}
    \begin{tabular}{c|ccccc}
    \hline\hline
    Member & CP & PrM & M & PsM & PSB\\ 
    \hline
    Central & 27 & 14 & 3 & 7 & 3\\
    Satellite & 13 & 7 & 3 & 4 & 1\\
    \hline
     \end{tabular}
    \caption{Number of central and satellite galaxies in the SIP and SIT classified in each merging stage. The galaxies were classified as centrals or satellites according to \citet{2015A&A...578A.110A}, where the galaxy A on the system is the central galaxy, and the other member galaxies are satellites.}
    \label{tab:Ngal_sc}
\end{table}

\subsubsection{Identification of post-merger galaxies}
\label{sec:psm}

This category is mainly composed of galaxies in the SIG and the SIM with signs of a recent interaction but not observable companion, as previously introduced. We also visually searched for shells signatures around since these are indicators of recent interactions or that galaxies underwent a nucleus merger process \citep{2019A&A...630A.112M, 2001AJ....122.1758B, 2023MNRAS.518.3261P}. In general, this category might be complex to identify, especially in the case of early-type galaxies, since the low-surface brightness features might not be visible in the optical SDSS images. Therefore we consider other observational aspects that might be related to a recent merger event. 

Some studies claimed that lenticular or S0 galaxies might be the result of a major merger event between two spiral galaxies of unequal masses \citep{2001Ap&SS.276..847B,2012A&A...547A..48E,2015A&A...573A..78Q,2015A&A...579L...2Q,2018A&A...617A.113E}. However there are other mechanisms involved in the formation of lenticular galaxies, for example internal secular evolution and environmental processes, in particular for the formation of intermediate to low mass lenticulars \citep{2017A&A...604A.105T}. Therefore, a selection based on optical morphology might introduce a strong bias. On the other hand, some works argue that a high percentage of post$-$starburst or PSB galaxies (sometimes also referred to as E$+$A or K$+$A galaxies, i.e., galaxies that recently experienced an episode of intense star formation that was rapidly truncated) might be due to a recent interaction that suddenly quenched their star formation \citep{2005MNRAS.357..937G,2011ApJ...741...77S,2022MNRAS.516.4354W,2022MNRAS.517L..92E}. Taking into account that we are working with isolated galaxies, we consider that the most probable reason for the presence of PSB spectral signatures in these galaxies might be due to a recent past interaction \citep{2021ApJ...919..134S}.

\begin{figure}
    \centering
\includegraphics[width=\columnwidth]{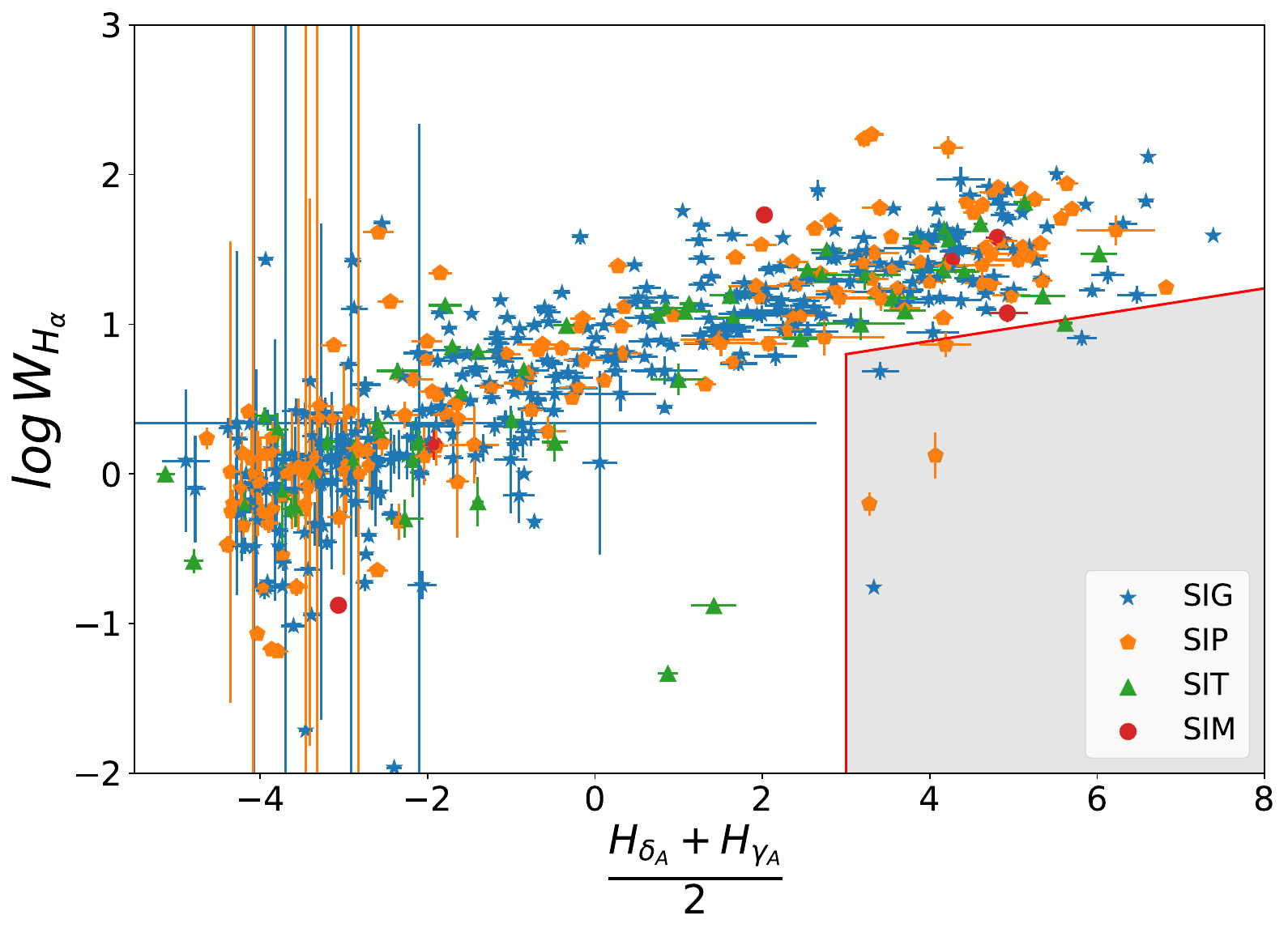}
    \caption[PSB diagnostic diagram for SIG, SIP, SIT, and SIM galaxies]{PSB diagnostic diagram (spectral index distributions $H_{\alpha}$ equivalent width vs. $\frac{H_{\delta_A}+H_{\gamma_A}}{2}$
    ) for SIG, SIP, SIT, and SIM galaxies with MaNGA data considering the spectra associated to the central spaxel, and their associated error bars. The values for galaxies in each sample are presented with symbols and colours according to the legend. The red solid line and area indicate the PSB region. 
    }
    \label{fig:Isol_psb}
\end{figure}

\begin{figure*}
\centering 
    \includegraphics[width=0.43\textwidth]{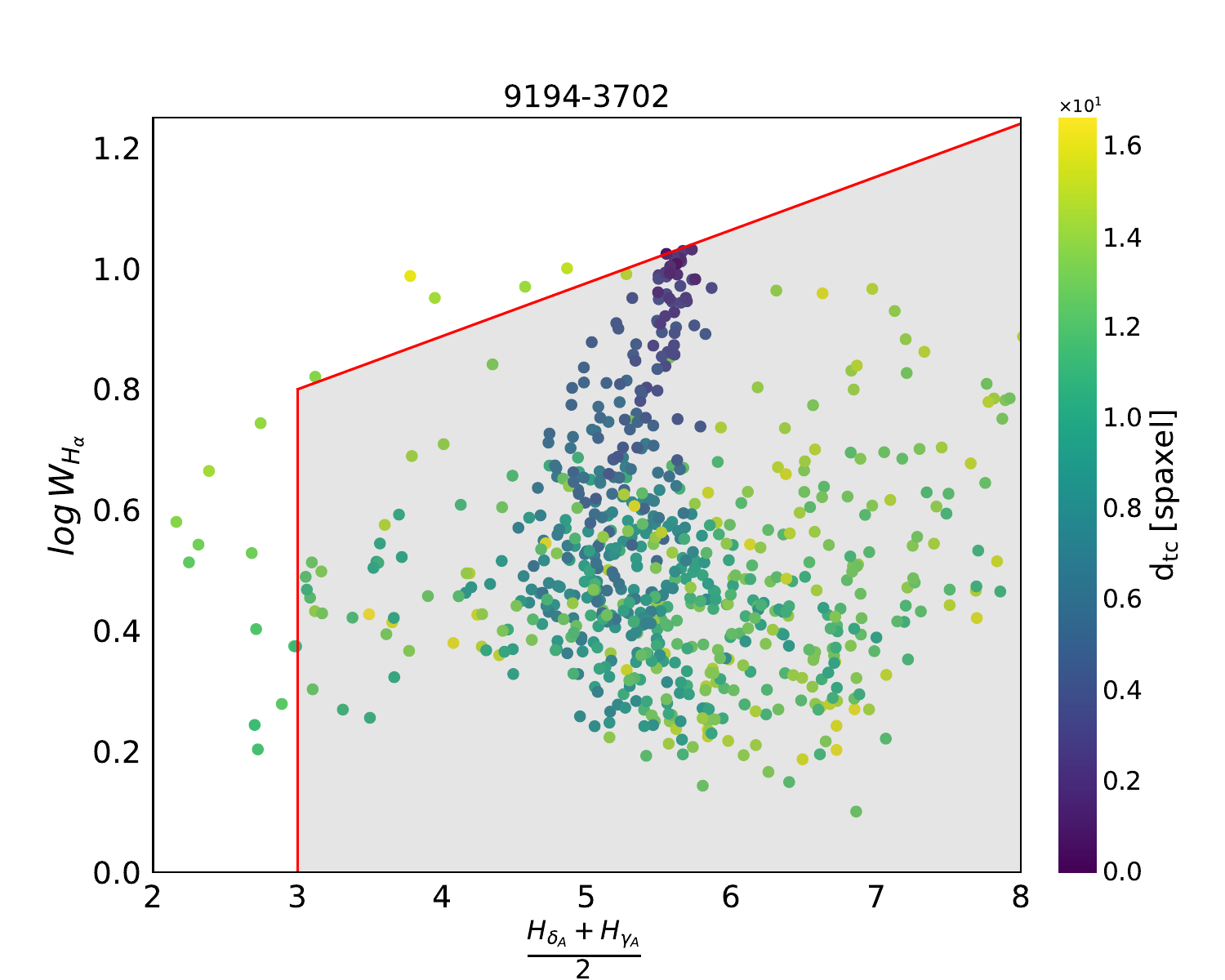}
    \includegraphics[width=0.43\textwidth]{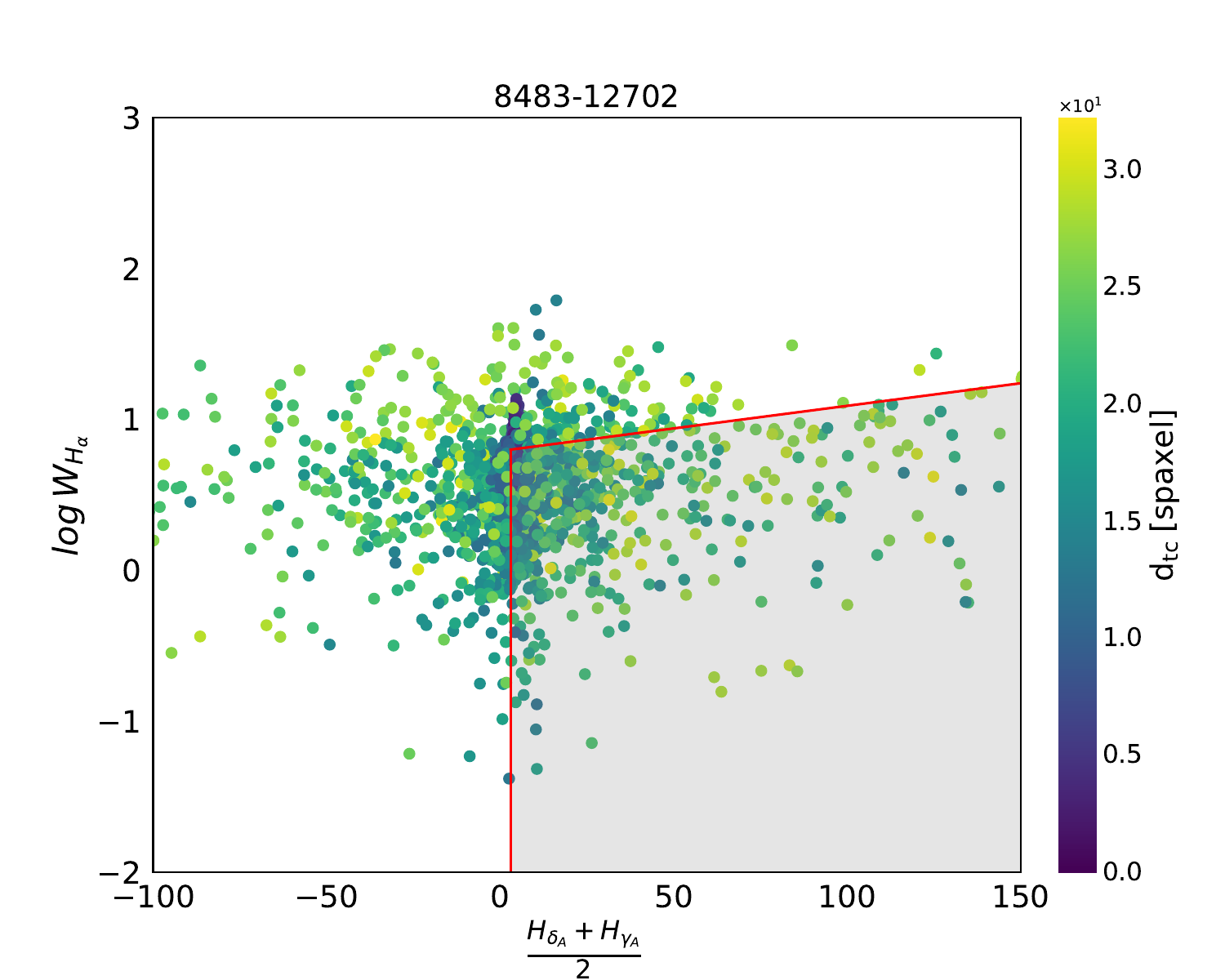} \\
    \includegraphics[width=0.43\textwidth]{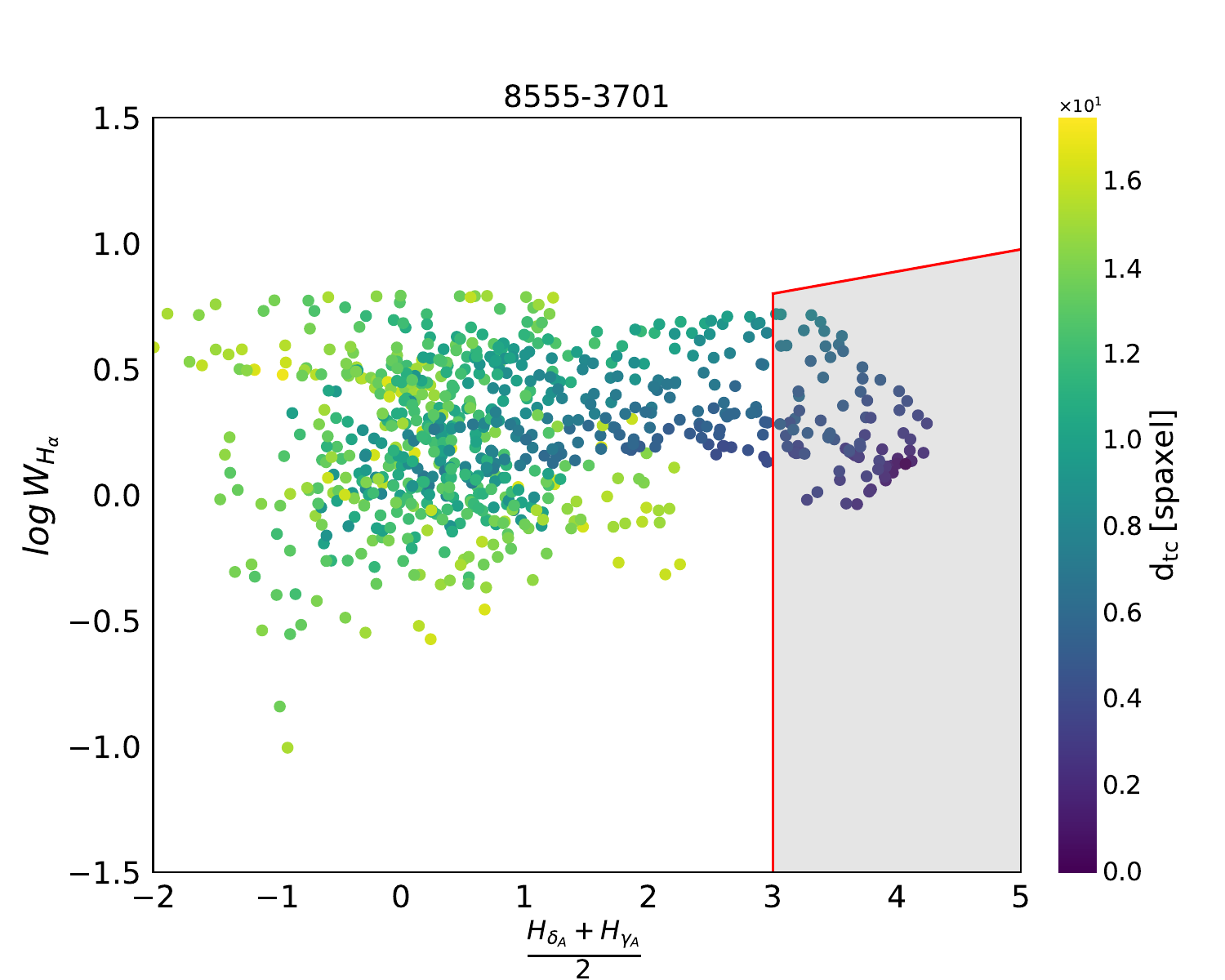}
    \includegraphics[width=0.43\textwidth]{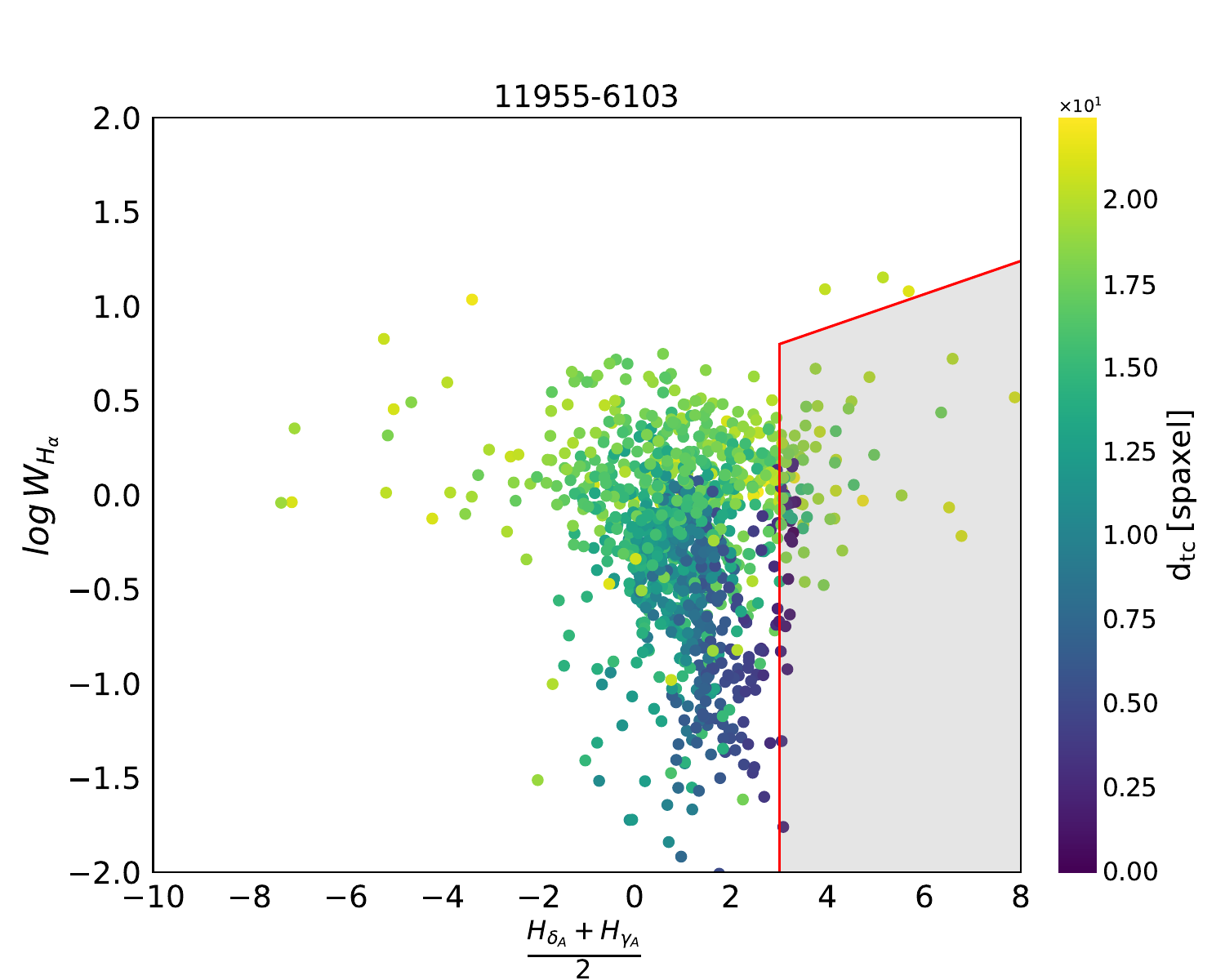}\\
    \includegraphics[width=0.43\textwidth]{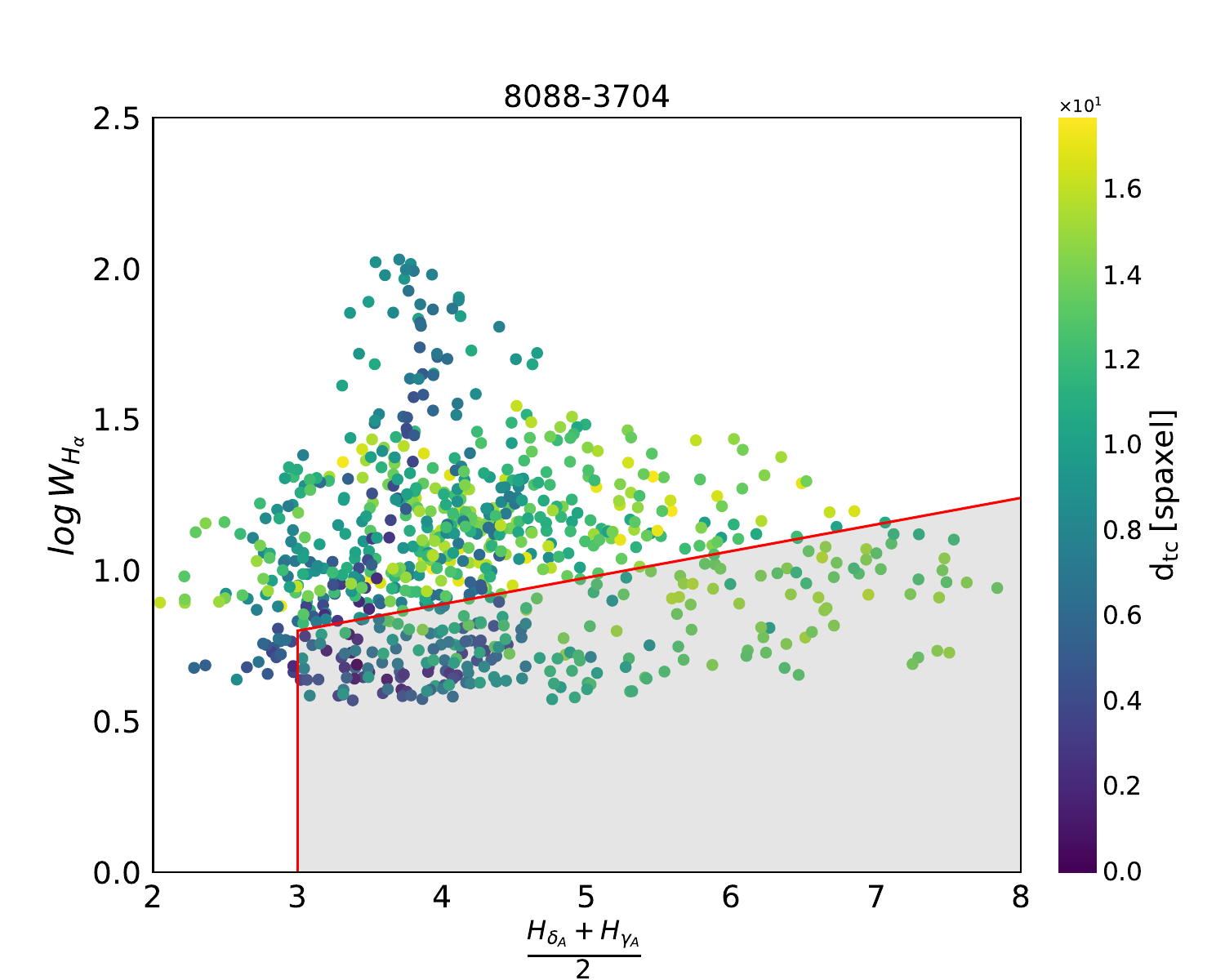}
    \includegraphics[width=0.43\textwidth]{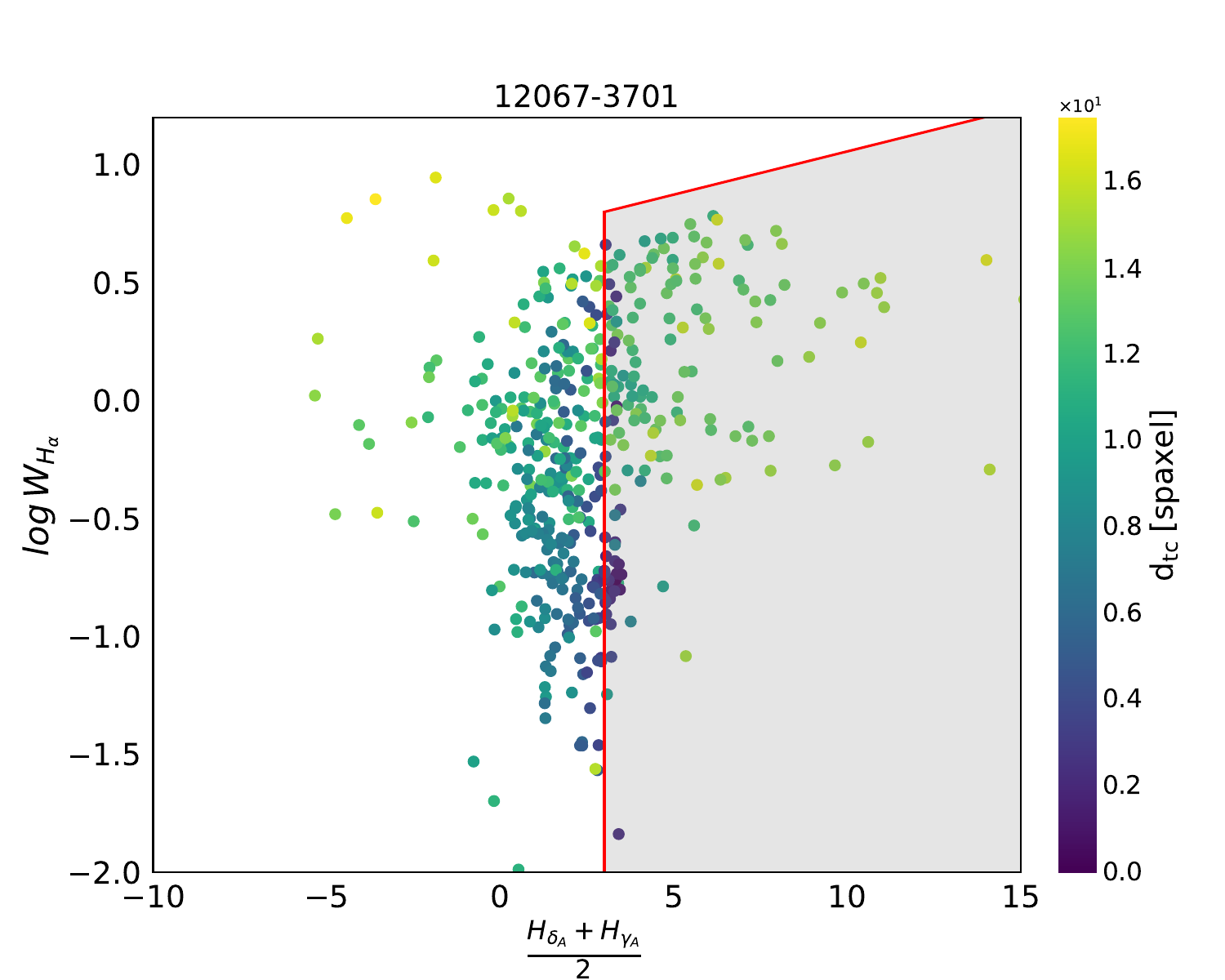}
    \includegraphics[width=0.43\textwidth]{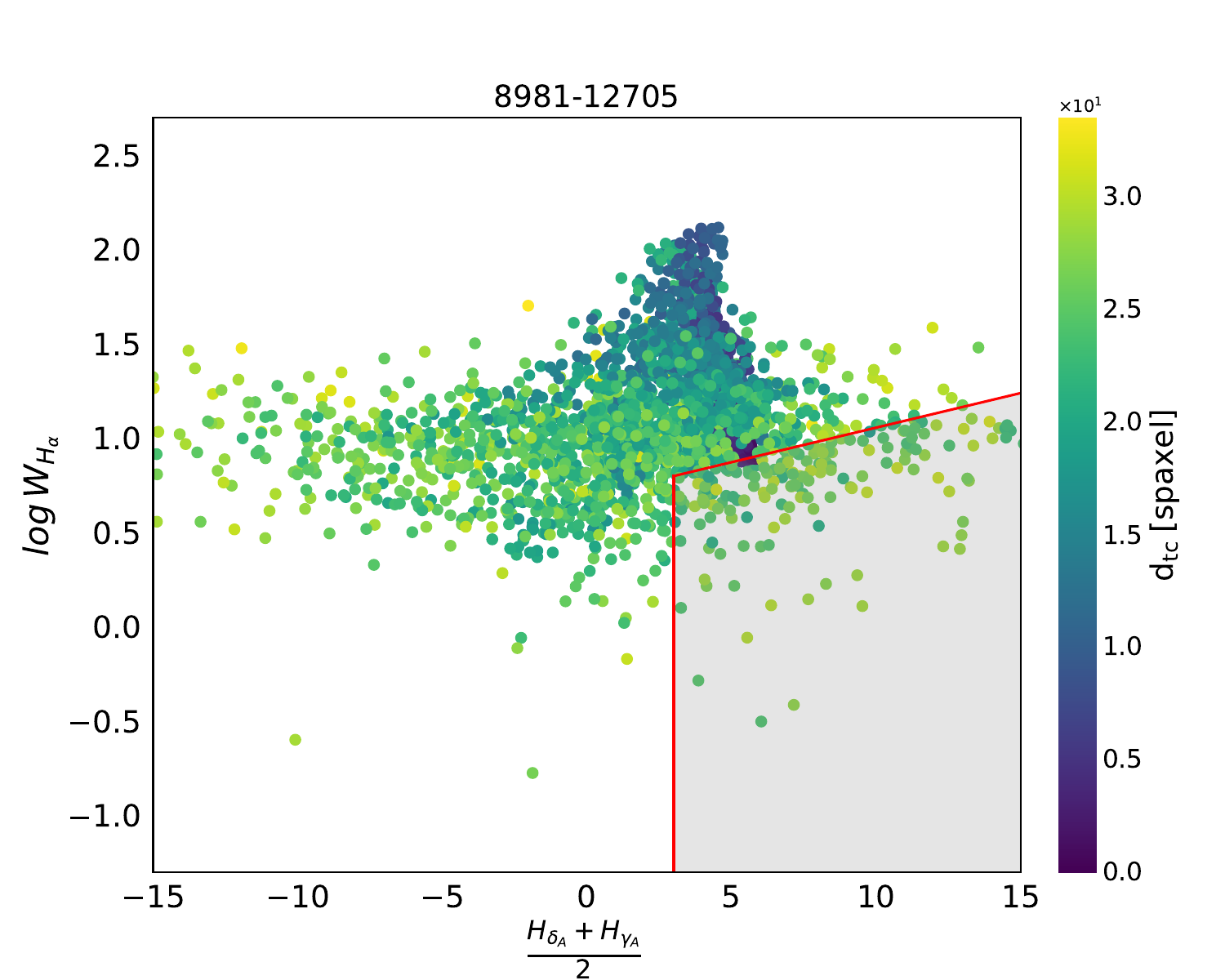}
    \caption[Spatially resolved PSB diagram for PSB candidate galaxies.]
    {Spatially resolved PSB diagram ($H_{\alpha}$ equivalent width vs. $\frac{H_{\delta_A}+H_{\gamma_A}}{2}$ indices) for the seven PSB candidate galaxies in our samples, colour coded by the distance to the central spaxel as indicated in the colour bar. The red solid line and shaded gray area demarcate the PSB region.}
    \label{fig:spatial_psb}
\end{figure*}

We identify PSB galaxies using traditional selection methods consistent on the analysis of the $H_{\alpha}$ equivalent width (in \AA) with respect to the summed stellar continuum indices $H_{\delta_A}$ and $H_{\gamma_A}$ diagnostic diagram \citep{2020MNRAS.498.1259Z}. These spectroscopic measurements are  available in the MaNGA data products. For simplicity, we consider the products from the central spaxel (usually corresponding to the galaxy nucleus) to identify candidate PSB galaxies, and then use the spatially resolved information to analyse them in more detail. For homogeneity, we reproduced the diagnostic diagram for all the galaxies in our samples (see Fig.~\ref{fig:Isol_psb}), including galaxies in the SIP and SIT as they might have suffered a previous merger event (not related to an interaction with the current galaxies in the system). The PSB candidate galaxies are those enclosed in the shaded area of the diagram, known as the PSB region \citep{2003PASJ...55..771G,2004ApJ...602..190Q,2005MNRAS.360..587B}. 
We found seven galaxies (three SIG galaxies, three SIP galaxies, and one SIT galaxy): 8088-3704, 12067-3701, 8981-12705, 8483-12702, 8555-3701, 11955-6103, and 9194-3702. Figure~\ref{fig:spatial_psb} shows the spatially resolved PSB diagnostic diagrams for these galaxies. The PSB emission is mainly present in the center of the galaxies but we also found some spatial distribution outer the center, which is also expected and in agreement with similar analysis on MaNGA galaxies \citep{2019MNRAS.489.5709C,2023MNRAS.523..720L}. We therefore included these seven galaxies in the PsM category.


\subsection{Galaxy morphology}
\label{sec:morpho}

The classification of the galaxies in different merging stage is independent of their morphology, i.e., if the galaxy is early- or late-type. Our visual classification is based on the presence (or absence) of a close companion and/or interaction features (tidal tails and shells, for instance). In order to take into account the morphological type of the galaxies in our sample, as a function of their merging stage, we use the morphological classification in \citet{2018MNRAS.476.3661D}, hereafter DS18, which provides deep learning-based morphological classification for $\sim$670\,000 galaxies using SDSS data. DS18 provided a T-Type classification, in a continuous scale, related to the Hubble morphological type, after using the visual morphological classification from the \textit{Galaxy Zoo} project,\footnote{\texttt{\url{www.galaxyzoo.org}}} in particular from the Galaxy Zoo 2 \citep[GZ2,][]{2013MNRAS.435.2835W} catalogue and the morphological classification by \citet{2010ApJS..186..427N}, to train and test their deep learning models. The T-Type parameter ranges from -3 to 10, where T-Type~$\leq$~0 corresponds to early-type galaxies (i.e. elliptical and lenticular galaxies), and positive values to late-type galaxies, with T-Type~=~10 for irregular galaxies. 

We find that 134 galaxies in our sample have morphology classification in DS18. The remaining 3 galaxies are early-type, based on the visual revision of the optical SDSS images. We only consider if the galaxies are classified as late-type (including irregulars) or early-type (considering ellipticals and lenticulars) using DS18 morphology. DS18 also provided a parameter $\rm P_{merger}$ as the probability to present merger morphology, and the probability to be a lenticular (or S0) galaxy ($\rm P_{S0}$), where a value of $\rm P_{S0}$\,>\,0.5, for a galaxy with T-Type~$\leq$~0, indicates that it is a lenticular galaxy.


\subsection{Spectro-photometric SED fitting}
\label{sec:SED}

We combined photometric and spectroscopic data to derive the spectral energy distribution (SED) of the 137 galaxies in our sample. The SED of a galaxy is a powerful tool for constraining key physical properties of the unresolved stellar populations, as their star formation history (SFH). We used the Code Investigating
GALaxy Emission\footnote{\url{https://cigale.lam.fr/}} \citep[CIGALE;][]{2005MNRAS.360.1413B,2009A&A...507.1793N,2019A&A...622A.103B} which allows us to perform the spectro-photometric SED fitting. CIGALE models the SED of galaxies considering energy balance, i. e. taking into account the UV/optical dust attenuation, to derive star formation rate (SFR), recent star-formation, age, stellar masses, and dust attenuation in galaxies, among other physical properties and parameters of the star formation history (SFH), following a Bayesian analysis \citep[e.g.,][]{2011A&A...533A..93B, 2011A&A...525A.150G,2014A&A...571A..72B,2017A&A...608A..41C,2018A&A...613A..13Y,2022A&A...663A..50B}. The results of the spectro-photometric SED fitting for each spaxel allow us to explore the spatially resolved SFH through the SFR, and other properties, as the stellar mass.

\begin{figure*}
    \centering
    \includegraphics[width=0.31\textwidth, trim={1.5cm 0 0.8cm 0},clip]{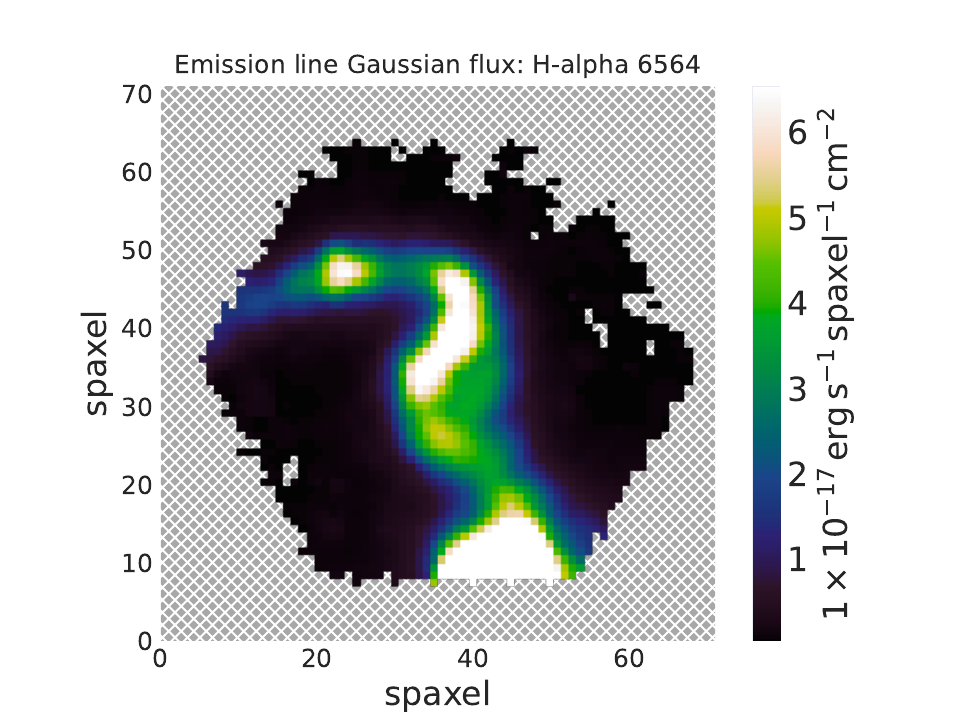} 
    \includegraphics[width=0.31\textwidth, trim={1.5cm 0 0.5cm 0},clip]{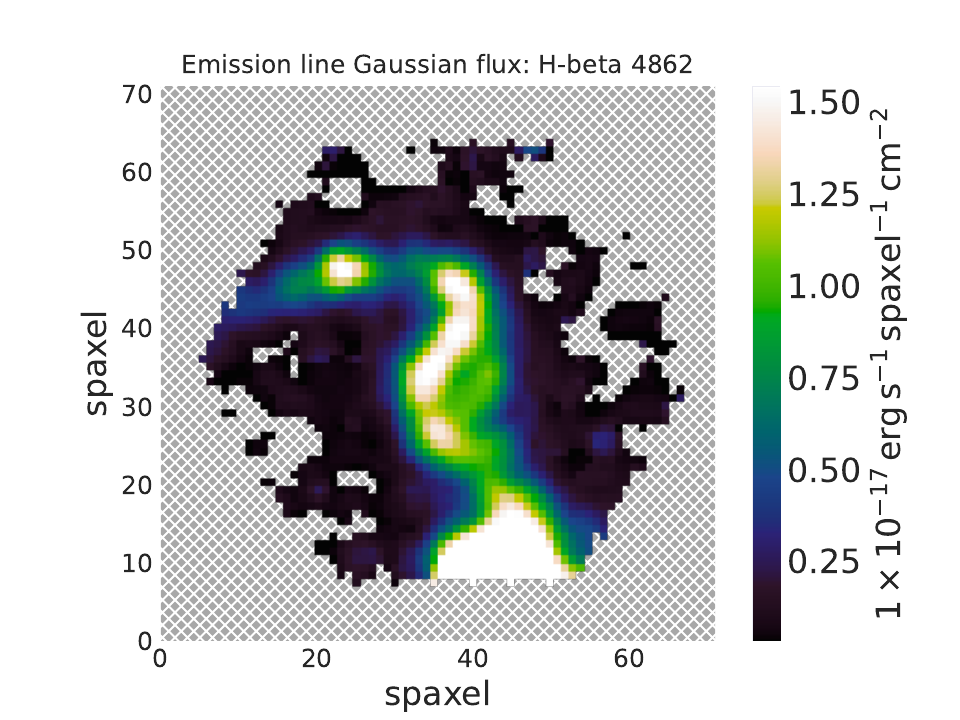} 
    \includegraphics[width=0.33\textwidth, trim={1cm 0 0 0},clip]{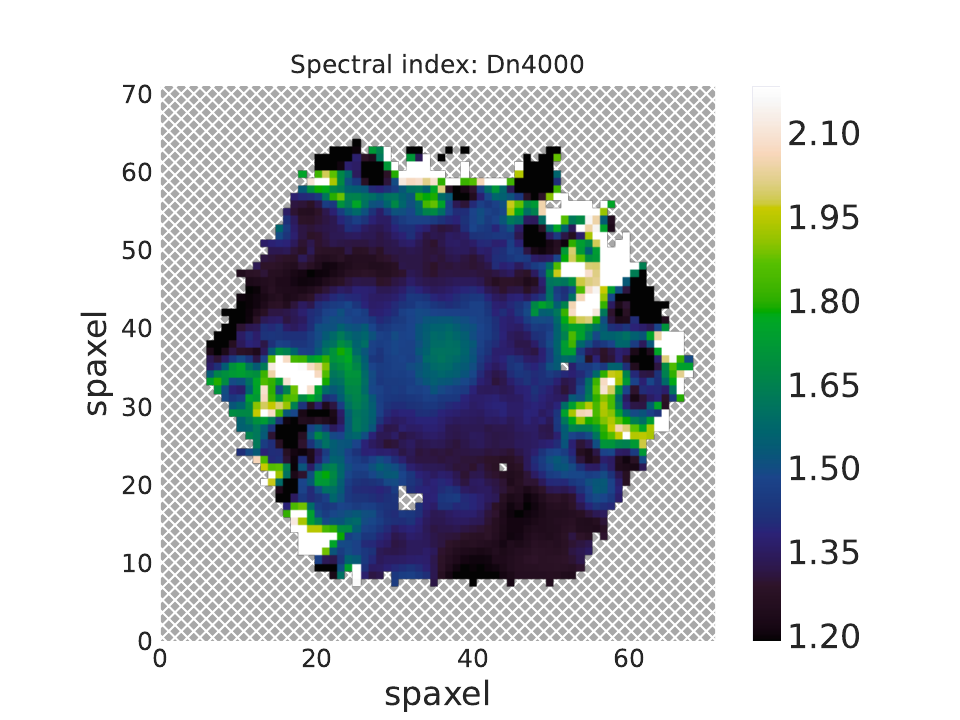} 
    \caption{MaNGA DAP products used as spectroscopic inputs in CIGALE to estimate physical properties from SED fitting for the galaxy 8241-12705 as example. From left to right: $H_{\alpha}$ emission line map, $H_{\beta}$ emission line map, and $D_{n}4000$ spectral index map.}
    \label{fig:dapsed}
\end{figure*}

\begin{figure}
    \centering
    \includegraphics[width=\columnwidth]{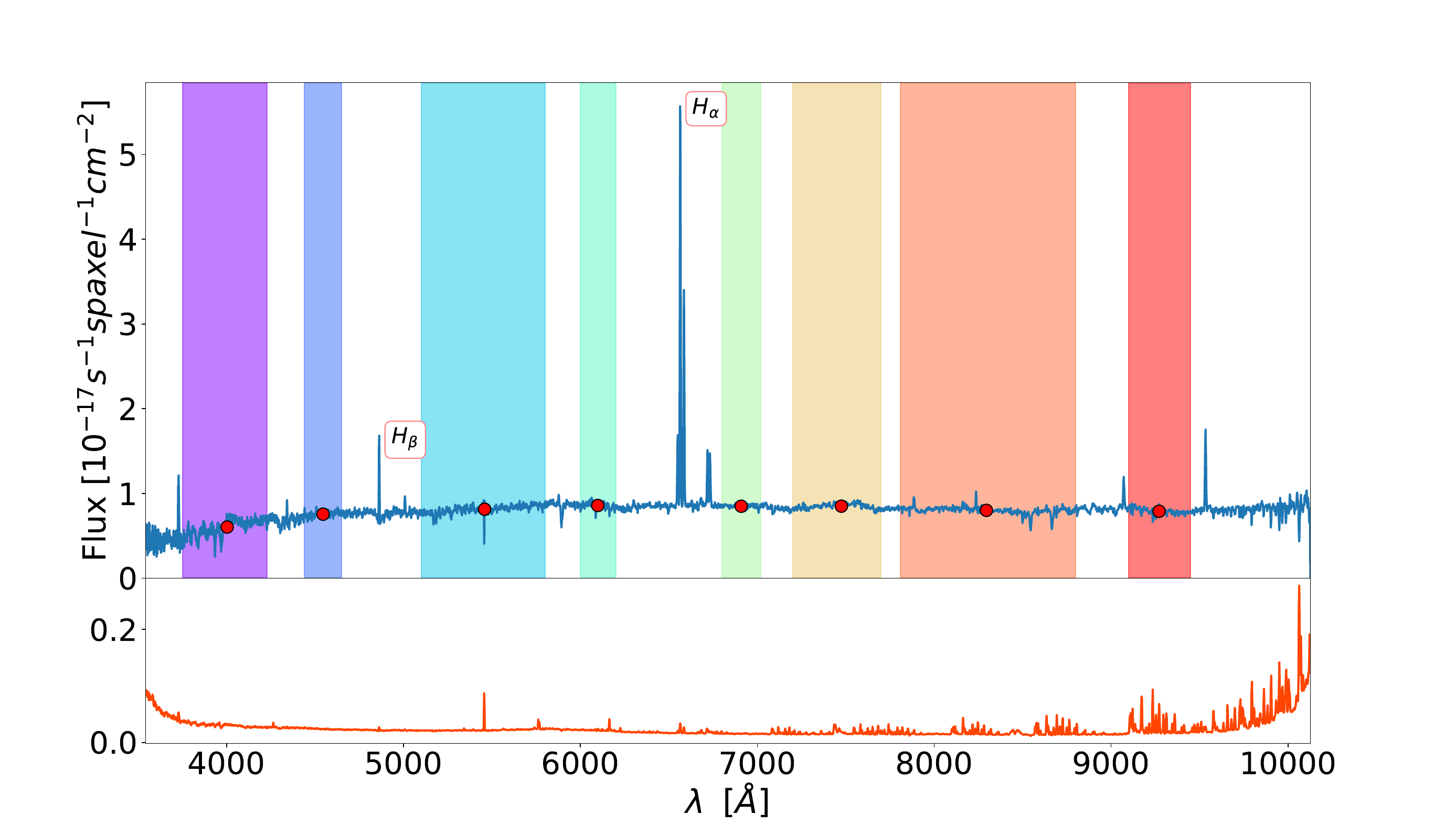} \\
    \includegraphics[width=\columnwidth]{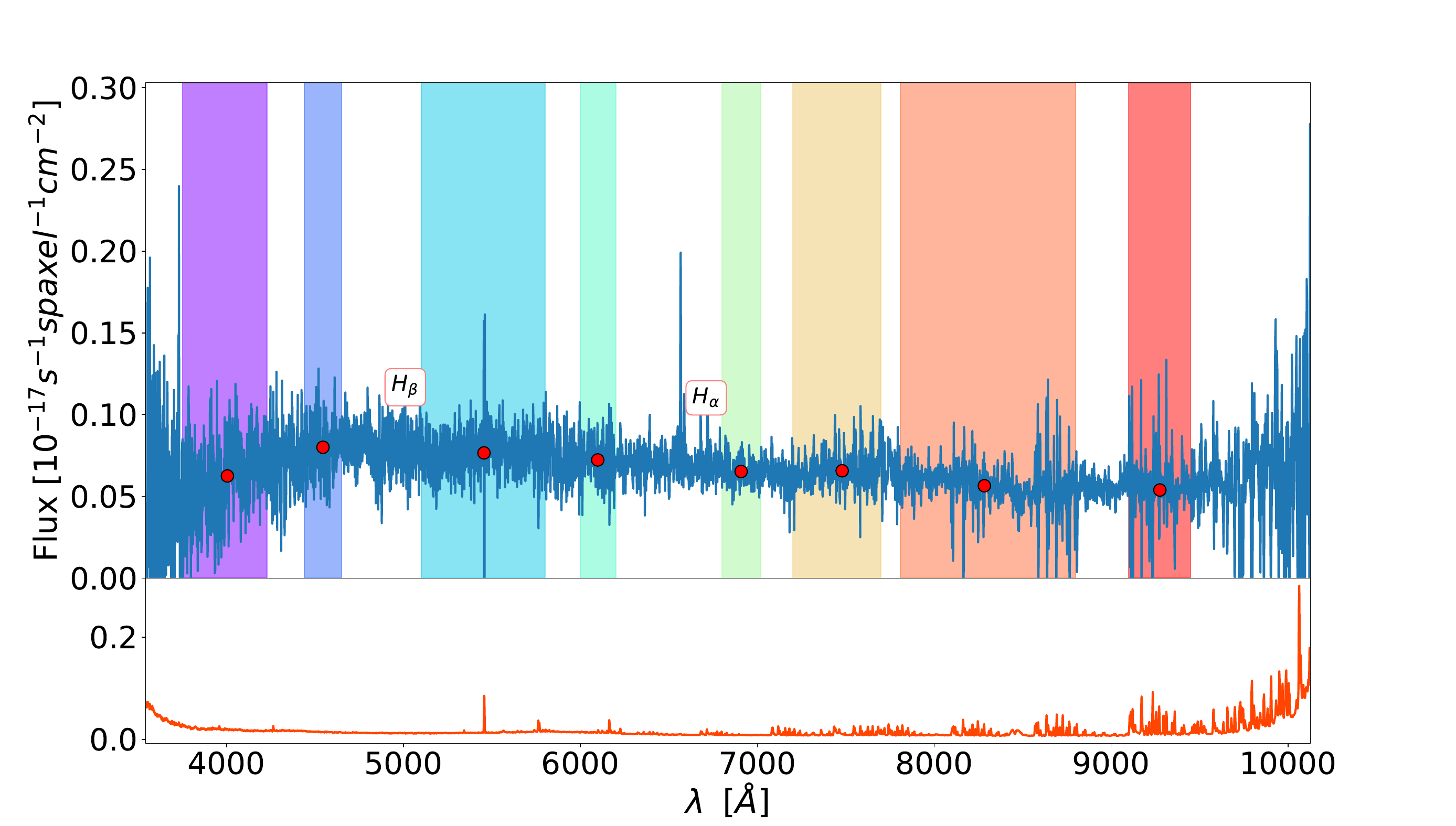} 
    \caption[Custom filters over a high and low SNR MaNGA spectra]{Custom filters over a high and low SNR MaNGA spectra. We show the spectra in two different spaxels for the MaNGA galaxy 12483-12704, one with high SNR (upper panel) and another with low SNR (lower panel) selected from the MaNGA DAP datacube, for comparison. In both panels, the spectra is represented in blue solid line covering the full range of the MaNGA spectra, from 3600 to 10000 \AA. The corresponding error is represented by orange line in the sub-panels. Colour regions show the custom filter wavelength range, with purple, blue, light blue, cyan, green, yellow, orange, red, corresponding to the M3992, M4542, M5446, N6097, N6908, O7473, O8281, and O9265 filters, respectively. The central wavelength corresponding to each filter is represented by red circles.}
    \label{fig:filters-spec}
\end{figure}

\begin{figure*}
\centering 
    \includegraphics[width=0.23\textwidth]{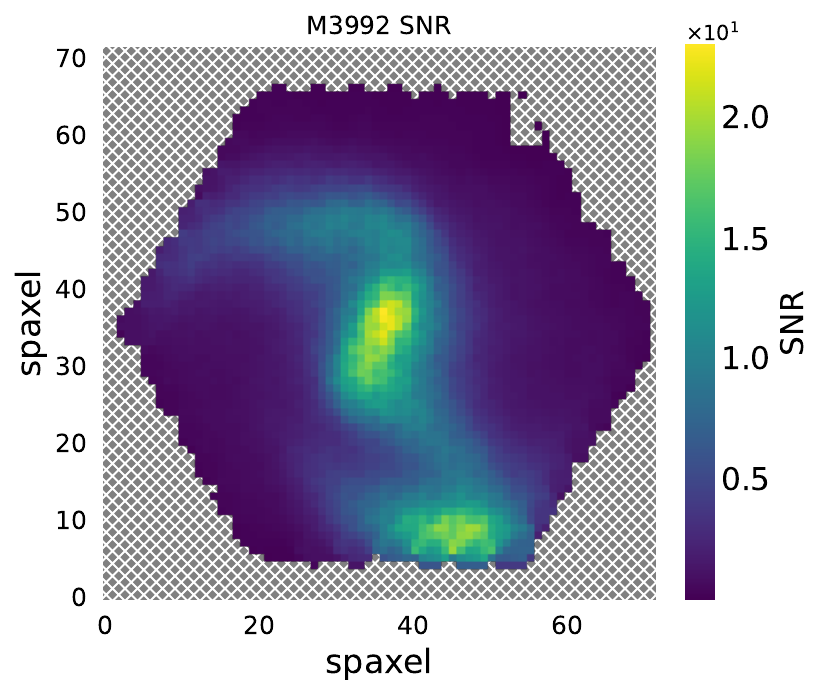}
    \includegraphics[width=0.23\textwidth]{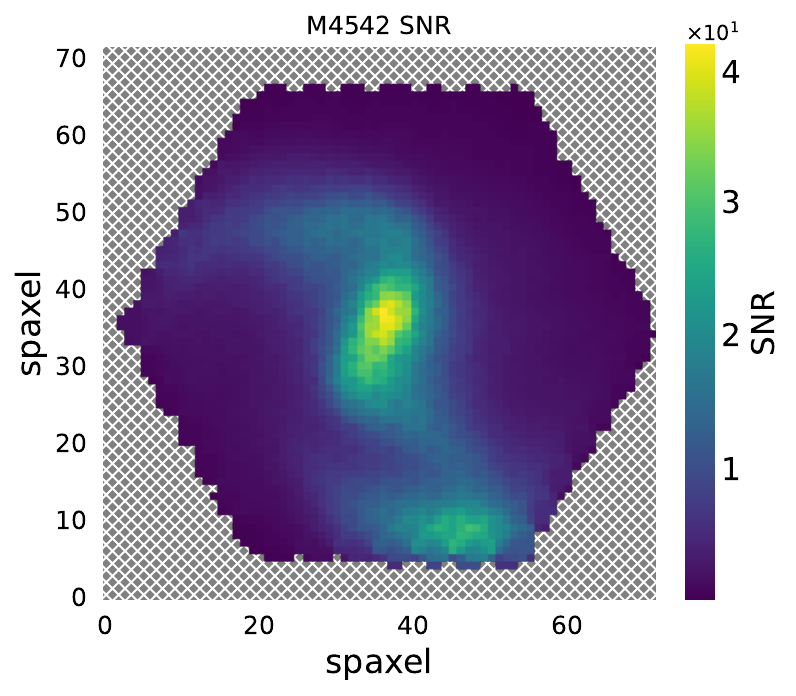}
    \includegraphics[width=0.23\textwidth]{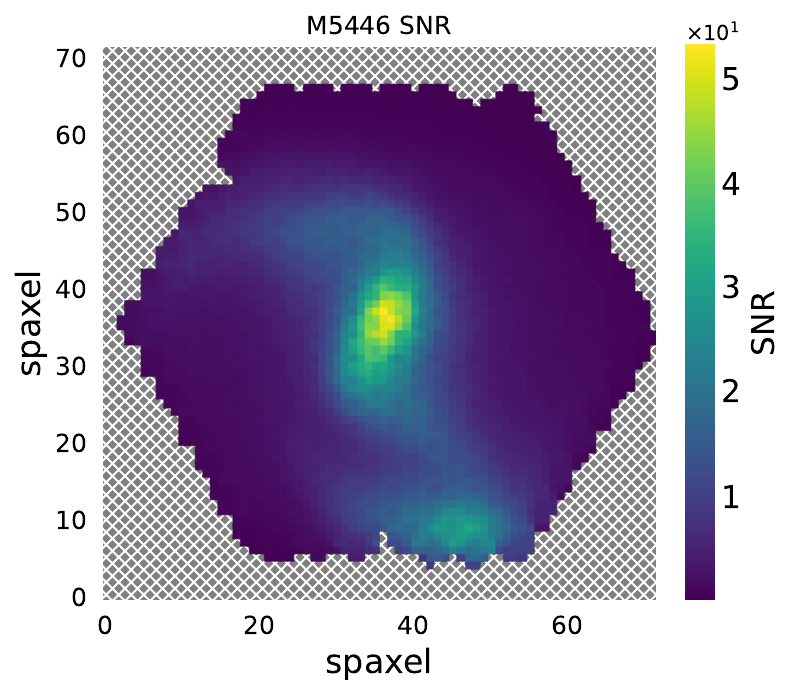}
    \includegraphics[width=0.23\textwidth]{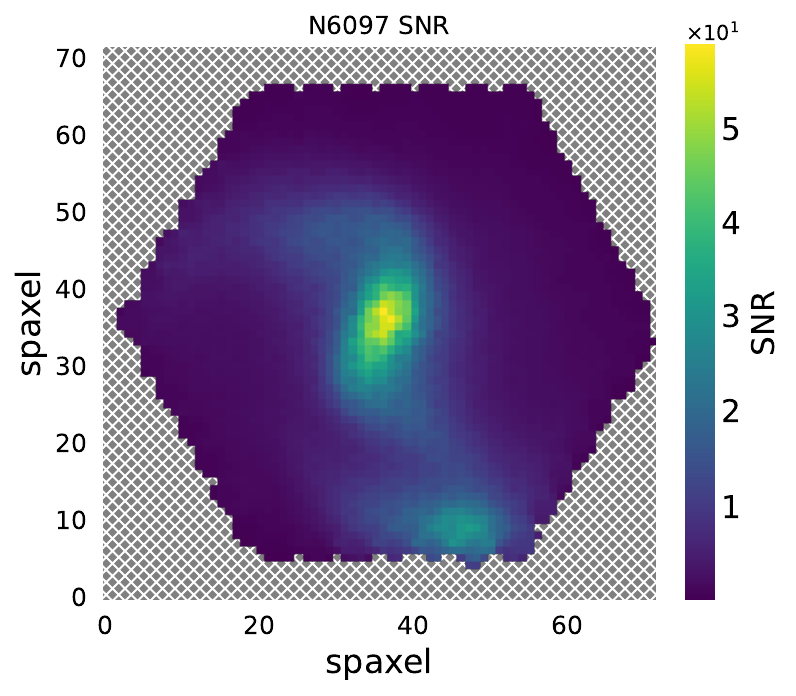}\\
    \includegraphics[width=0.23\textwidth]{N6097_SNR.pdf}
    \includegraphics[width=0.23\textwidth]{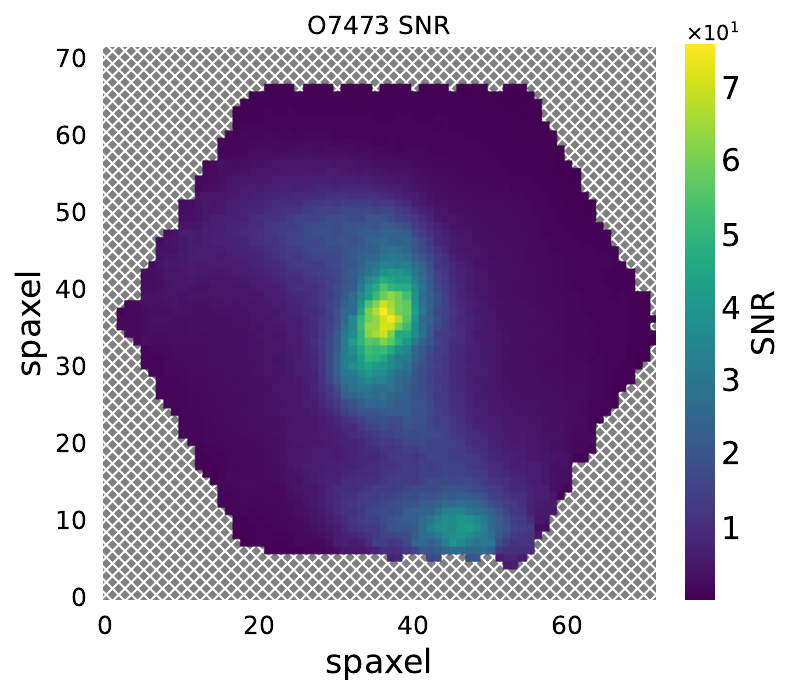}
    \includegraphics[width=0.23\textwidth]{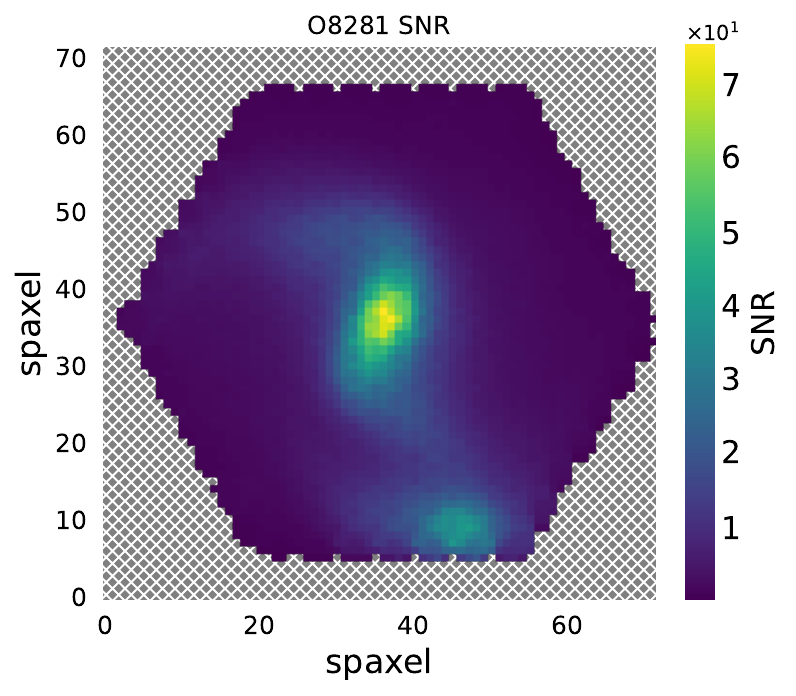}
    \includegraphics[width=0.23\textwidth]{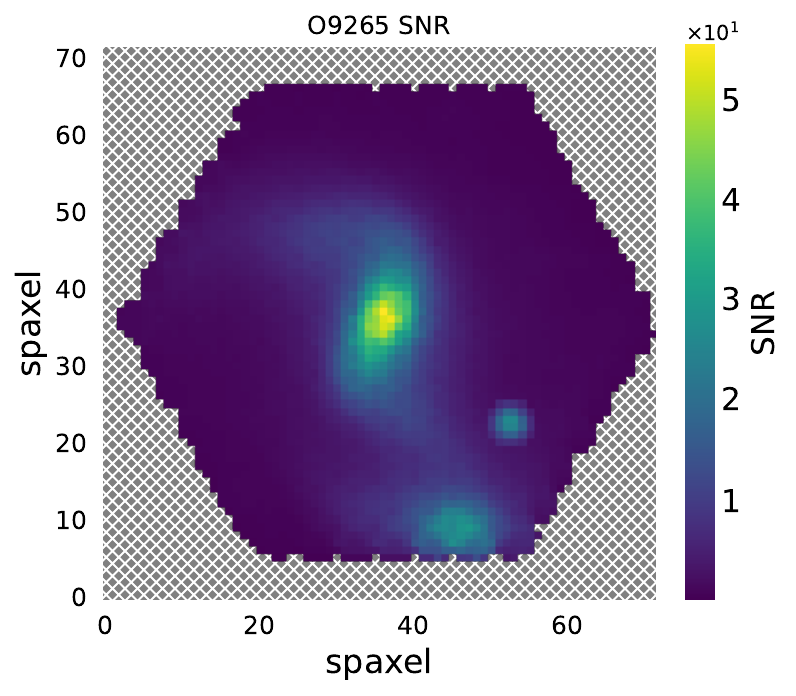}
    \caption[Signal-to-noise ratio maps for each custom filter.]
    {Signal-to-noise ratio maps for each custom filter for the merger galaxy 8241-12705. From upper left to lower right, SNR maps for filters M3992, M4542, M5446, N6097, N6908, O7473, O8281, and O9265, respectively.}
    \label{fig:snr}
\end{figure*}

To get the required CIGALE input data we use the Marvin tool. From the extensive MaNGA data products, we selected the maps of the $H_{\alpha}$ and $H_{\beta}$ emission lines, as well as the $D_{n}4000$ break spectral index, as defined in \citet{1999ApJ...527...54B}, for a more accurate estimation of the physical properties. Figure~\ref{fig:dapsed} shows these products for an example merger galaxy, 8241-12705, in our sample. In combination with spectroscopy, we defined custom photometric filters to account for the continuum, avoiding the main emission lines in the optical range of the MaNGA spectra, for the computation of the stellar mass. We defined our filters using a rectangle response function with a width providing enough signal to noise ratio, even in the external regions of galaxies covered by the MaNGA FoV. The information about each filter is summarised in Table~\ref{tab:filters}. An example of the selected filters on a random selected high and low S/N ratio spaxel in the FoV of the merger galaxy 12483-12704 is shown in the upper and lower panels of Fig.~\ref{fig:filters-spec}, respectively. Fig.~\ref{fig:snr} shows a map of the S/N ratio for each filter for the same galaxy.

\begin{table}
    \centering
    \setlength{\arrayrulewidth}{0.1mm}
    \begin{tabular}{ccc}
    \hline\hline
      Filter & Wavelength range & Central wavelength\\ 
       & (Å) & (Å) \\
    \hline
    M3992 & 3750 - 4230 & 3992\\
    M4542 & 4440 - 4650 & 4542\\
    M5446 & 5100 - 5800 & 5446\\
    N6097 & 6000 - 6200 & 6097\\
    N6908 & 6800 - 7020 & 6908\\
    O7473 & 7200 - 7700 & 7473\\
    O8281 & 7810 - 8800 & 8281\\
    O9265 & 9100 - 9450 & 9265\\
    \hline
     \end{tabular}
    \caption[List of the custom filters defined for the SED fitting.]{List of the custom filters defined for the SED fitting. The columns correspond to: (1) filter name; (2) minimum and maximum wavelength considered in each filter, in Angstroms; (3) effective wavelength within the considered wavelength range, commonly known as central wavelength, in Angstroms.}
    \label{tab:filters}
\end{table}

We chose a delayed SFH that considers a recent burst/quench episode as defined in \citet{2017A&A...608A..41C}. This is the simplest SFH model for interacting galaxies as it easily accounts for the presence of a recent variation in the SFR due to a starburst episode or a rapidly transition to a quench state \citep{2016A&A...585A..43C}. This module is provided in CIGALE as \texttt{sfhdelayedbq} \citep{2019A&A...622A.103B} following the equation:

\begin{equation}
    SFR(t)\propto \left\{\begin{matrix}
    \quad\frac{t}{\tau^2} \times \exp{-t/\tau}\qquad t\le t_0\\ 
    \;\;r_{SFR}\times SFR(t=t_0)\quad t> t_0
    \end{matrix} \right. \;\;\;,
\end{equation} 

where $t_0$ is the age when is allowed, a rapid increment/decrease in the star formation history, $r_{SFR}$ is defined as the ratio of the SFR after/before the $t_0$, and $\tau$ represent the time at which the SFR peaks.


In the creation of the SED models for the SFH, we used the BC03 models on single stellar populations by \citet{2003MNRAS.344.1000B}, considering a Salpeter initial mass function (IMF) \citet{1955ApJ...121..161S}, with a stellar metallicity value of Z=0.02 (solar metallicity) for the modelling of an unattenuated stellar emission, and continuum and line nebular emission. We considered a modified \citet{2000ApJ...533..682C} starburst attenuation model as \texttt{dustatt\_modified\_staburst}, with a \citet{2002ApJS..140..303L} curve  extended between the Lyman break and 150 nm, and with a flexible attenuation curve \citep{2009A&A...507.1793N}. 


Since we do not have data in the mid- or far-infrared (i. e., no band sensitive to dust emission), and to speed up the computation time of the spaxel-by-spaxel SED fitting, we do not select any dust emission model in the computation of the modelled SEDs. Note that, in addition to the synthetic bands, we also fit H$_\alpha$, H$_\beta$ and D4000. In this case CIGALE uses H$_\alpha$/H$_\beta$ as a tracer of the attenuation, and therefore it serves a similar role as the dust would in the construction of the modelled spectra. A total of 1749600 SED models were created using the parameters and values summarised in Table~\ref{tab:SEDparams}. With this configuration, the mean computational time to perform the spaxel-by-spaxel SED fitting is $\sim$\,2\,hours/galaxy \footnote{Using the PROTEUS supercomputer of the Institute Carlos I for Theoretical and Computational Physics (iC1).} (for about 6850 total spectra for 137 galaxies).

\begin{table*}
\centering
\begin{tabular}{ccc}
    \hline \hline
    Model & Parameter & value\\ 
    \hline
    SFH & $\rm Age$ (Myr) & 11000, 12000, 13000 \\ 
     & $\tau_{\rm main}$ (Myr) & 1000, 3000, 5000, 7000, 9000 \\
     & $\rm Age_{bq}$ (Myr) & 20, 50, 100, 300 \\
     & $r_{\rm SFR}$ & 0, 0.25, 0.5, 0.75, 1, 1.5, 2, 5, 10 \\ 
    \hline
    Dust attenuation & $\rm E(B-V)_{\rm lines}$ (mag) & 0.005, 0.01, 0.025, 0.05, 0.075, 0.10, \\ & &  0.15, 0.20, 0.25, 0.30, 0.35, 0.40, \\ & & 0.45, 0.50, 0.55, 0.60, 0.65, 0.70 \\
     & $\rm E(B-V)_{\rm factor}$  & 0.25, 0.5, 0.75 \\
      & UV bump wavelength (nm) & 217.5\\
      & UV bump width (nm) & 35.0 \\
      & UV bump amplitude & 0.0, 1.5, 3.0 \\
& $\Delta \delta$  & -1.2, -1.1, -1.0,-0.9,-0.8,-0.7,-0.6,-0.5,\\ & &  -0.4,-0.3,-0.2,-0.1, 0,0.1,0.2 \\ 
    \hline
\end{tabular}
\caption[Parameters used in CIGALE to model the SFH]{Parameters used in CIGALE to model the SFH (upper rows) and dust attenuation (lower rows). Meaning of the parameters: a) SFH: $\rm Age$ --  Age of the main stellar population in the galaxy, in Myr; $\tau_{\rm main}$ -- $e$-folding time of the main stellar population model, in Myr; $\rm Age_{bq}$ -- Age of the burst/quench episode, in Myr; and $r_{\rm SFR}$ -- Ratio of the SFR after/before $\rm Age_{bq}$, values larger than one correspond to an enhancement of the SFR whereas values lower than one will correspond to a decrease; b) Dust attenuation: $E(B-V)_{\rm lines}$ -- Colour excess of the nebular lines light for both the young and old population; $E(B-V)_{factor}$ -- Reduction factor to apply on $\rm E(B-V)_{\rm lines}$ to compute $E(B-V)_{s}$ the stellar continuum attenuation. Both young and old populations are attenuated by $E(B-V)_{s}$; UV bump wavelength -- Central wavelength of the UV bump in nm; UV bump width -- Width (FWHM) of the UV bump in nm; UV bump amplitude -- Amplitude of the UV bump. For the Milky Way: 3; $\Delta \delta$ -- Slope delta of the power law modifying the attenuation curve.}
    \label{tab:SEDparams}
\end{table*}

\subsection{Analysis of the nuclear activity}
\label{sec:WHAN}

For an analysis of the nuclear activity in galaxies, it is useful to use diagnostic emission line diagrams. These are an accurate empirical method to classify galaxies in terms of their emission in relation with the different mechanisms for gas ionisation. The most commonly used diagnostic diagrams for nearby galaxies are the BPT diagrams \citep{1981PASP...93....5B,2006MNRAS.372..961K}. Based on emission line ratios [O III]/$H_{\beta}$ versus [N II]/$H_{\alpha}$, [S II]/$H_{\alpha}$, and [O I]/$H_{\alpha}$, BPT diagrams allow to classify galaxies into galaxies with a star forming nuclei (SFN), AGN Seyfert galaxies or low-ionization (nuclear) emission-line galaxies, or LI(N)ERs, and passive galaxies. 

MaNGA spectroscopy allows to reproduce spatially resolved diagnostic diagrams, considering the information from each spaxel. However, using BPT diagrams, the spectra only becomes classified if it meets the criteria in all three diagrams. Even selecting the most relaxed criterion ([O III]/$H_{\beta}$ versus [N II]/$H_{\alpha}$), it depends on the emission in four lines, where [O III] and $H_{\beta}$ might be weak, limiting the area where we could analyse the galaxy. A way to overcome this issue is using  the [N II]/$H_{\alpha}$ WHAN diagram \citep{2011MNRAS.413.1687C}, which is able to analyse weak line galaxies, providing a more complete view of the diverse activity at different regions of the MaNGA galaxies that could not be explored using the BPT diagram due, for example, to the absence of absorption lines. 

The WHAN diagnostic diagram identifies the weak AGNs from fake AGNs, named as retired galaxies (RGs) from LI(N)ERs, allowing a more complete analysis of regions where the heating of the ionised gas is the result of old stars, rather than star-formation or AGN activity. We can also classify stellar activity into pure star-forming (PSF) or passive galaxies (PG), slightly different from the RGs, where PG is a closely related with RGs, being a extreme case of stellar activity. \citet{2011MNRAS.413.1687C} classified the different categories respect to values of $\log[NII]/H_{\alpha}$, $W_{H_{\alpha}}$, and $W_{NII}$, in the following way:

\begin{itemize}
    \item Pure star-forming galaxies (PSF): $\rm log[N II]/H_{\alpha}\,<\,-0.4$ and $W_{H_{\alpha}}\,>\,3\,\AA$;
    \item Strong AGN (sAGN): $\rm log[N II]/H_{\alpha}\,>\,-0.4$ and $\rm W_{H_{\alpha}}\,>\,6\,\AA$;
    \item Weak AGN (wAGN): $\rm log[N II]/H_{\alpha}\,>\,-0.4$ and $\rm 3\,<\,W_{H_{\alpha}}\,<\,6\,\AA$;
    \item Retired galaxies (RG): $\rm W_{H_{\alpha}}\,<\,3\,\AA$;
    \item Passive galaxies (PG): $\rm W_{H_{\alpha}}\,< \,0.5\,\AA$ and $\rm W_{NII}\,< \,0.5\,\AA$.
\end{itemize}

\begin{figure}
    \centering
    \includegraphics[width=\columnwidth]{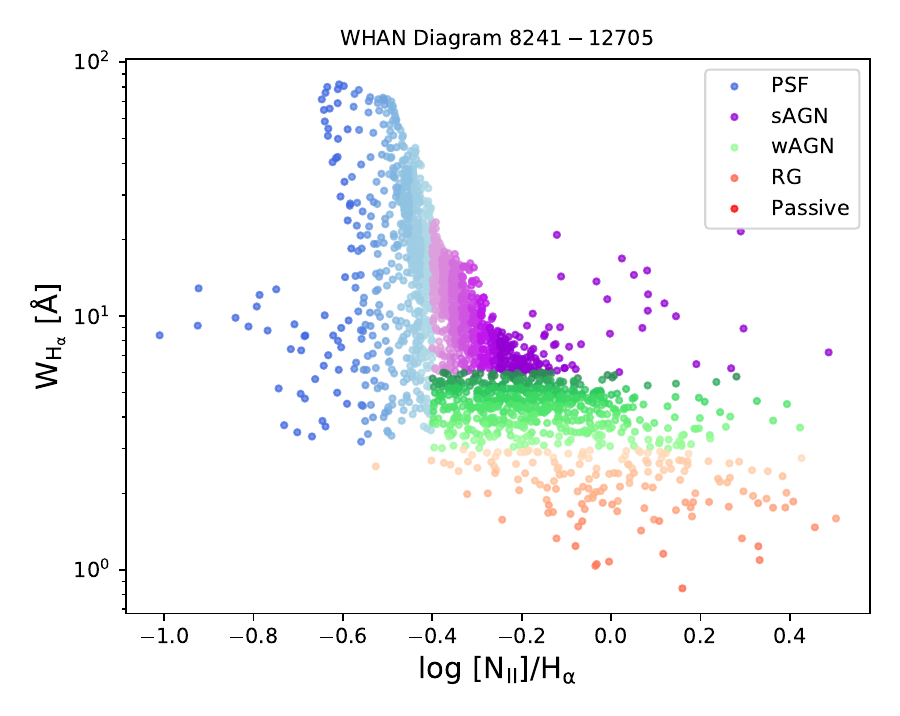} \\
    \includegraphics[width=\columnwidth]{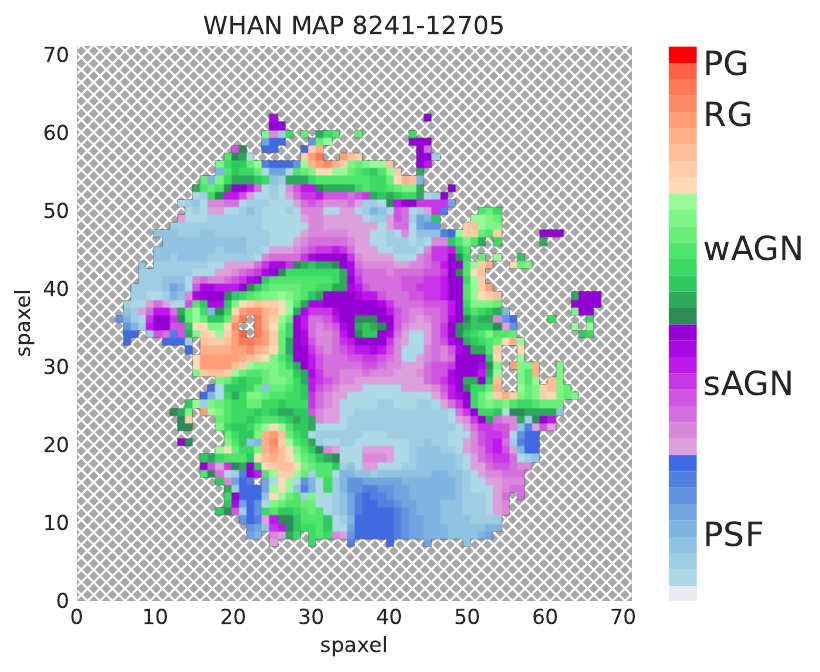} 
    \caption{Spatially resolved WHAN diagnostic diagram (upper panel) and map (lower panel) for the merger galaxy 8241-12705. Each colour, and color gradient, correspond to a WHAN category according to the legend and colorbar, respectively. The WHAN diagram and map has been created using the public tool for the visualization of the spatially resolved WHAN diagnotic diagram for galaxies in the MaNGA survey described in Appendix~\ref{sec:WHAN-tools}.} 
    \label{fig:whan-example}
\end{figure}

Both, BPT and WHAN diagrams can be computed using the Marvin tools. In particular, we have developed the visualisation tool described in Appendix~\ref{sec:WHAN-tools}, which uses the functionality already implemented in Marvin, but it is specifically designed to even explore the intermediate areas between adjoining WHAN categories, allowing us to investigate transition processes in the diverse stellar population among different regions of the galaxies. We show the WAHN diagnostic diagram and map for galaxy 8241-12705 as example in Fig.~\ref{fig:whan-example}. In combination with the spatially resolved SFR from SED fitting, we used the resolved WHAN diagrams to investigate if there is any relation between AGN activity and star-formation in our sample of close pairs/mergers. The level of agreement between the star formation from SED fitting and WHAN diagrams is presented in Sect.\ref{Sec:stellarprop}.

\section{Results}
\label{Sec:merger-res}

\subsection{Spatially resolved stellar populations} 
\label{Sec:stellarprop}

\begin{figure*}
    \centering
    \includegraphics[width= .3\textwidth]{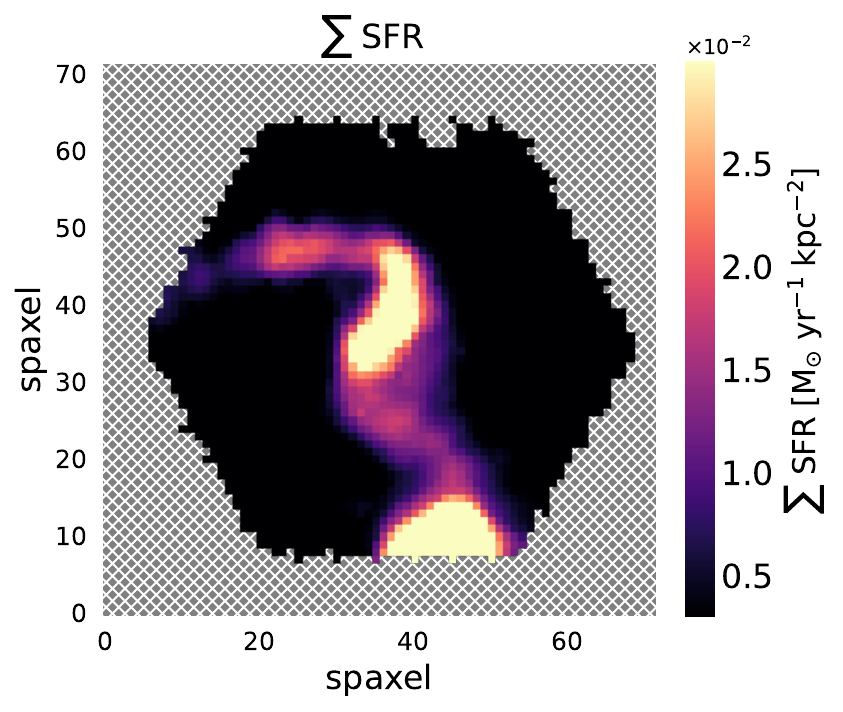} 
    \includegraphics[width= .3\textwidth]{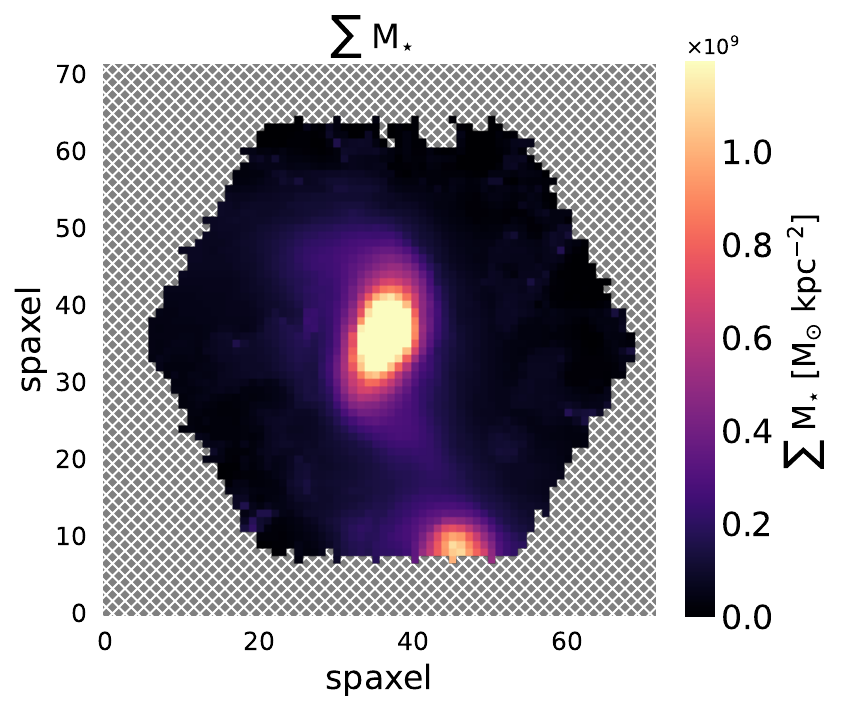} 
    \includegraphics[width= .3\textwidth]{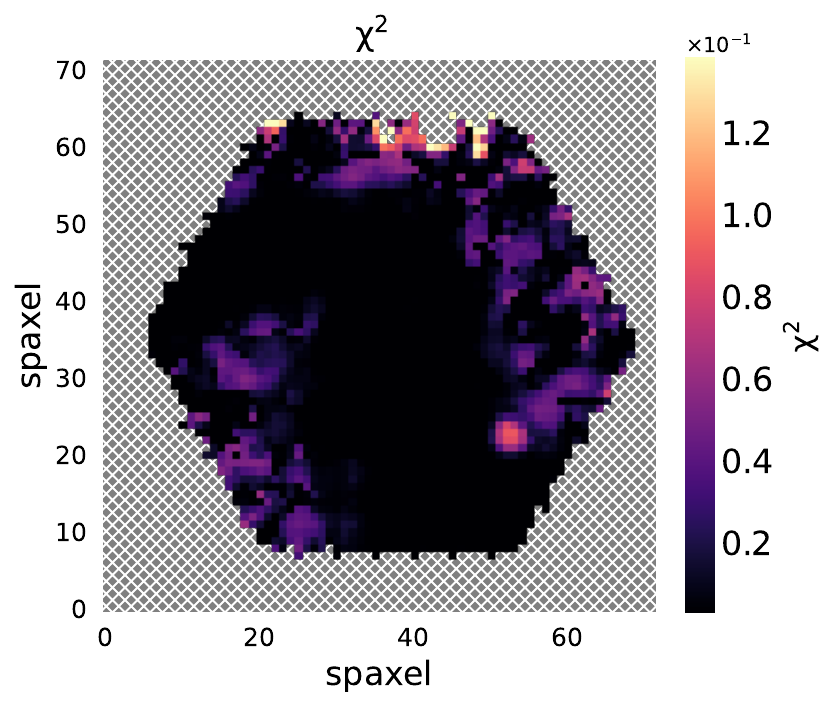} 
    \caption{Spatially resolved $\Sigma$\,SFR (left panel), $\Sigma$\,M$_\star$ (middle panel), and $\chi^2$ (right panel) maps from the results of the spectro-photometric SED fitting with CIGALE for the galaxy 8241-12705. The scale of the $\chi^2$ map is normalised to the maximum value.}
    \label{fig:SED-example}
\end{figure*}

We used the results from the spectro-photometric SED fitting with CIGALE to create SFR and stellar mass surface densities maps ($\Sigma \;SFR$ and $\Sigma\;
 M_\star$, respectively). 
We also created maps of the resulting $\chi^2$, which provides a view of the statistical distribution of the goodness of the fit of the models we created to the observations \citep{2019A&A...622A.103B}, so we can identify if there is any problem with the fitting. We can detect if the maps we are getting are reliable (lower $\chi^2$ values and no doubtful structures are shown in the maps), or if we need to increase the parameter space when creating the models. An example of the maps generated from the results of the SED fitting are shown in Fig.~\ref{fig:SED-example}. In comparison with the S/N maps in Fig.~\ref{fig:snr}, the spaxels with the smallest $\chi^2$ values (<\,0.02) have typically S/N~20. An example of the SED fitting products, among others, for a galaxy in each merger stage is also shown in the Appendix~\ref{sec:products}. The resulting SFR and M$_\star$ from SED fitting will be used to construct the SFR-$M_{\star}$ diagram for galaxies in our sample, as a function of the merging stage, as explained in the next section. In addition, to explore stellar population properties as age or nuclear activity, we used the $D_n$(4000) parameter and the results of the WHAN diagrams for all galaxies in our sample. 

\begin{figure}
    \centering
    \includegraphics[width=\columnwidth, trim={1.5cm 1cm 1.cm 0},clip]{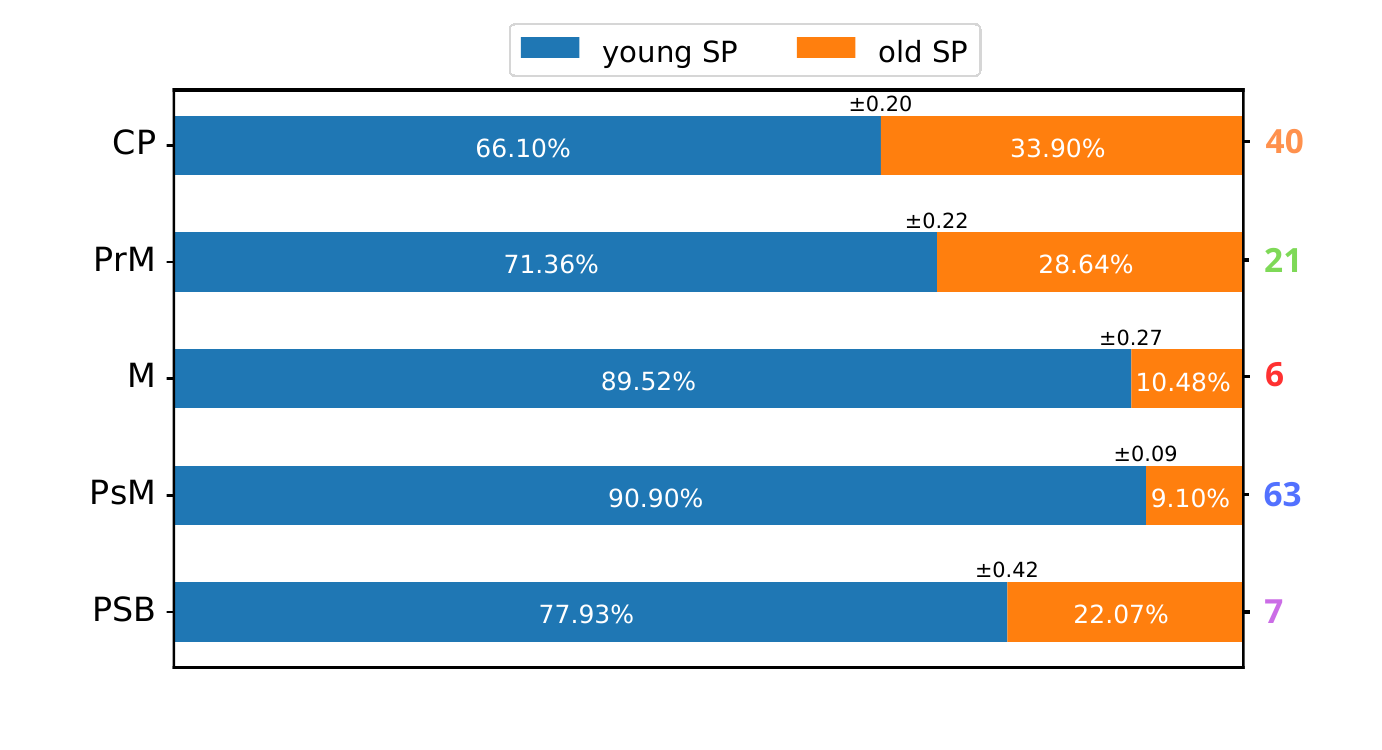}  
    \caption[]{Fraction of young and old spaxels for galaxies in  each merger stage. We use the $D_n$(4000) parameter and the divisory value at $D_n$(4000)~=~1.67 \citep[as in][]{2006MNRAS.370..721M} to separate between young ($D_n$(4000)\,$\leq$\,1.67, in blue) and old ($D_n$(4000)\,$>$\,1.67, in orange) spaxels. The number of galaxies in each merger stage are indicated in the right side, with colored text as in Fig.~\ref{fig:d4000radial}.}
    \label{fig:mean-age}
\end{figure}

\begin{figure}
    \centering
    \includegraphics[width=\columnwidth, trim={1.5cm 0.5cm 2.5cm 1.5cm},clip]{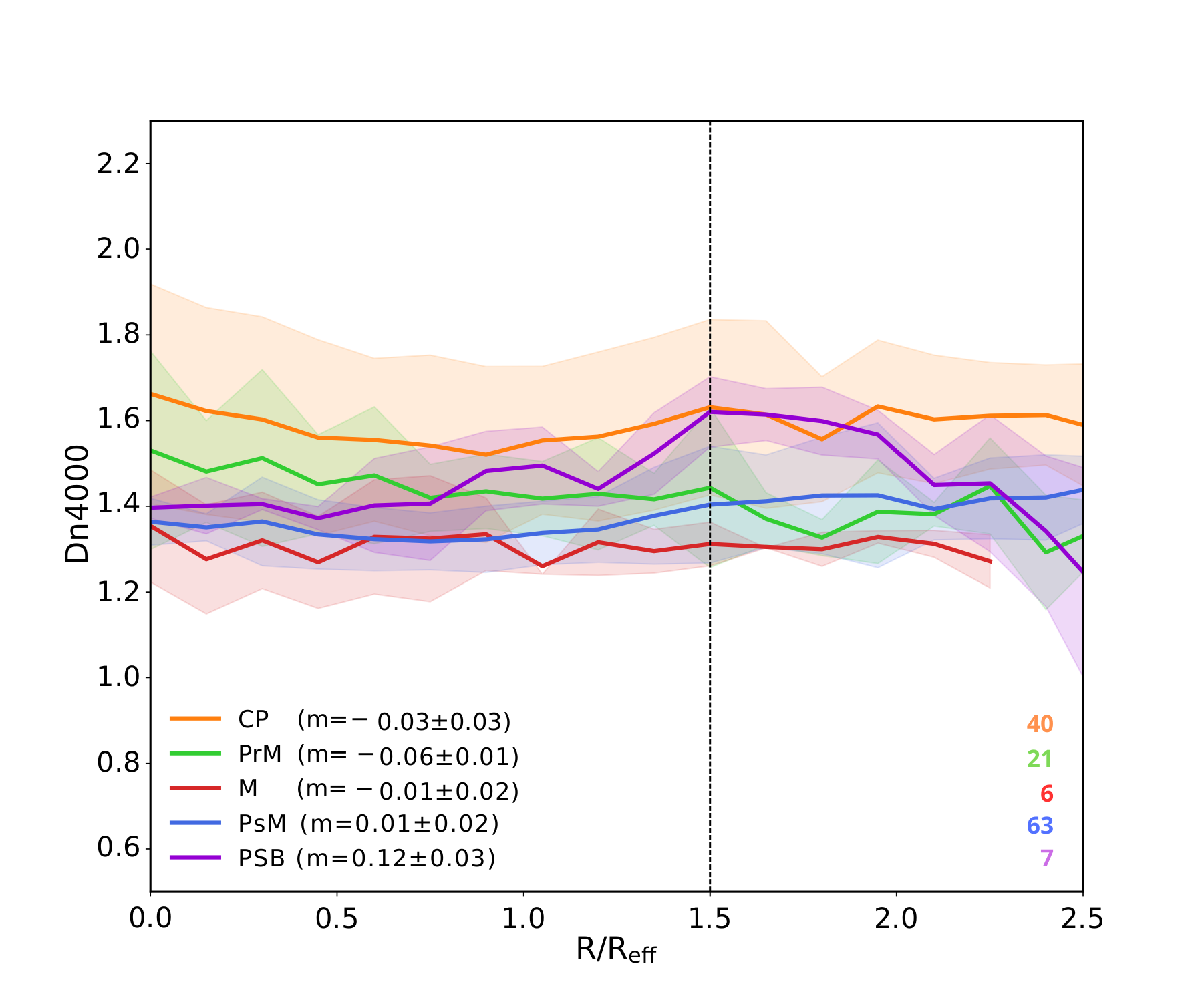}
    \caption[Dn4000 mean profile]{Mean radial profiles of the $D_n$(4000) parameter with respect to $R/R_{eff}$  for all the galaxies in each merger stage. The radial profiles are represented in different colours for each merger stage according to the legend. Uncertainties are represented in the same colours at 1$\sigma$ level. The slope, m, of the linear fit to the mean radial profile up to 1.5$R/R_{eff}$ (marked with a black dashed vertical line), with its corresponding uncertainties, are included in the legend for each merger stage. The number of galaxies in each merger stage are indicated in the lower right corner, with colored text as in the legend.}
    \label{fig:d4000radial}
\end{figure}

Besides using the $D_n$(4000) maps from the MaNGA Data Analysis Pipeline (DAP) as an input for CIGALE, we also use this parameter as the stellar population age indicator, which is small for younger stellar populations, and large for older stellar populations. We use the divisory value at $D_n$(4000)~=~1.67 following \citet{2006MNRAS.370..721M}. Figure~\ref{fig:mean-age} shows the mean fraction of young ($D_n$(4000)\,$\leq$\,1.67) and old ($D_n$(4000)\,$>$\,1.67) spaxels for each merger stage. We observe that, in general, merger and post-merger galaxies have a larger fraction of spaxels with young stellar population, even galaxies in the PsM-PSB category. In addition, Fig.~\ref{fig:d4000radial} shows the mean radial profile of $D_n$(4000) for each merger stage, with the results of the linear fit within 1.5\,R/R$_{eff}$\footnote{Most of the galaxies in this study (79\%) belong to the MaNGA primary sample, while 21\% belong to the secondary sample.}.

\begin{figure}
    \centering
    \includegraphics[width=\columnwidth, trim={1.5cm 1.2cm 1.5cm 1.5cm},clip]{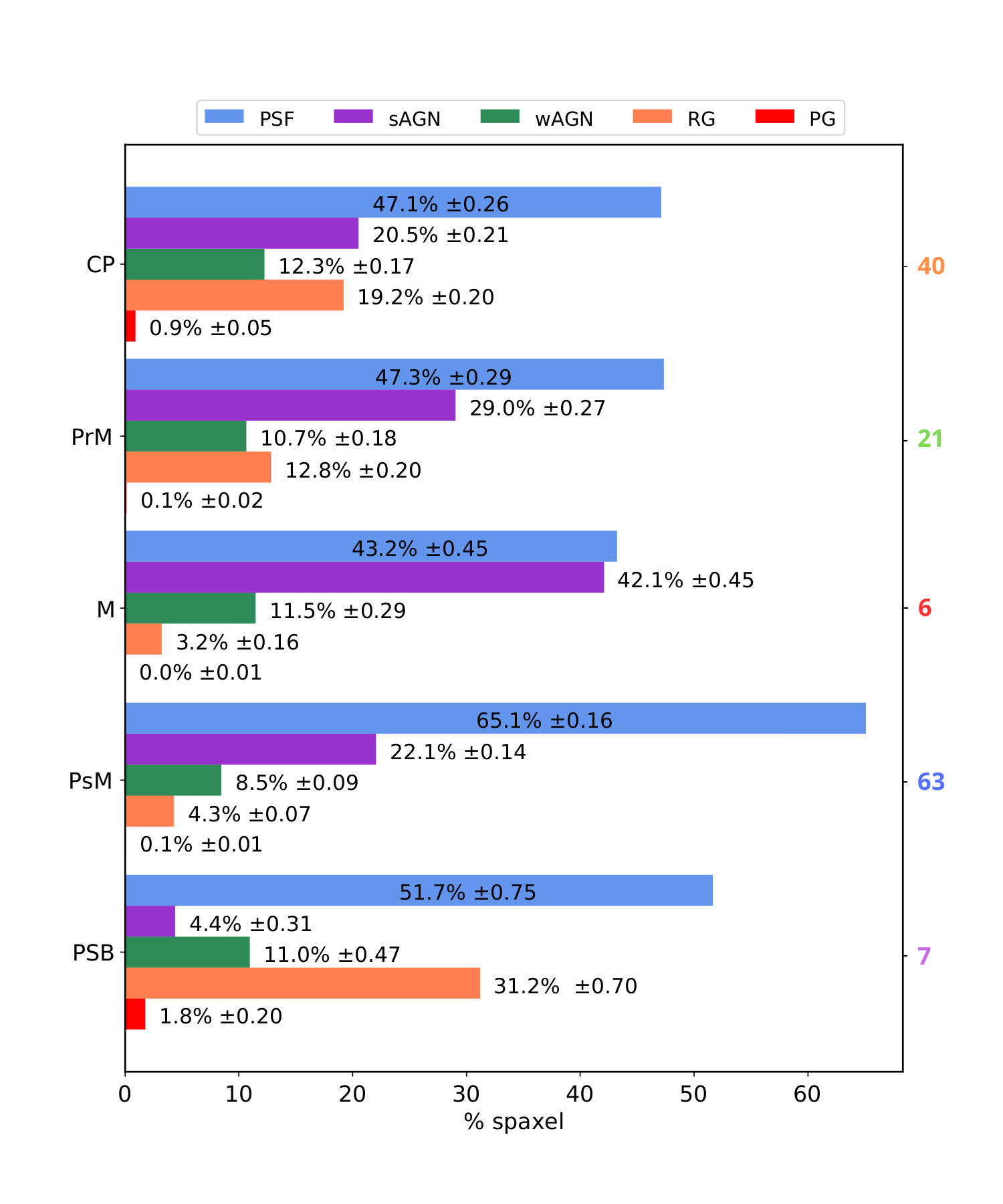}  
    \caption[Percentage of spaxels for each WHAN category for each merger stage]{Percentage of spaxels for each WHAN category for each merger stage. Blue colour bars represent the fractions of pure star-forming (PSF) spaxels, purple colour bars represent the fractions of strong AGN (sAGN) spaxels, green colour bars represent the mean fractions of the weak AGN (wAGN) spaxels, orange colour bars represent the fraction of spaxels classified as retired galaxy-like emission (RG), and red colour bars represents the fraction of spaxels classified as passive galaxy-like emission (PG), according to the WHAN diagnostic diagram. The number of galaxies in each merger stage are indicated in the right side, with colored text as in Fig.~\ref{fig:d4000radial}.}
    \label{fig:whan-mean-fraction}
\end{figure}

Similarly, Fig.~\ref{fig:whan-mean-fraction} shows the percentage of spaxels in each WHAN category, for each merger stage. PsM and PsM-PSB galaxies show the largest fraction of PSF spaxels ($\sim$\,65.1\% and $\sim$\,51.7\%) in comparison to M ($\sim$\,43.2\%), CP ($\sim$\,47.2\%), and PrM ($\sim$\,47.3\%). In the merger category, the percentage on AGN spaxels (sAGN + wAGN) is $\sim$\,10\% higher to the percentage of PSF spaxels (43.2\% versus 53.6\%). In comparison, the fractions of PSF, AGN, and RG spaxels in CP galaxies is about one third in each case. The spatially resolved WHAN diagrams, and WHAN emission maps, for a galaxy in each merger stage are also shown in the Appendix~\ref{sec:products}. 

Figure~\ref{fig:ssfr-whan} shows the level of agreement between the star formation obtained from SED fitting and from WHAN diagrams. The distribution of the specific star-formation rate (sSFR, defined as SFR/$M_\star$) surface density for spaxels corresponding to each WHAN category has some level of overlap. However, in general, PSF classified spaxels present highest values of the sSFR surface density, followed by the sAGN, wAGN, and RG classifed spaxels, respectively. For all merger stages, the median values of the sSFR surface density are discernible at least at one sigma level for sAGN and RG WHAN classifications, while there is a large overlap for PSF and sAGN spaxels, specially for galaxies in the merger and post-merger stages. This is understandable considering the mix of processes occurring on galaxies at these stages.

\begin{figure}
    \centering
    \includegraphics[width=\columnwidth]{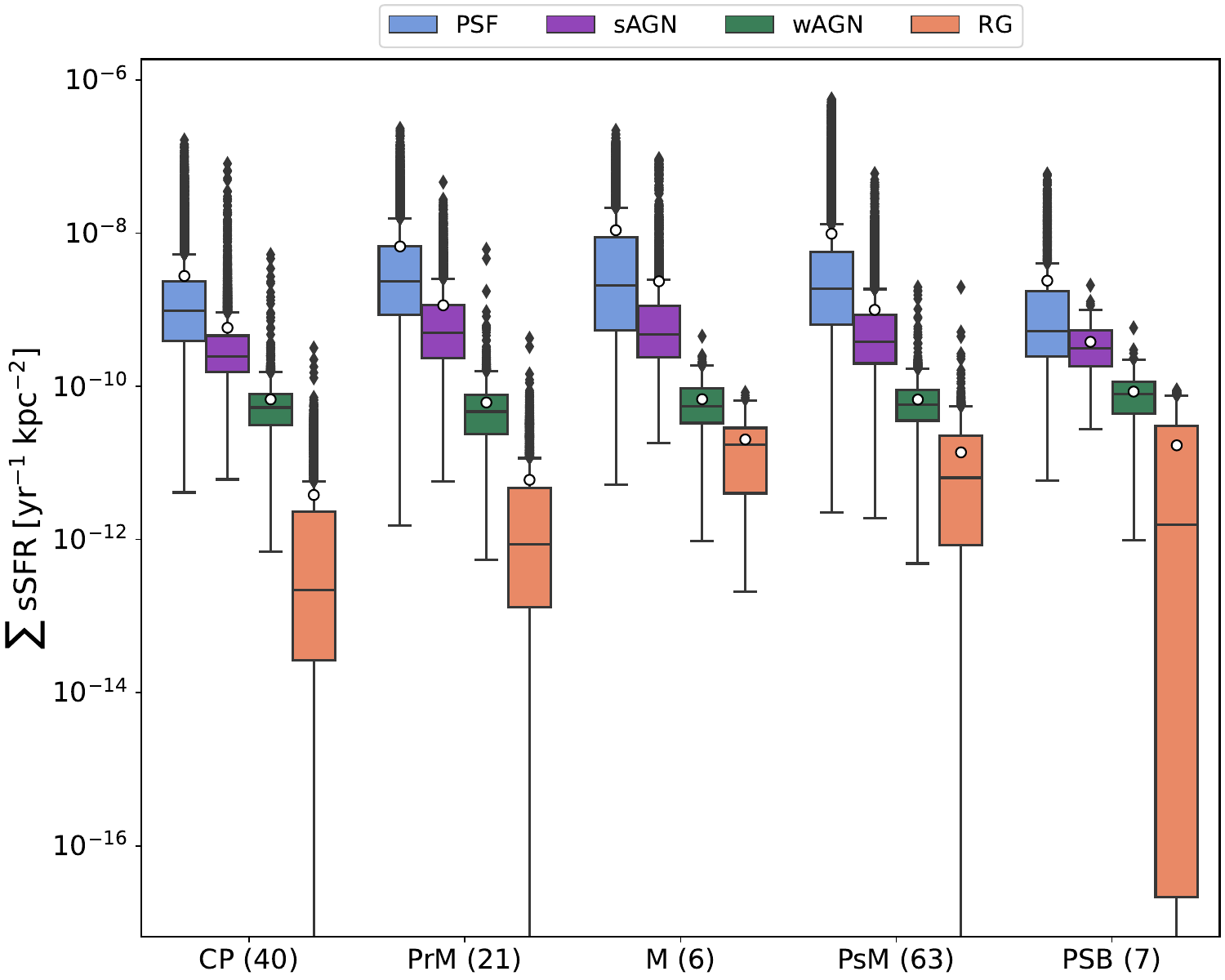}
    \caption[Specific Star Formation Rate (sSFR)]{Distribution of the sSFR surface density ($\rm \Sigma sSFR$) for spaxels in each WHAN category for each merger stage (the number of galaxies in each category is indicated in parentheses). The distributions are shown in form of box plots, with a different colour for each WHAN category as in Fig.~\ref{fig:whan-mean-fraction}. The median values of the distributions are represented by a horizontal line within the box plots, while the mean values are represented by a white circle. Outlier values (larger than 1.5$\times$ the inter-quartile range) are represented by black diamonds. The distribution for passive galaxies is not presented since the values of the $\rm \Sigma sSFR$ are null.}
    \label{fig:ssfr-whan}
\end{figure}

\subsection{Integrated SFR--M\texorpdfstring{$_\star$}{star} diagram}
\label{sec:mer_res_int}

\begin{figure}
    \centering
    \includegraphics[width=\columnwidth, trim={0 0 8cm 6cm},clip]{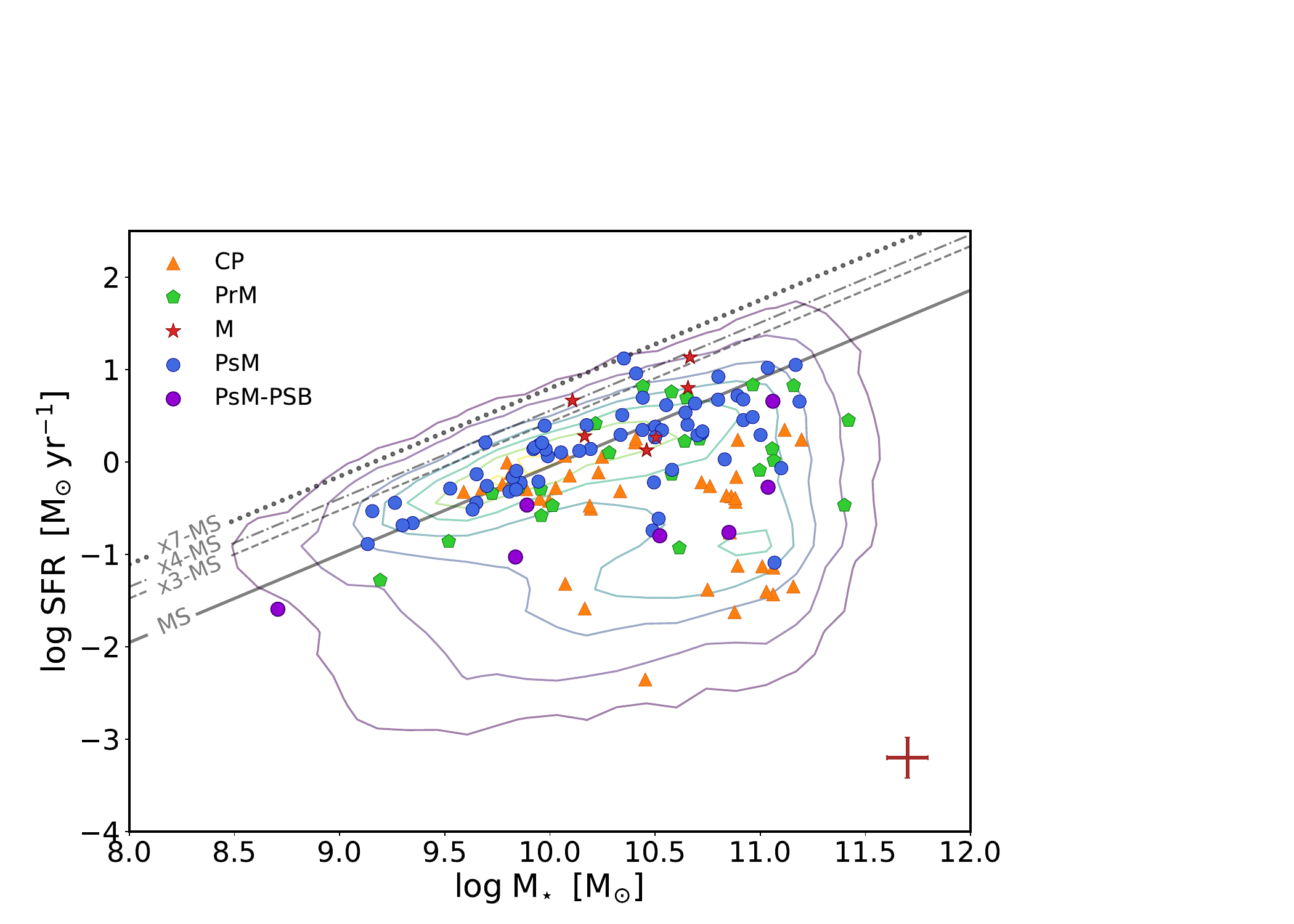}
    \caption[Integrated $SFR-M_{\star}$ diagram.]{Integrated $SFR-M_{\star}$ diagram for the 137 interacting galaxies in our sample. Galaxies classified as close pairs (CP) are represented by orange triangle, galaxies classified as pre-mergers (PrM) are represented by green hexagonal marks, merger (M) galaxies are represented by red stars, and galaxies classified as post-merges (PsM) are represented by blue and purple circles in the case of absence or presence of PSB emission, respectively. In the lower left corner we present a representative error, given by the mean value of the error of the integrated $\rm log\,SFR_{mean}\,=\,0.223\,M_{\odot}yr^{-1}$ and the integrated $\rm log\,M_{\star\,mean}\,=\,0.097\,M_{\odot}$. As reference, contour background correspond to the integrated properties for all MaNGA galaxies from the Pipe3D value added catalogue \citep{2016RMxAA..52..171S,2016RMxAA..52...21S}, and the gray continuous line correspond to the ``nurture" free main sequence for star-forming galaxies derived by Argudo-Fernández et al. 2025, submitted. The dashed lines mark the locations with 3, 4, and 7 times above the main sequence (along the SFR axis), respectively.}
    \label{fig:SFR-M-psb}
\end{figure}

\begin{figure}
    \centering
    \includegraphics[width=\columnwidth, trim={0 0 8cm 6cm},clip]{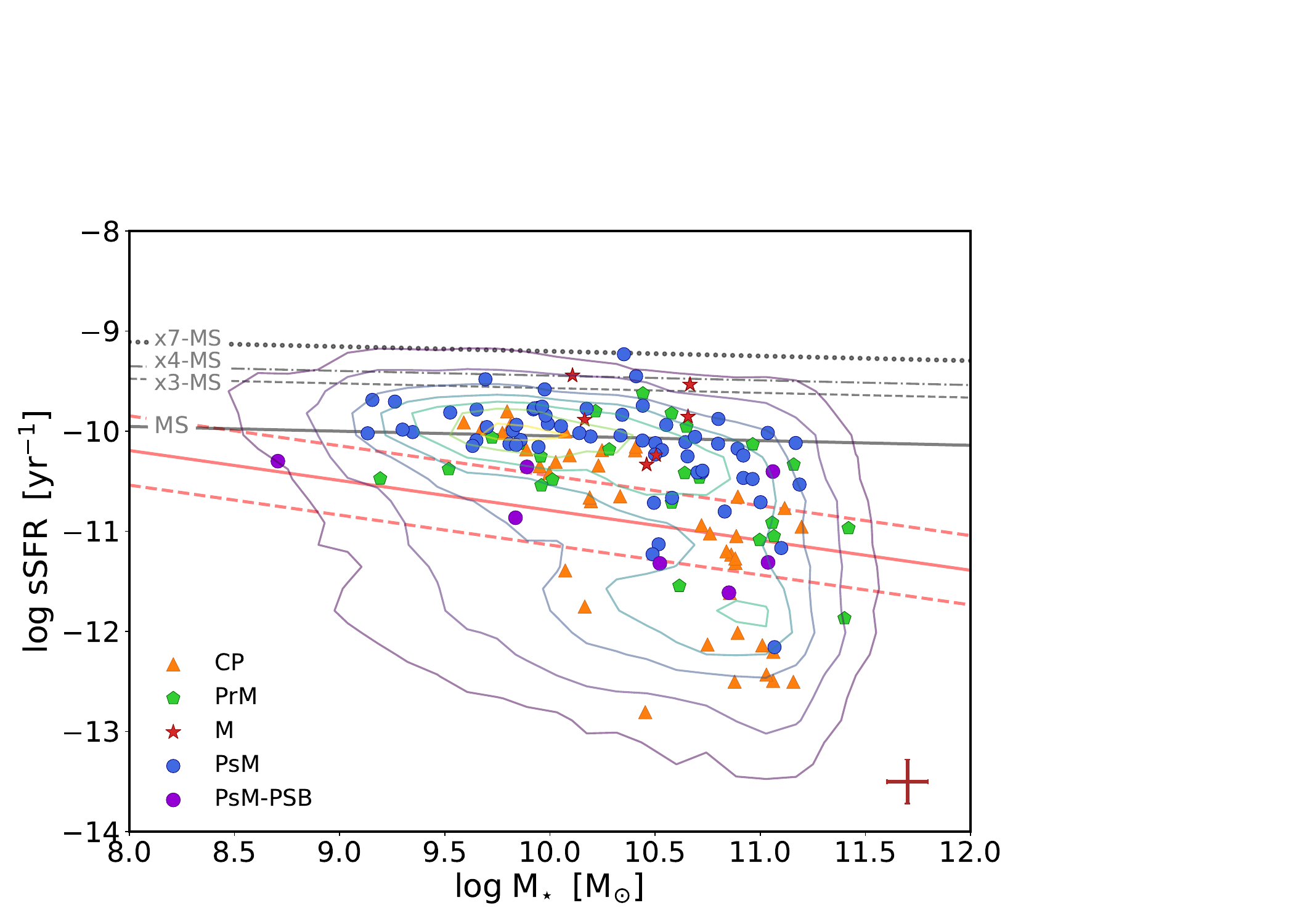}
    \caption[Integrated $sSFR-M_{\star}$ diagram.]{Similarly as Fig.~\ref{fig:SFR-M-psb}, integrated $sSFR-M_{\star}$ diagram for the 137 interacting galaxies in our sample. Galaxies classified as close pairs (CP) are represented by orange triangle, galaxies classified as pre-mergers (PrM) are represented by green hexagonal marks, merger (M) galaxies are represented by red stars, and galaxies classified as post-merges (PsM) are represented by blue and purple circles in the case of absence or presence of PSB emission, respectively. In the lower right corner we present a representative error, given by the mean value of the error of the integrated $\rm log\,sSFR_{mean}\,=\,0.24\,yr^{-1}$ and the integrated $\rm log\,M_{\star\,mean}\,=\,0.097\,M_{\odot}$. As reference, contours lines correspond to the integrated properties for all MaNGA galaxies from the Pipe3D value added catalogue \citep{2016RMxAA..52..171S,2016RMxAA..52...21S}. The gray continuous line correspond to the ``nurture" free main sequence for star-forming galaxies derived by Argudo-Fernández et al. 2025, submitted. The dashed lines mark the locations with 3, 4, and 7 times above the main sequence (along the SFR axis), respectively. The red continuous line correspond to the fit to separate star forming from quenched galaxies in the Pipe3D sample, with its corresponding 1$\sigma$ to delimit the green valley (red dashed lines).} 
    \label{fig:sSFR-M-psb}
\end{figure}

\begin{figure}
    \centering
    \includegraphics[width=\columnwidth, trim={0 0 1cm 1.3cm},clip]{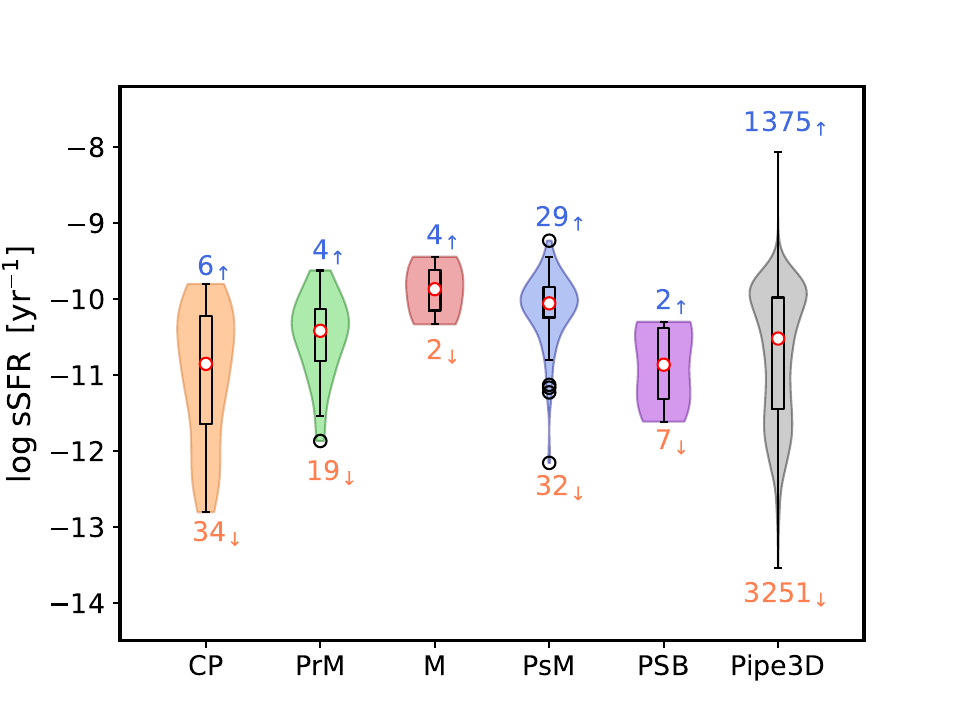}
    \caption[]{Distribution of the integrated sSFR in the form of violin plots, for each merger stage. The colored area determines the density distribution for close pairs (CP, in orange), pre-mergers (PrM, in green), mergers (M, in red), non-PSB post-mergers (PsM, in blue), PSB post-mergers (PSB, in purple). For reference, the values of the integrated properties for all MaNGA galaxies (10010 galaxies) from the Pipe3D value added catalogue  \citep{2016RMxAA..52..171S,2016RMxAA..52...21S} has been added (in grey). The inner box on each violin plot is a representation of interquartile range of the median (red dot) and its 95$\%$ of confidence intervals. We also show the outliers points of the distributions (if any) as black open circles, which represent the atypical values for that category. The number of galaxies above and below the MS in Fig.~\ref{fig:sSFR-M-psb} is indicated above and below the violin plots, respectively, while the total number of galaxies considered in each merger stage is presented in Table~\ref{tab:Ngal}.}
    \label{fig:mean-SFR}
\end{figure}

For each galaxy in our sample, we computed the integrated SFR and stellar mass considering the values in each spaxel to create an integrated SFR-$M_{\star}$ diagram. Figure~\ref{fig:SFR-M-psb} shows the values for the interacting/mergers galaxies in our sample as function of their merger stage, including a differentiation for PSB galaxies within the PsM category. For reference, we added the values of the integrated SFR and stellar masses for all the MaNGA sample from the Pipe3D value added catalogue \citep{2016RMxAA..52...21S,2016RMxAA..52..171S}, as background contours, and the fit to the main sequence found by Argudo-Fernández et al. 2025, submitted, for SIG star-forming galaxies (grey dashed line). Galaxies in our sample are well-distributed in the diagram, populating the main sequence and the quenched region. In particular, galaxies in the CP category can be found in both regions, PrM galaxies are scattered among the main sequence, with only one case in the quenched region, while M galaxies are found in the main sequence, with stellar masses within $10.1\;\log(M_{\odot})\lesssim M_{\star}\lesssim 10.7\;\log(M_{\odot})$. Finally, PsM galaxies are found in two regions, where non-PSB galaxies are mainly located in the main sequence, while PsMs classified as PSBs can be found in the lower region of the main sequence, in the green valley, and in the quenched region. To quantify these differences we also estimate the integrated sSFR and explore its relation with stellar mass using the sSFR--$M_{\star}$ diagram (see Fig.~\ref{fig:sSFR-M-psb}\footnote{To separate the main sequence from the quenched area in Fig.~\ref{fig:sSFR-M-psb}, we followed a similar methodology as in \citet{2014ApJ...788...29L} on MaNGA galaxies, by fitting two Gaussians to the corresponding sSFR distribution at different stellar mass bins, using Pipe3D data products \citep{2016RMxAA..52..171S,2016RMxAA..52...21S}, and choosing the sSFR values at the given bin as the ones where both Gaussians intersect. From our fit, we defined the green valley as the area within 1$\sigma$.}). 
The density distributions of the integrated properties are shown in Fig.~\ref{fig:mean-SFR} using violin plots. We use the distributions of the sSFR to explore if there is any enhancement of star formation at any merger stage. In addition, the median values and uncertainties for each merging stage are presented in Table~\ref{tab:sfr}. We discuss the differences of the distributions for each merger stage in Sect.~\ref{Sec:dis-sfrmass}.

\begin{figure*}
    \centering 
    \includegraphics[width=\textwidth, trim={2cm 0.5cm 7cm 0},clip]{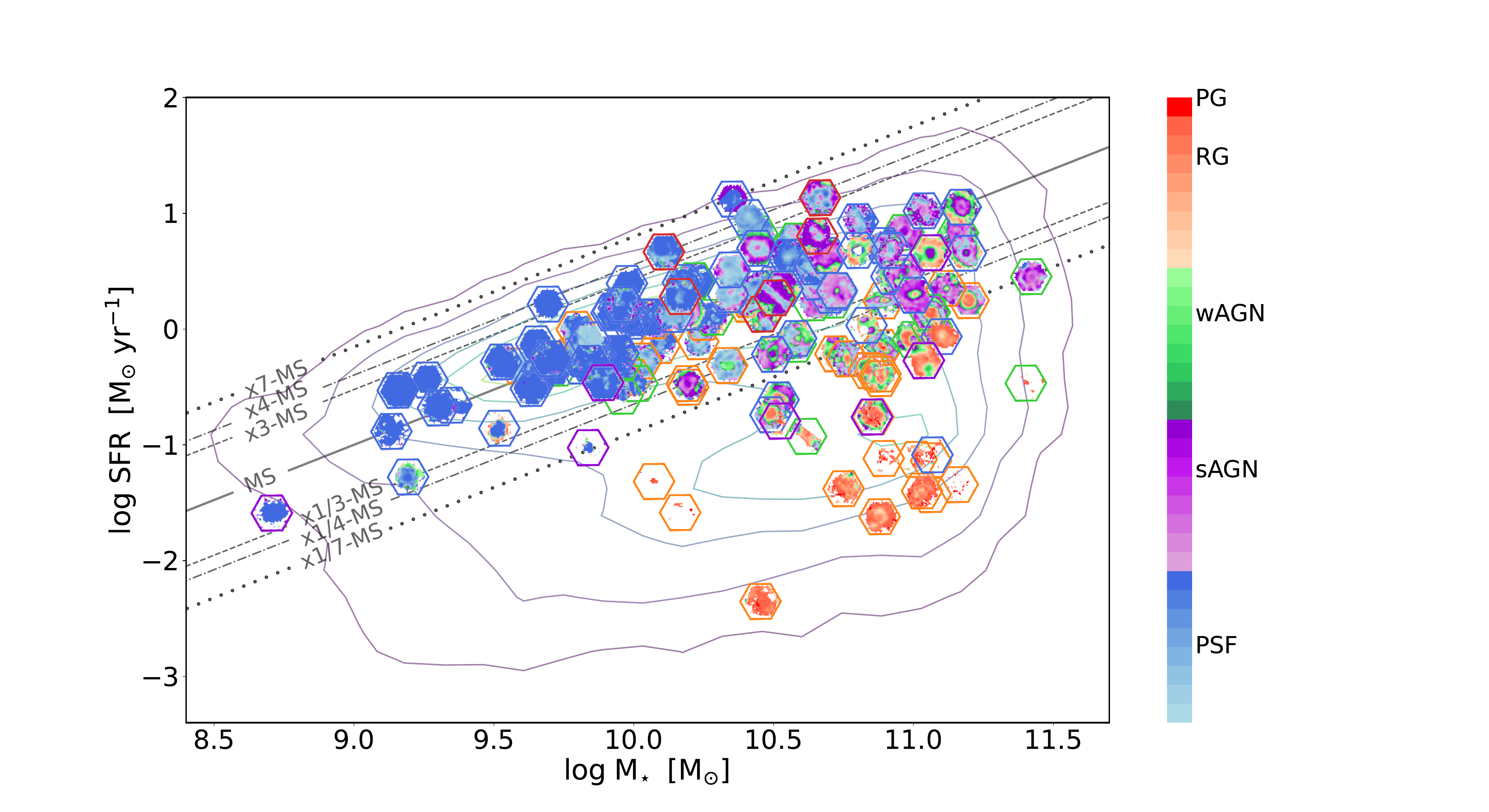}
    \caption[Integrated $SFR-M_{\star}$ diagram in terms of WHAN diagnostic diagrams.]{Integrated $SFR-M_{\star}$ diagram for the 137 galaxies in the sample as in Fig.~\ref{fig:SFR-M-psb}, with the point values as WHAN maps respectively for each galaxy. The colour bar represents the WHAN categories based on \citet{2011MNRAS.413.1687C}. The coloured  hexagons indicate the merger stage for each galaxy in the diagram following the colour scheme in Fig.~\ref{fig:SFR-M-psb}. The dashed lines mark the locations with 3, 4, and 7 times above the main sequence (along the SFR axis), and 1/3, 1/4, and 1/7 times below the main sequence, respectively.}
    \label{fig:SFR-M-whan}
\end{figure*}

\begin{figure*}
    \centering
    \includegraphics[width=\textwidth, trim={2cm 0.5cm 7cm 0},clip]{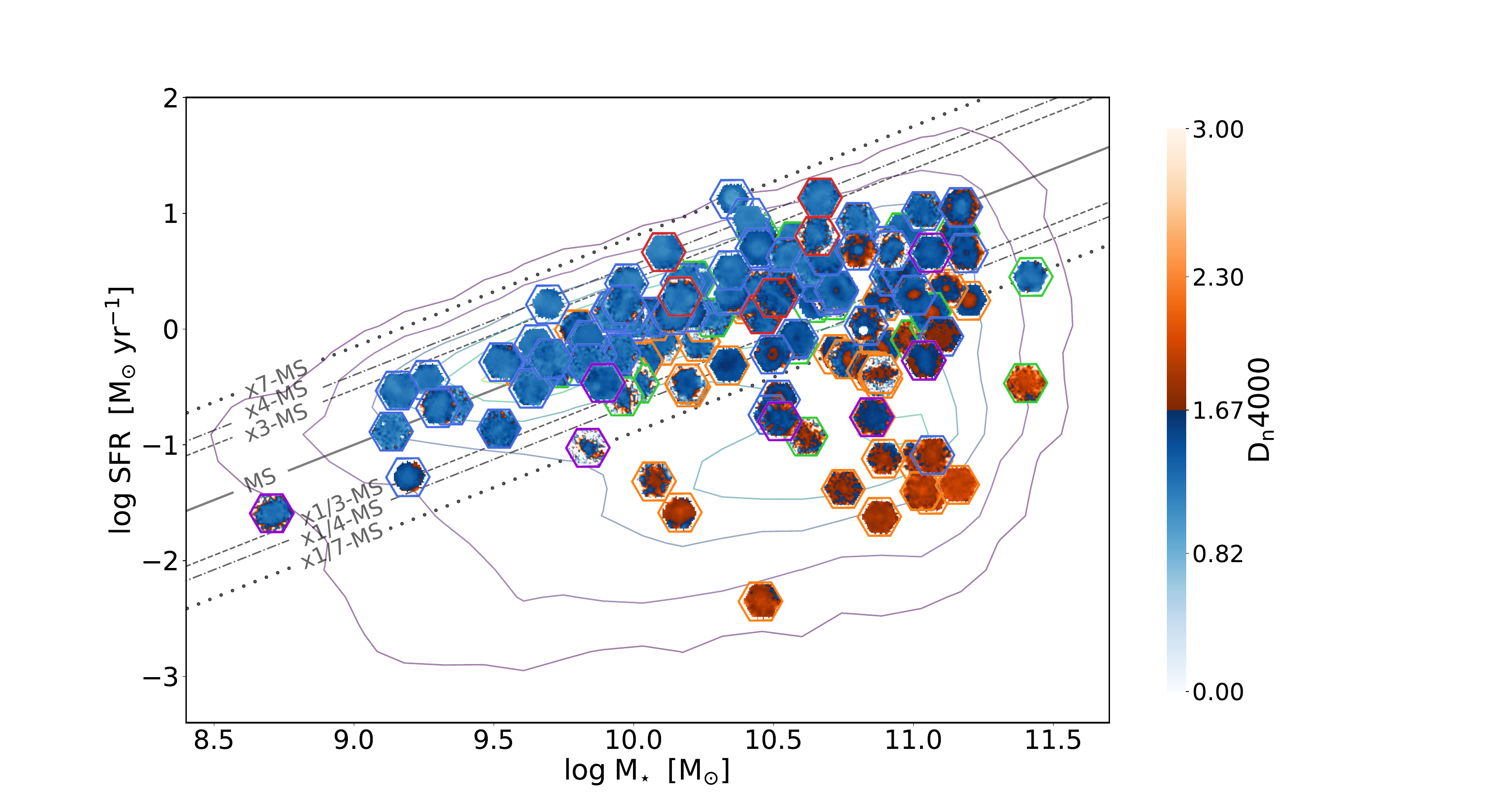}
    \caption[Integrated $SFR-M_{\star}$ diagram in terms of age of the stellar populations.]{Integrated $SFR-M_{\star}$ diagram for the 137 galaxies in the sample as in Fig.~\ref{fig:SFR-M-whan}, but using maps of the age of the stellar populations instead of WHAN maps. The maps are divided into young blue degraded colour (for spaxels with $D_n$(4000)\,$\leq$\,1.67) and old orange degraded colour (for spaxels with $D_n$(4000)\,$>$\,1.67) according to their $D_n4000$ parameter, respectively for each galaxy.}
    \label{fig:SFR-M-D4000}
\end{figure*}

To relate the properties of the stellar populations for galaxies in our sample, with their location in the SFR-$M_\star$ diagram, we also create different versions of the diagram as function of the WHAN emission (see Fig.~\ref{fig:SFR-M-whan}), and the $D_n 4000$ spatial distribution, divided into young ($D_n$(4000)\,$\leq$\,1.67) and old ($D_n$(4000)\,$>$\,1.67) stellar populations (as shown in Fig.~\ref{fig:SFR-M-D4000}). We discuss our results in Sect.~\ref{Sec:merger-dis}.

\begin{table}
    \centering
    \setlength{\arrayrulewidth}{0.1mm}
    \begin{tabular}{ccc}
    \hline\hline
     (1) & (2) & (3) \\
     Category & SFR & sSFR  \\ 
      & log [M$_{\odot}$\,yr$^{-1}$] & log [yr$^{-1}$] \\
    \hline
    CP  & $-$0.34\,$\pm$\,0.98 & $-$10.85\,$\pm$\,1.42\\
    PrM & 0.10\,$\pm$\,0.75 & $-$10.42\,$\pm$\,0.79 \\
    M   & 0.47\,$\pm$\,0.50 & $-$9.87\,$\pm$\,0.53 \\
    PsM & 0.16\,$\pm$\,0.53 & $-$10.09\,$\pm$\,0.46 \\
    PSB & $-$0.76\,$\pm$\,0.54 & $-$10.86\,$\pm$\,0.94 \\
     \hline
     \end{tabular}
    \caption{Median values and uncertainties (given by the interquartile range) of the physical properties derived from CIGALE for each merger stage category. The columns correspond to: (1) merge stage category: CP for close pairs, PrM for pre-merger galaxies, M for mergers, PsM for post-merger galaxies without post-starburst spectral signatures, and PSB for post-merger galaxies with post-starburst spectral signatures; (2) star formation rate (SFR), in log M$_{\odot}$\,yr$^{-1}$; and (3) specific star formation rate (sSFR), in log yr$^{-1}$.}
    \label{tab:sfr}
\end{table}

\section{Discussion}
\label{Sec:merger-dis}

\subsection{Classification of merger stage}
\label{Sec:dis-merger-stage}

The use of physical properties maps in combination with diagnostic maps, as well as emission maps, provides a comprehensive and complete picture of the physical processes that are occurring within galaxies. Thus it helps us to understand how the star-formation/quenching and AGN processes arise or evolve during galaxy interactions. Note that we found a discrepancy in the data provided by the WHAN maps, where they indicate strong AGN (sAGN) ionization features in areas where this emission is unlikely to occur and would be more likely that the emission comes from stellar populations that are in quenching process. This is shown in Fig.~\ref{fig:ssfr-whan} as the overlapping distributions of sSFR surface density, however the median values are distinguishable at 1$\sigma$ level. The maps also allowed us to perform a quality control check, where some galaxies were re-classified or removed. We removed some galaxies from the analysis in case the MaNGA FoV was too small, centered only on the bulge of the galaxy, or did not have good coverage of the galaxy. The final number of galaxies in each category is shown in Table~\ref{tab:Ngal}. 

We also compared our visual classification with other independent methods. We found that all galaxies classified as mergers in our sample have a mean probability $\rm p_{merger}$\,=\,0.78 to present merger morphology according to DS18. 
There are 6 M galaxies in our sample with MaNGA data, which is consistent with the expected number of merger galaxies in isolated systems \citep{2023A&A...669A..23G}. Note that the selection criteria for the M category consider strong interactions but also when the two overlapping galactic nuclei are still distinguishable in the optical image, which allows the selection of dry mergers (i.e., merger of early-type galaxies, with a lower content of cold gas). However, this type of mergers are less common in low-density environments \citep{2010ApJ...718.1158L}. None of the 6 M galaxies is composed by two early type galaxies.

Other authors use non-parametric image predictors, as Gini, M20, and Concentration–Asymmetry–Clumpiness (CAS) parameters \citep{2003ApJS..147....1C, 2004AJ....128..163L, 2016MNRAS.456.3032P}, to help in the identification of tidal tails and other interaction features \citep{2019ApJ...872...76N,2023MNRAS.523.4164H,2023MNRAS.522....1N}. Taking advantage of the MaNGA DAP maps, we have computed these parameters, applied to the MaNGA FoV in the wide r-band, presented in Appendix~\ref{sec:CAS-param}. The comparison of these parameters for the galaxies in our sample, with respect to the distribution of the same parameters for all MaNGA galaxies, is shown in Fig.~\ref{fig:CAS_r}. We do not observe any clear trend in these parameters for our galaxies with respect to their merging stage. It might be due to the fact that MaNGA FoV usually covers from 1.5 to 2.5 effective radius, and interaction features are usually found at galaxy outskirts. However, we have repeated the same analysis with the same morphology parameters for MaNGA galaxies provided by \citep{2023MNRAS.522....1N} using r-band imaging data from SDSS-DR16. We also do not observe any clear relation with the parameters and the merger stage, however the galaxies used in this study have shown in general higher A values than most of MaNGA galaxies, probably due to asymmetries that can be observed at galaxy outskirts in the SDSS imaging. Using both methods, we also observed some trends for galaxies in the PsM-PSB category. PsM-PSB galaxies are generally concentrated in regions with lower clumpiness values (S\,<\,0.1), and have lower inverse concentration index than galaxies at fixed M20 values. This means that light is more concentrated in the inner region of post-merger galaxies with spectral post-starburst features than galaxies with similar morphology, where the distribution of the light is also smoother. Moreover, 2 of the 7 galaxies classified as PsM-PSB are lenticular galaxies, with a probability to be an S0 galaxy $\rm P_{S0}$\,>\,0.5 (mean $\rm P_{S0}$\,=\,0.8) in DS18, which is in agreement with the CAS parameters for these galaxies. We also find 13 CP galaxies with lenticular morphology, which are distributed in the same region of the CAS parameters where PsM-PSB galaxies are located (as shown in Fig.~\ref{fig:CAS_r}). The incidence of lenticular galaxies in the PsM-PSB category is similar to CP galaxies, both larger than the other merger stages\footnote{The percentage of lenticular galaxies in each merger stage according to DS18 is: CP (36.5\%), PrM (4.7\%), M (0\%), PsM (9.5\%), and PSB (28.7\%).}. We conclude that isolated galaxies that have recently undergone a major merger event, which triggered PSB activity, present lenticular morphology, and therefore their morphology might be the result of the merger process. This formation scenario is in agreement with recent studies for the pathways for the formation of S0 galaxies, where merger-triggered formation becomes a more efficient mechanism in lower density environments \citep{2020MNRAS.492.2955C,2022MNRAS.515..201C,2024arXiv240905064C}.  

\subsection{Spatially resolved stellar populations}
\label{Sec:dis_maps}

After analysing the spatially resolved SFR and diagnostic diagrams introduced in Sect.~\ref{Sec:merger-met}, with the MaNGA data products as described in Sect.~\ref{Sec:merger-res}, we found some relations between the physical properties we considered in this study and their merging stage.

In general, the spatially resolved SFR and AGN activity found for CP galaxies (i. e. galaxies without visible signs of interaction) and their observed properties are in agreement with what is expected according to the intrinsic characteristics of their morphological type. The distribution of sSFR is comparable with all MaNGA galaxies, showing a bimodal distribution (see Fig.~\ref{fig:mean-SFR}). Early-type CP galaxies show no evidence of an enhancement of star formation or nuclear activity in general, while in late-type CP galaxies there is some moderate star formation in the disk and spiral arms, normally associated with regions with young populations, and lower SFR in the inner region (bulge), dominated by older populations. When looking at the SFR maps for CP galaxies, we find that in some cases (9 galaxies) the SFR is distributed forming a ring shape within the galaxy disk, while the galaxy does not look like a ringed galaxy in the colour image. This indicates that star formation is happening at similar galactocentric distance, which is related to secular evolution. Even if CP galaxies have close physically bound neighbours, there are no appreciable interaction features, neither in their visual morphology nor in their spectral properties. The CP sample can be therefore used as a ground base to compare with the galaxies in more advanced merger stage. 

PrM galaxies are in an initial stage of interaction, with an appreciable interchange of baryonic matter between the galaxies in form of bridges and tidal tails, which may also present recent star formation. 
The ionisation emission due to recent star formation is weak in these regions, as shown in the SFR maps, however it is shown as AGN emission in the WHAN map. In this regard, WHAN diagrams might have a limitation when there is weak emission or recent star-formation, missclassifying the emission as AGN.  
The distributions of SFR and sSFR for galaxies in this merger stage suggest a slight increment (about 0.4\,dex) of the star-formation activity with respect to the PrM (as shown in Fig.~\ref{fig:mean-SFR}).

In the M category we observe more violent interactions between the galaxies, where there are indications of strong star-formation activity in the nucleus, with two galaxies well above the main sequence (up to about 4 times the main sequence, as shown in Fig.~\ref{fig:SFR-M-psb}), and more intense star-formation emission in bridges and tidal tails in comparison to PrM galaxies. We find that the 6 galaxies in this category are distributed around the MS. We can already appreciate the effect of mergers, with a noticeable increment of the star-formation activity for M and PSM (non PSB) galaxies with respect to the other merger stages, as shown in Fig.~\ref{fig:mean-SFR}. These galaxies also present flat $D_n$(4000) radial profiles with similar values (see Fig.~\ref{fig:d4000radial}), indicating that the stellar populations during the M and PsM stage are mixed throughout the galaxy, as expected.
The mean fraction of spaxels with younger stellar populations is also higher than in PrM galaxies (about 10\% more, as shown in Fig.~\ref{fig:mean-age}). The recent star-formation regions, typically with $D_n$(4000)\,<\,1.2, \citep[for which a mean stellar age younger than 150\,Myr is expected,][]{2006MNRAS.370..721M}, are well differentiated with respect to areas where star formation processes have already occurred over longer periods of time. 
A more accurate estimation of the age would provide some insight into the timing of interactions between galaxies. However the maps of the $Age_{bq}$ and $r_{SFR}$ parameters of the SFH, which are related to a recent burst/quench episode, usually show some doubtful structures with respect to emission line maps. A better characterisation of these parameters would allow us to relate the observed recent star formation episodes with the expected dynamical time of interactions in comparison to simulations, which would help us to better understand the interaction process \citep{2008MNRAS.383...93B, 2008MNRAS.391.1137L, 2018A&A...614A..66S}. However this would require to increase the parameter space for the SED fitting, which would significantly increase the computation time, and would be therefore limited to the study of individual systems.

In the PsM category there is a broad variety of galaxies in terms of star formation. On one hand, there are PsM (non PSB) galaxies where star formation is distributed throughout the galaxy. Usually these galaxies are blue in the color images, mainly with flocculent spiral morphology (see Fig.\ref{fig:SFR-M-img}). Their WHAN diagrams are mainly dominated by pure star-forming emission (see Figures \ref{fig:whan-mean-fraction} and \ref{fig:SFR-M-whan}), which is also recent according to their SFH and their $D_n$(4000) parameter (the mean fraction of young stellar population is slightly higher than in M galaxies, as shon in Fig.~\ref{fig:mean-age}). We consider that these galaxies might have recently undergone a minor merger process. In fact, about 60\%\footnote{There is available classification information for 60 galaxies, out of 68 classified as PsM (non PSB). Of them, 28 are classified as post minor merger event, and 12 as post-major merger event.} of these galaxies are classified as post-merger (both minor or major) by \citet{2024ApJ...963...53C}, who identify mergers in the MaNGA dataset using the methodology in \citet{2023MNRAS.522....1N}.
On the other hand, PsM-PSB galaxies in general do not have an appreciable disk with spiral arms, but are in transition to a spheroidal galaxy. Four out of seven galaxies in this category show shell-like structures in their outskirts, appreciable in their SDSS colour images, which are clear indicators of a recent merger event \citep{2023MNRAS.518.3261P}. Most of the stellar populations in these galaxies are young and the mass is mainly distributed in the bulge of the galaxy, with the outer regions being mostly older and quenched. PsM-PSB galaxies are therefore in the process of quenching outside-in. 

As shown in Fig.~\ref{fig:whan-mean-fraction}, the fractions of PSF and sAGN in the M category are similar, and also overlapping high sSFR values in Fig.~\ref{fig:ssfr-whan}. In case that interaction enhance star-formation and/or nuclear activity, this increment happens at the merger stage, both proceeding the subsequent quenching process. However, in the PsM category (for galaxies without post-starbust emission), the fraction of PSF spaxels doubles the fraction of AGN spaxels. This can be interpreted in two ways: AGN plays a secondary role on quenching galaxies after a merger event, or AGN feedback may not has had time to quench star-formation yet.

\subsection{Integrated SFR--M\texorpdfstring{$_\star$}{star} diagram}
\label{Sec:dis-sfrmass}

As pointed out in Sect.~\ref{sec:mer_res_int}, in the SFR-$M_{\star}$ diagram presented in the Fig.~\ref{fig:SFR-M-psb} we can appreciate how the galaxies in our sample are distributed along the main sequence region, the green valley, and the quenched area. These regions are well drawn using the integrated SFR-$M_{\star}$ values for all MaNGA galaxies from the Pipe3D value added catalogue \citep{2016RMxAA..52..171S,2016RMxAA..52...21S}. For reference, we added the ``nurture" free main sequence for star-forming galaxies found by Argudo-Fernández et al. 2025, submitted, using aperture-corrected SFR estimates from SDSS spectra following the methodology of \citet{2017A&A...599A..71D}. We observe that most of the galaxies in each category are mainly found following the main sequence. In particular, M and PSM (non PSB) galaxies have, in general, higher sSFR than galaxies in other merger stages (and higher than the median values for MaNGA galaxies), as shown in Fig.~\ref{fig:mean-SFR}. This is an indication of an enhancement of star formation during the interaction process.
Note that we integrated all the good spaxels (i.e. using the default DAP masks definitions) within the MaNGA FoV, therefore for some galaxies we might have to consider the contribution of the companion galaxy. In fact PrM galaxies show higher values of the integrated $M_{\star}$ than expected, due to the contribution of the companion, as can be seen in Fig.~\ref{fig:prm_maps}. Because of this, PrM galaxies appear in the lower region of the main sequence when we expected to find them well located in the main sequence, considering their moderate-to-high $\Sigma$\,SFR values. In fact, Fig.~\ref{fig:mean-SFR} shows that, when normalising by stellar mass, the enhancement of star-formation activity (about 1\,dex higher) occurs in an advanced stage of the merger process (M and PsM galaxies). This result is in agreement with \citet{2024A&A...691A..82C}, who reported SFR for $\sim$600 mergers not belonging to denser structures in the nearby universe ($z\,<\,0.1$), based on photometric SED fitting. There is no much difference between galaxies in early stage of the interaction process (i.e., PrM galaxies) and control galaxies (CP), even comparable with the values for PsM-PSB galaxies. 

When analysing PsM galaxies, we found two different distributions in the SFR-$M_{\star}$ diagram, which are also related to the presence (or absence) of PSB emission. Non-PSB PsM galaxies are well located in the main sequence. These galaxies mainly present clumpy spiral morphology and pure star-formation emission dominating the WHAN maps (as shown in Fig.~\ref{fig:whan-mean-fraction}). We suggest that the interaction features found in these galaxies are due to a recent minor merger event, which increased their SFR but not necessarily triggered any subsequent quenching process or main structural morphology transformation. 
On the other hand, PsM-PSB galaxies are distributed from below the main sequence to the quenched area, from more centrally located PSB emission to more spatially distributed PSB emission, respectively. We found that PsM-PSB galaxies with higher integrated SFR values present PSB features concentrated in the central region but still not expanded to outer regions of the galaxy. The extension of the PSB emission is therefore directly related to the quenching process and the percentage of quenched area in these galaxies. The spatially distribution of PSB emission has been explored in previous MaNGA galaxies. In particular \citet{2019MNRAS.489.5709C} stated that central and non-central PSB distributed galaxies are not simply different evolutionary stages of the same event, where central PSB galaxies would be mainly caused by a significant disruptive event, as a major merger, while non-central PSB emission is caused by disruption of gas fuelling to the outer regions. In contrast, we find that PSB emission in PsM galaxies would be mainly associated with a previous interaction, and the fact that we find these galaxies at different locations of the SFR-Mass diagram is because these transforming processes after a recent galaxy interaction could happen slowly on isolated environments. 

For a complete analysis of the distribution of interacting galaxies in the SFR-$M_{\star}$ diagram, we reproduced them and replaced each point by their WHAN map\footnote{Computed using the tool described in Appendix~\ref{sec:WHAN-tools}.} (see Fig.~\ref{fig:SFR-M-whan}), and the maps of the $D_n$(4000) parameter, separating the stellar populations between into young ($D_n$(4000)\,$\leq$\,1.67) and old ($D_n$(4000)\,$>$\,1.67), as shown in Fig.~\ref{fig:SFR-M-D4000}, and additionally using the SDSS three-colour image of the galaxies (see Fig.~\ref{fig:SFR-M-img} in the Appendix). When analysing the three diagrams altogether we observe that, as expected, pure star-forming dominated galaxies, with young stellar populations, are found in the main sequence. These are observed from stellar masses between $\rm 9.2\,\,M_{\odot}\, \lesssim\,\log\,M_{\star}\,\lesssim\,10.6\,M_{\odot}$, from clumpy spirals to blue/red spiral galaxies, as well as starburst galaxies. For higher values of $\log\,M_{\star}\,\gtrsim \,10.9 $ we can find galaxies with shells features. Galaxies in the green valley present both, early- and late-type morphologies. Galaxies in the quenched area present early-type morphology. These are, in general, CP or PsM-PSB galaxies. We found that, in comparison with close pairs, the PSB galaxies are lenticular galaxies, while close pairs are ellipticals. This result is in agreement with studies that claimed that lenticular galaxies might be the result of a merger event \citep{2001Ap&SS.276..847B,2012A&A...547A..48E,2015A&A...573A..78Q,2015A&A...579L...2Q,2018A&A...617A.113E}. We would need more statistics to state that this is a characteristic morphology for post-merger post-starburst galaxies. 

The $SFR-M_{\star}$ mass diagram in terms of WHAN diagnostic diagrams in Fig.~\ref{fig:SFR-M-whan}, shows an stratification as a function on both, integrated stellar mass and integrated SFR. Galaxies in the main sequence are dominated by star-formation up to $\rm log(M_{\star})\,\lesssim\,10.5\;M_{\odot}$. At higher stellar mass, the main sequence is dominated by galaxies with sAGN activity, with wAGN activity in the area below the main sequence. RG ionization values are found in the green valley, while the quenched area is dominated by passive galaxies. Note that the sAGN activity for high mass galaxies in the main sequence might be due to miss-classification of low star-formation emission, which is expected for evolved spirals that mainly populate that region of the diagram. With respect to the relation with the merging stage of the galaxies, about 46\% of spaxels in merger galaxies are AGN, as shown in Fig.~\ref{fig:whan-mean-fraction}. In comparison, about one third of the fraction of spaxels in close pairs are classified as AGN. This could be interpreted as an excess of AGN in mergers. However, when comparing at the same stellar mass range, as in Fig.~\ref{fig:SFR-M-whan}, we observe that there is no difference in the fraction showing AGN activity in close pairs (control sample) and strongly interacting galaxies. This is in agreement with previous studies that neither found a distinction on AGN activity in interacting galaxies with respect to isolated galaxies \citep{2021ApJ...923....6J,2023ApJ...942..107S}.

When analysing the $SFR-M_{\star}$ mass diagram in terms of the $D_n$(4000) parameter in Fig.~\ref{fig:SFR-M-D4000}, we found that, in comparison with quenched galaxies in CP galaxies, where quenching happens inside-out (slightly negative radial profiles, where inner regions are older, with $D_n$(4000)\,$>$\,1.67, while outer regions are younger, with $D_n$(4000)\,$\leq$\,1.67), the quenching process in PsM-PSB galaxies is happening outside-in (i.e., inner regions are younger, with $D_n$(4000)\,$\leq$\,1.67, while outer regions are older, with $D_n$(4000)\,$>$\,1.67) (with positive radial profile, m\,=\,0.12\,$\pm$\,0.03, as shown in Fig.~\ref{fig:d4000radial}). This might be an observational proof of the effect of interactions on the quenching process. In addition, the gradients of CP and PSB galaxies within 1\,$R_{eff}$ show opposite behavior, while from 1\,$R_{eff}$ to 1.5\,$R_{eff}$ follow a similar trend, ending at the same $D_n$(4000) value.  In comparison, PrM galaxies show opposite behavior to PSB galaxies within the full 1.5\,$R_{eff}$ range. This could indicate that the inner region is more affected by merger-enhanced star-formation than the outskirts, while the outskirts are more sensitive in the earlier stages of the interaction. We also found that PsM-PSB galaxies present lower sSFR values with larger fraction of PSB spaxels, being located at lower regions of the $SFR-M_{\star}$ diagram. Therefore, we are observing galaxies at different stage of their quenching process due to post-starburst activity. This is known to be a rapid episode in the evolution of galaxies, on time-scales of a few 100\,Myr \citep[<\,500\,Myr,][]{2019MNRAS.489.5709C}. 
The fact that we are finding galaxies under these processes might be connected to their environment, considering the slower evolution of galaxies in low density environments. In general, secular processes dominate the evolution of isolated galaxies, not having interaction with another galaxy for at least 5\,Gyr \citep{2007A&A...472..121V,2015A&A...578A.110A} and therefore their gas is being consumed more slowly, making galaxies in isolated systems optimal laboratories to better understand post-starburst quenching process.

\section{Summary and conclusions}
\label{Sec:merger-con}

In this work we have identified a sample of interacting galaxies in isolated environments \citep[isolated galaxies and physically bound isolated pairs and triplets from the catalogues compiled by][]{2015A&A...578A.110A} at different stages of the merging process. We have analysed their star-formation and AGN emission to explore whether there is an enhancement of this activity triggered by the interaction and how these are spatially distributed in the galaxies. Galaxies were classified as close pairs (CP), pre-mergers (PrM), mergers (M), and post-merger (PsM) galaxies, where we also classified PsM galaxies considering whether they have post-starburst (PSB) spectral signatures or not. 

We have constrained the SFH of the galaxies using spectro-photometric SED fitting to model spectra using CIGALE. For this we used MaNGA IFU spectra in the optical range, where we defined eight custom filters, selected in the continuum, to combine with $H_\alpha$ and $H_\beta$ emission lines and the $D_n4000$ spectral index, from the MaNGA Data Analysis Products (DAP). The integrated SFR and stellar mass, estimated from the SFR and stellar mass maps that we computed for each galaxy from CIGALE outputs, allow us to study the distribution of the galaxies in the SFR-$M_{\star}$ diagram and analyse their level of star-formation activity, with respect to their merging process stage. 

In addition, we analyse the connection to nuclear activity and stellar population age with the evolutionary stage of the galaxies (given by their position in the SFR-$M_{\star}$ diagram), considering also their merging stage. To characterise the AGN emission we used spatially resolved WHAN diagrams \citep[according to][]{2011MNRAS.413.1687C} using DAP data, created with a customised visualisation tool, for which we share the code. We used the $D_n$(4000) parameter as the stellar population age indicator following \citet{2006MNRAS.370..721M}, where, for each galaxy, we classify each spaxel between young ($D_n$(4000)\,$\leq$\,1.67) and old ($D_n$(4000)\,$>$\,1.67). 

We combined all this information to explore the role of galaxy mergers in the SFR-$M_{\star}$ plane evolution on isolated galaxies and galaxies in isolated systems, where we confine the merging process from other environmental processes present in denser environments that accelerate galaxy evolution.  

Our main findings are the following:

\begin{itemize}
\item In general, galaxies show some characteristic properties intrinsically related to each stage of the merger process. The effect of mergers is appreciable, with a noticeable increment of the star-formation activity (measured by their integrated sSFR) for M and PSM (non PSB) galaxies with respect to the other merger stages. The results support the scenario where star-formation activity is enhanced by major interaction.
    
\item In the merger stage, the fractions of PSF and sAGN are similar, with also overlapping high sSFR values. This indicates that, in case that interaction enhance star-formation or nuclear activity, this increment happens at the same time, both proceeding the subsequent quenching process. However the level of star-formation activity persists during the post-merger stage (for galaxies without post-starbust emission). This can be interpreted in two ways: AGN plays a secondary role on quenching galaxies after a merger event, or AGN feedback may not has had time to quench star-formation yet.
    
\item Galaxies with visible signs of interaction (PrM, M, and PsM) galaxies, are distributed along the main sequence of the SFR-M$_{\star}$ diagram up to about 4 times the main sequence. This supports the scenario where galaxy interactions trigger star-formation in galaxies. On the other hand, there is no differentiation between AGN activity in close pairs (control sample) and strongly interacting galaxies at same stellar mass. 
    
\item PsM (non PSB) galaxies are found on the main sequence, probably as result of a minor merger event, while PsM-PSB galaxies are found below the main sequence, on transition into the quenching region of the diagram. We also found that PsM-PSB isolated galaxies generally present lenticular morphology, with higher incidence than in CP galaxies. Therefore, isolated galaxies that have recently undergone a major merger event (which triggered PSB activity) present lenticular morphology, which might be as a result of the merger process.

\item In addition, we found that the quenching process in PsM-PSB galaxies is happening outside-in (i.e., inner regions are younger, with $D_n$(4000)\,$\leq$\,1.67, while outer regions are older, with $D_n$(4000)\,$>$\,1.67), being an observational proof of the effect of interactions on the quenching process. Using the SFR-M$_{\star}$ diagram for isolated galaxy mergers, we can observe galaxies at different stage of their quenching process due to post-starburst activity.

\item The fact that we find PsM-PSB galaxies at different locations of the SFR-M$_{\star}$ diagram is because the transforming processes occurring in galaxies after a recent major galaxy interaction might happen slowly on isolated environments, where the system evolves in a common dark matter halo without any perturbation of external galaxies.
\end{itemize}

A better characterisation of the parameters space for the components of the SFH, from the spectro-photometric SED fitting in these interacting isolated systems, would allow us to compare with the predictions of numerical models and simulations. In particular, to compare the dynamical times of the increment/quench of star-formation in relation to other properties of the galaxies (morphology, stellar-mass ratio, gas content, etc) to better quantify the role of the interaction on star-formation or AGN activity. In this regard, WHAN diagnostic diagrams are a promising tool to explore SFR/AGN/quench processes and transition regions between these along the disk of spiral and lenticular galaxies. However, we have found some limitations where low ionised emission is missclassified as AGN. It would be interesting to further explore the use of this diagram with higher resolution IFU data to better constraint the different sources of ionisation at different regions of the disk.

\begin{acknowledgements}
We thank our referee whose valuable comments have certainly contributed to improve and clarify this paper.

MAF and PVB acknowledge financial support by the DI-PUCV research project 039.481/2020, the research project PID2023-150178NB-I00 and PID2023-149578NB-I00, financed by MCIN/AEI/10.13039/501100011033, the project A-FQM-510-UGR20 financed from FEDER/Junta de Andaluc\'ia-Consejer\'ia de Transforamción Económica, Industria, Conocimiento y Universidades/Proyecto, by the grants P20\_00334 and FQM108, financed by the Junta de Andalucía (Spain), the Emergia program (EMERGIA20\_38888) from Consejería de Transformación Económica, Industria, Conocimiento y Universidades and University of Granada, and the Grant AST22-4.4, funded by Consejería de Universidad, Investigación e Innovación and Gobierno de España and Unión Europea - NextGenerationEU. MAF also acknowledges support from FONDECYT iniciaci\'on project 11200107. We are also grateful for the computing resources and related technical support provided by PROTEUS, the supercomputing center of Institute Carlos I in Granada, Spain. MB gratefully acknowledges support from the ANID BASAL project FB210003 and from the FONDECYT regular grant 1211000. This work was supported by the French government through the France 2030 investment plan managed by the National Research Agency (ANR), as part of the Initiative of Excellence of Universit Côte d'Azur under reference number ANR-15-IDEX-01.

This research made use of \textsc{astropy}, a community-developed core \textsc{python} ({\tt http://www.python.org}) package for Astronomy \citep{2013A&A...558A..33A}; \textsc{ipython} \citep{PER-GRA:2007}; \textsc{matplotlib} \citep{Hunter:2007}; \textsc{numpy} \citep{:/content/aip/journal/cise/13/2/10.1109/MCSE.2011.37}; \textsc{scipy} \citep{citescipy}; and \textsc{topcat} \citep{2005ASPC..347...29T}. This research made use of \textsc{astrodendro}, a Python package to compute dendrograms of Astronomical data (http://www.dendrograms.org/).

This research has made use of the NASA/IPAC Extragalactic Database, operated by the Jet Propulsion Laboratory of the California Institute of Technology, un centract with the National Aeronautics and Space Administration.

Funding for SDSS-III has been provided by the Alfred P. Sloan Foundation, the Participating Institutions, the National Science Foundation, and the U.S. Department of Energy Office of Science. The SDSS-III Web site is http://www.sdss3.org/. The SDSS-IV site is http://www/sdss/org. Based on observations made with the NASA Galaxy Evolution Explorer (GALEX). GALEX is operated for NASA by the California Institute of Technology under NASA contract NAS5-98034. This publication makes use of data products from the Wide-field Infrared Survey Explorer, which is a joint project of the University of California, Los Angeles, and the Jet Propulsion Laboratory/California Institute of Technology, funded by the National Aeronautics and Space Administration.
\end{acknowledgements}

\bibliographystyle{aa} 
\bibliography{Bib} 

\appendix

\section{Spatially resolved maps}
\label{sec:products}

\begin{figure*}
    \centering
    \includegraphics[width=\textwidth]{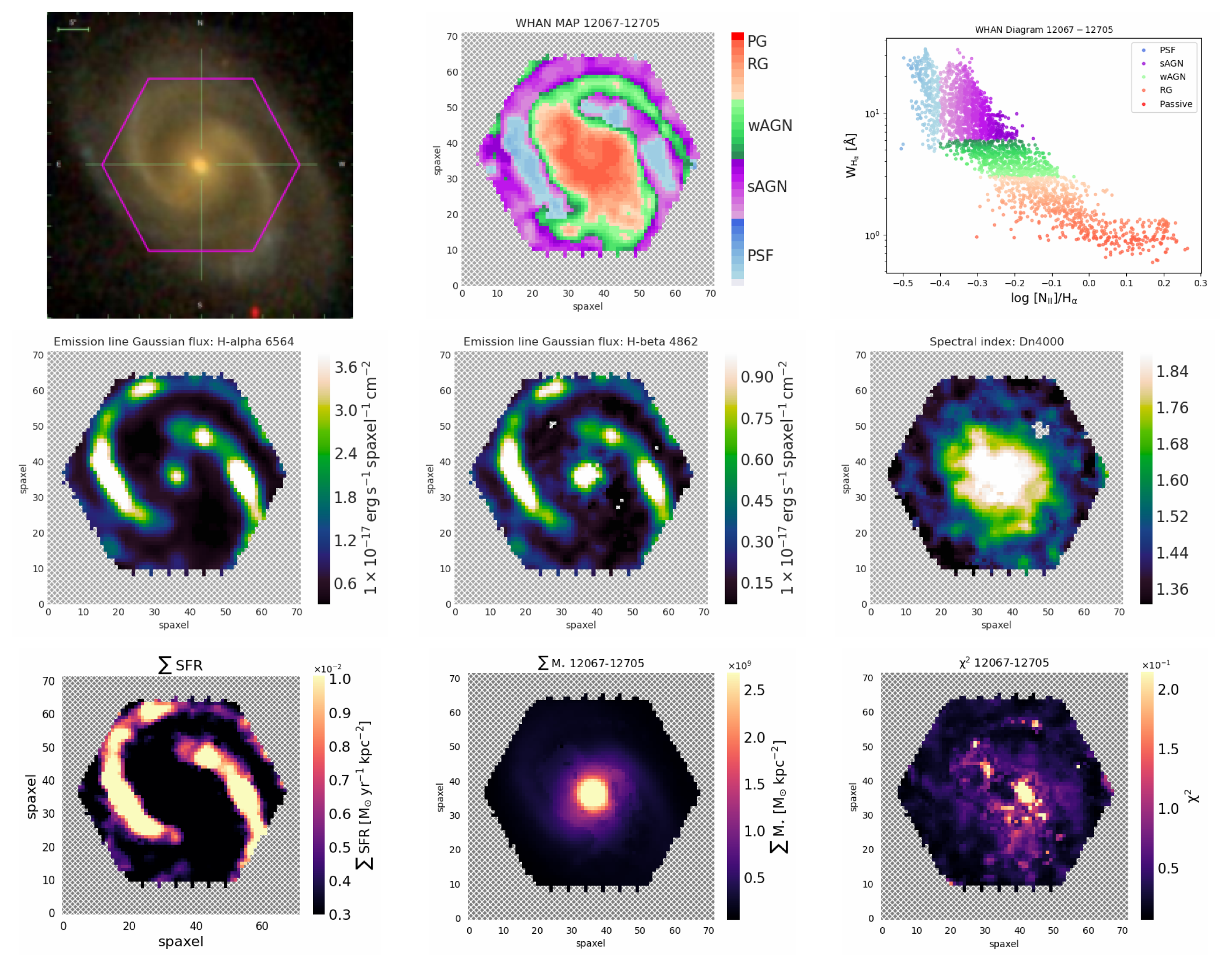}
    \caption[Analysis maps for the CP galaxy 12067-12705]{Analysis maps for the CP galaxy 12067-12705. From left upper to right lower panels: SDSS colour image with the MaNGA FoV, WHAN diagnostic diagram map, WHAN diagnostic diagram scatter plot, $H_{\alpha}$ emission line map, $H_{\beta}$ emission line map, spectral index $D_n 4000$ map, SFR CIGALE result map, $M_{\star}$ CIGALE result map, and $\chi ^2$ CIGALE result map.}
    \label{fig:cp_maps}
\end{figure*}

\begin{figure*}
    \centering
    \includegraphics[width=\textwidth]{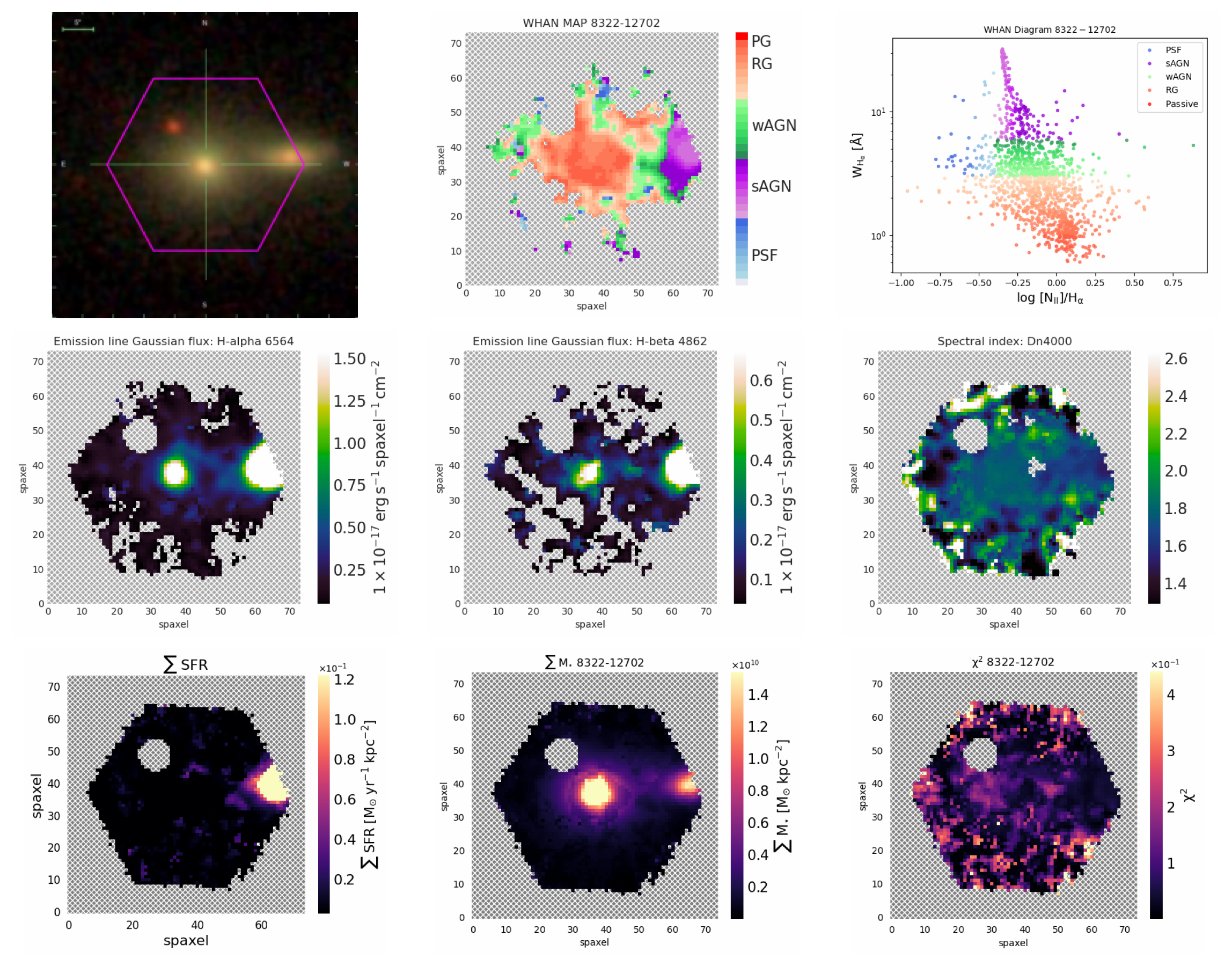}
    \caption[Analysis maps for the PrM galaxy 8322-12702]{Analysis maps for the PrM galaxy 8322-12702. From left upper to right lower panels: SDSS colour image with the MaNGA FoV, WHAN diagnostic diagram map, WHAN diagnostic diagram scatter plot, $H_{\alpha}$ emission line map, $H_{\beta}$ emission line map, spectral index $D_n 4000$ map, SFR CIGALE result map, $M_{\star}$ CIGALE result map, and $\chi ^2$ CIGALE result map.}
    \label{fig:prm_maps}
\end{figure*}

\begin{figure*}
    \centering
    \includegraphics[width=\textwidth]{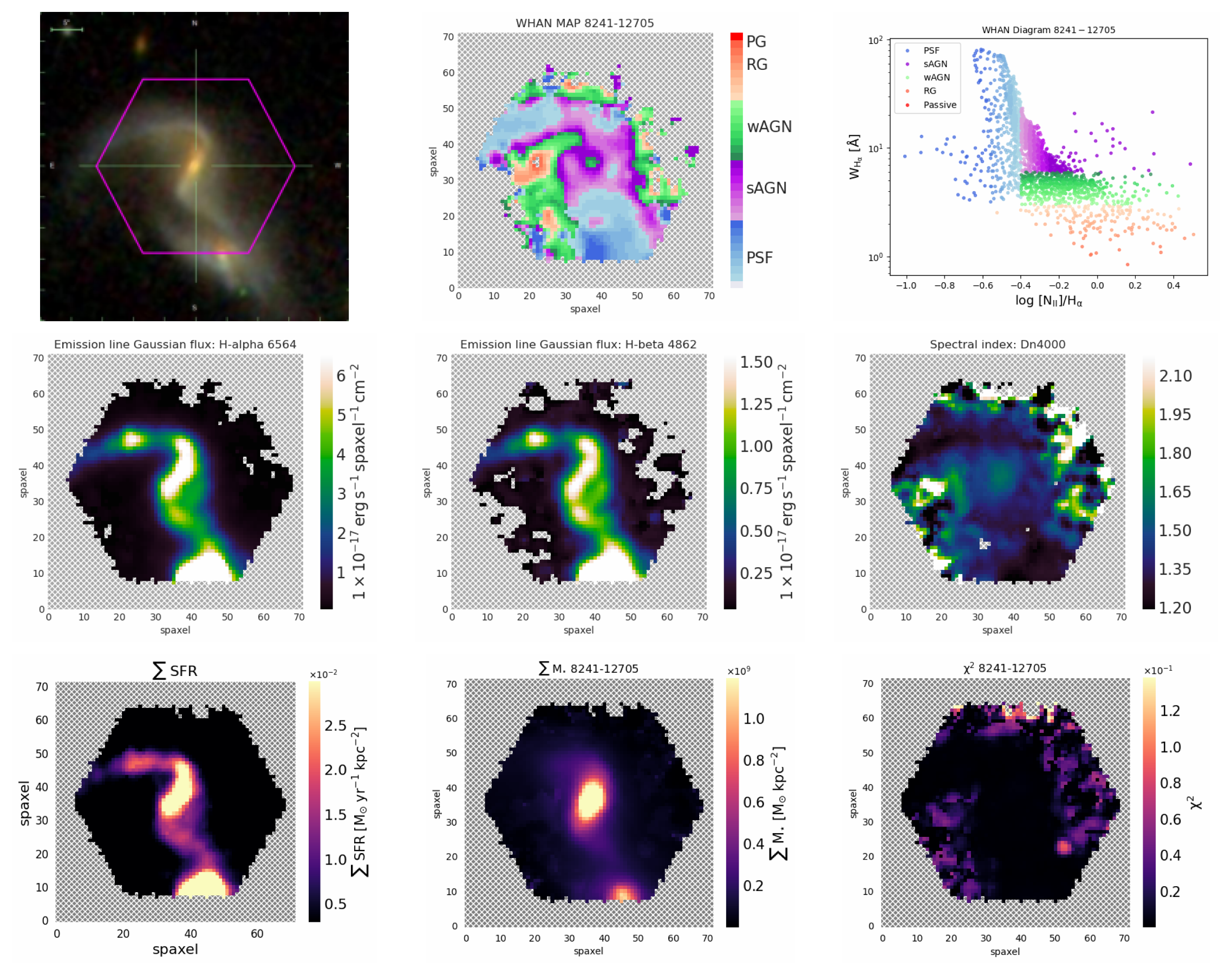}
    \caption[Analysis maps for the M galaxy 8241-12705]{Analysis maps for the M galaxy 8241-12705. From left upper to right lower panels: SDSS colour image with the MaNGA FoV, WHAN diagnostic diagram map, WHAN diagnostic diagram scatter plot, $H_{\alpha}$ emission line map, $H_{\beta}$ emission line map, spectral index $D_n 4000$ map, SFR CIGALE result map, $M_{\star}$ CIGALE result map, and $\chi ^2$ CIGALE result map.}
    \label{fig:m_maps}
\end{figure*}

\begin{figure*}
    \centering
    \includegraphics[width=\textwidth]{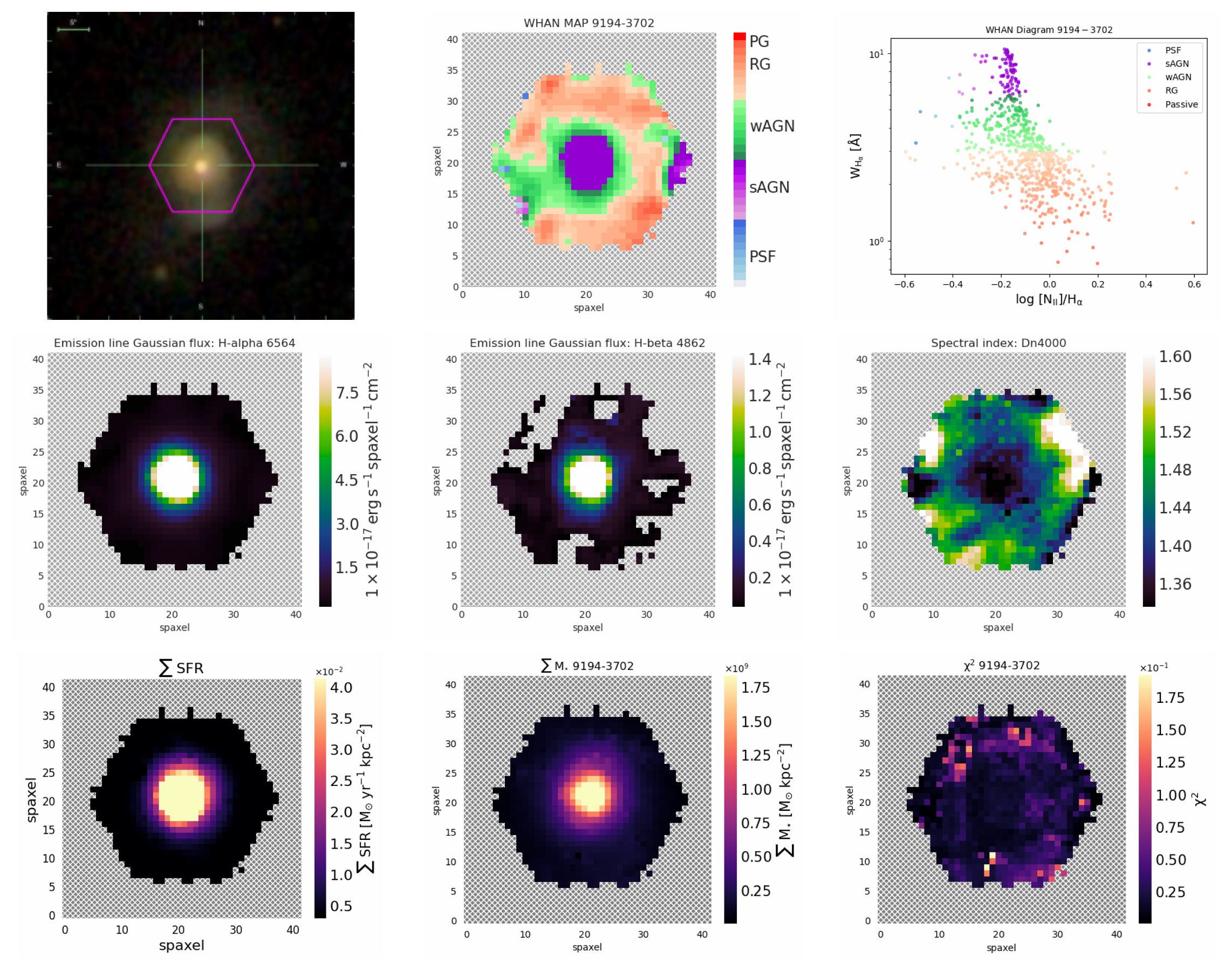}
    \caption[Analysis maps for the PsM galaxy 9194-3702]{Analysis maps for the PsM galaxy 9194-3702. From left upper to right lower panels: SDSS colour image with the MaNGA FoV, WHAN diagnostic diagram map, WHAN diagnostic diagram scatter plot, $H_{\alpha}$ emission line map, $H_{\beta}$ emission line map, spectral index $D_n 4000$ map, SFR CIGALE result map, $M_{\star}$ CIGALE result map, and $\chi ^2$ CIGALE result map.}
    \label{fig:psm_maps}
\end{figure*}

For this study, we selected a total of 12 products per galaxy. We collected this information for the 137 galaxies in our sample. We show an example of these maps/diagrams for one representative galaxy of each merger stage in Figures \ref{fig:cp_maps}, \ref{fig:prm_maps}, \ref{fig:m_maps}, and \ref{fig:psm_maps}, for a CP, PrM, M, and PsM galaxy, respectively. 

Using the Marvin tool we selected the SDSS three-colour image of the galaxy, with the maps of the spectral properties we provide as inputs for CIGALE $H_{\alpha}$ and $H_{\beta}$ emission lines and $D_n\, 4000$ spectral index, and also the map of the velocity dispersion in the $H_{\alpha}$ line to consider the kinematics of the galaxies.

Among the results of the CIGALE SED fitting, we created maps of the SFR and stellar mass surface densities ($\sum SFR$ and $\sum M_\star$, respectively). We also selected the parameters age of the last burst/quench ($Age_{bq}$) and the ratio of the SFR after/before $Age_{bq}$, plus a map of the resulted $\chi^2$ to help us to identify any problem with the fitting.

We include the spatially resolved WHAN diagram and its corresponding map for each galaxy, using the visualisation tool described in Appendix~\ref{sec:WHAN-tools}.

\section{Visualisation tools}
\label{sec:WHAN-tools}

Marvin, a tool specifically designed to visualise and analyse MaNGA data, has the capability of generating BPT diagrams for a particular galaxy. By default, a spaxel only becomes classified if it meets the criteria in all three diagrams. Even selecting the less strict criterion (based on the [OIII]/H$_\beta$ versus [NII]/H$_\alpha$ diagram), it depends on the emission in four lines, where [OIII] and /H$_\beta$ might be weak, limiting the area where we could analyse the galaxy. The BPT diagrams also limit the classification of the stellar activity to three categories: the star-formation type (SF), the Seyfert type, and LI(N)ERs type. This last category could also contain some kind of passive activity. It is also possible to create a WHAN diagram for a particular galaxy with Marvin, where the code is provided as a proposed Marvin Science Case Exercise (i.,e., it is not implemented as a functionality). Using only two emission lines (H$_\alpha$ and [NII], which are easily observable in spectra), it is very useful for classifying regions with weak emission-lines that cannot be classified using BPT diagrams. It also identifies weak AGN from fake AGN, named retired galaxies (RGs) from LI(N)ERs, allowing a more complete analysis of regions where the heating of their ionised gas is the result of old stars, rather than star formation or AGN activity. 

Taking advantage of the huge potential of the WHAN diagram to investigate galaxy evolution processes at the different stellar population and at galaxy components level, we have developed a tool to visualise the spatially resolved WHAN diagram in MaNGA galaxies\footnote{The code is available at \texttt{\url{https://github.com/PauloVB72/WHAN_MaNGA-map}}.}. Based on the WHAN diagnostic diagram, the tool allows us to comprehensively explore the nuclear activity in MaNGA galaxies. Our function is designed to even explore the intermediate areas between adjoining WHAN categories, allowing us to investigate transition processes in the diverse stellar population among different regions of the galaxies. To do this, we use different colours to move gradually from one category to the other, according to their values of W$_{H_\alpha}$ and/or [NII]/H$_\alpha$, translating this information to a map of the galaxy. With this new technique we can study the spatially resolved nuclear activity in MaNGA galaxies, and it can be analysed in more detail than with previous methods. This function allows exploring the AGN/star-formation/quenching activity in the different galaxy components, which may help to investigate, for instance, how the quenching process occurs in galaxies, if there is any relation between Active Galaxy Nuclei (AGN) activity or star-formation with the local environment in close pairs/mergers, whether nuclear activity is concentrate in the inner regions or it is more extended, etc. This function can be easily incorporated in Marvin. In the panels of figures in Appendix~\ref{sec:products}, we show the WHAN diagnostic diagrams and maps for an example galaxy in each merging stage in this work.

The code is 100\% based on free software (MIT License),
making extensive use of the Python language. We have devel-
oped the code under a Linux and Windows platform, and it may be straightforwardly ported to any other operating system (Mac, FreeBSD, etc).

\section{Galaxy morphology and CAS parameters}
\label{sec:CAS-param}

In this section we compute non-parametric image predictors Gini, M20, and Concentration–Asymmetry–Clumpiness (CAS) parameters \citep{2003ApJS..147....1C, 2004AJ....128..163L, 2016MNRAS.456.3032P}, which quantitatively determine the morphological characteristics of a galaxy, to compare with the analysis in \citet{2023MNRAS.523.4164H} and \citet{2023MNRAS.522....1N}, who used these parameters to identify merger galaxies, as well as their merging process stage.

\begin{figure*}
    \centering
    \includegraphics[width=\textwidth, trim={4cm 3.5cm 5cm 2cm},clip]{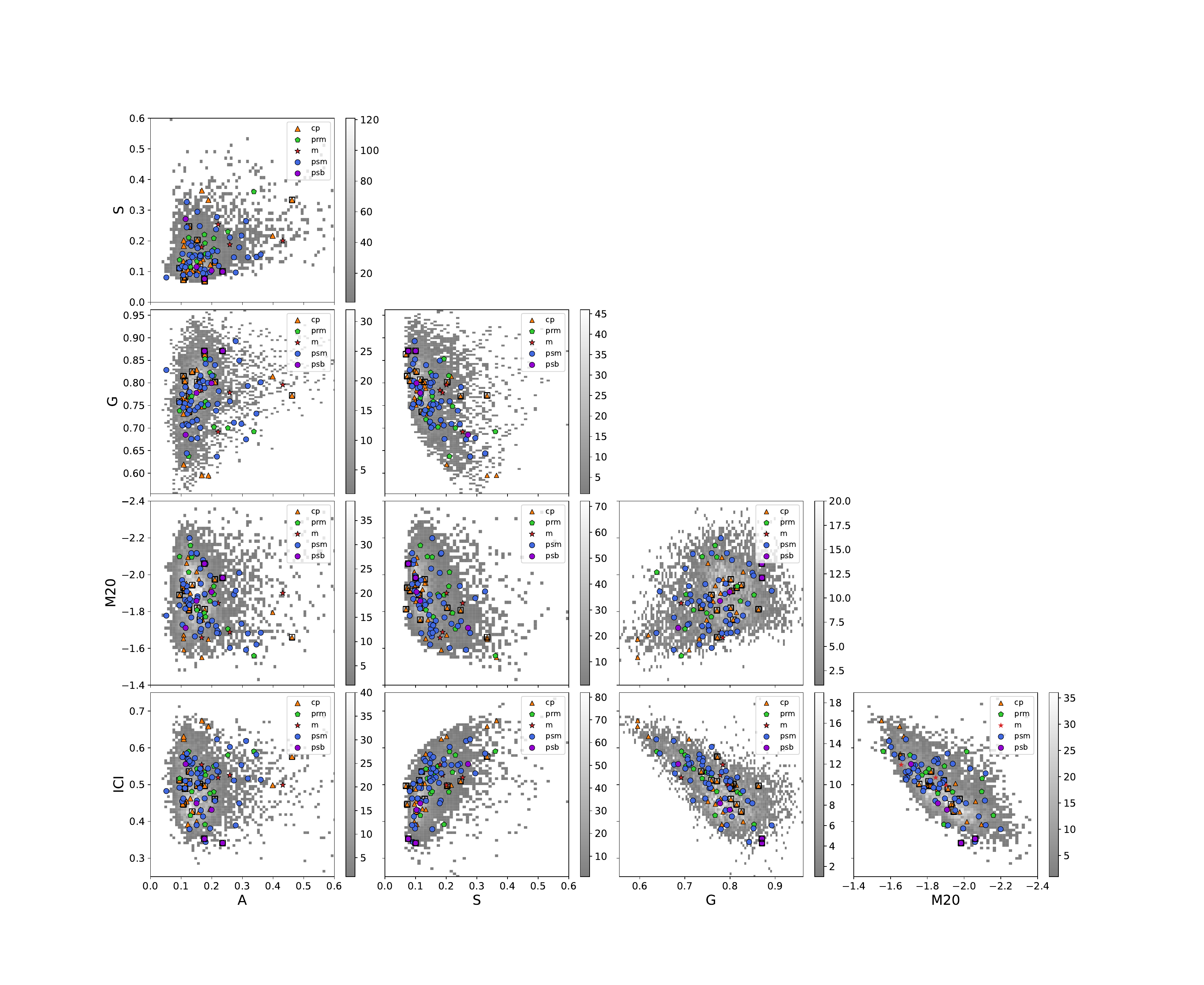}
    \caption{Comparison between non-parametric image morphological parameters for galaxies in our sample, with different markers and colour-coded as a function of their merger stage according to the legend. Additionally, galaxies presenting lenticular morphology according to DS18 are surrounded by black square. From up-to-down and left-to right: Clumpiness (S),  Gini index (G), Moment of light (M20), inverse concentration index (ICI), and asymmetry (A). Background points correspond to the 2D distribution of the same parameters for all MaNGA galaxies, coloured by density of objects.}
    \label{fig:CAS_r}
\end{figure*}

\begin{figure*}
\centering 
    \includegraphics[width=1\textwidth]{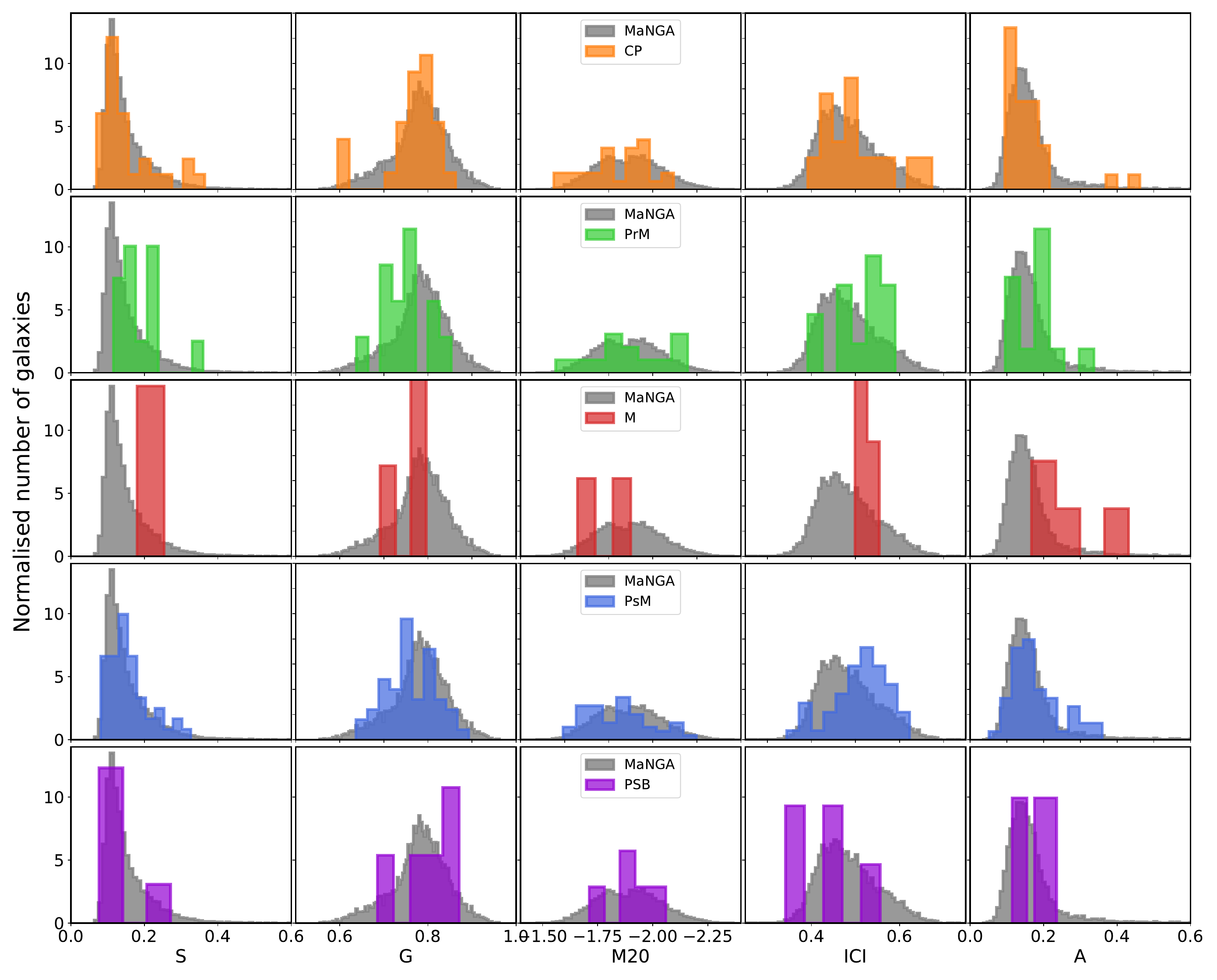}
    \caption[Comparison between non-parametric image morphological parameters for galaxies in our sample]
    {Distribution of the non-parametric image morphological parameters for galaxies in our sample, with different colour as a function of their merger stage, according to the legend. From left-to right: Clumpiness (S),  Gini index (G), Moment of light (M20), inverse concentration index (ICI), and asymmetry (A). The distribution of the same parameters for all MaNGA galaxies is shown (in grey) for reference.}
    \label{fig:CAShist}
\end{figure*}

In this work, we tested these parameters computed for the galaxies in our sample, in comparison with all MaNGA galaxies, applied to the MaNGA FoV, using the optical wide band maps (r-band) provided by the MaNGA DAP. The correlation between the parameters is shown n Fig.~\ref{fig:CAS_r}, while the distribution of the parameters for each merger stage is presented in Fig.~\ref{fig:CAShist}.

\begin{figure*}
    \centering
    \includegraphics[width=\textwidth, trim={3cm 0.5cm 4cm 2cm},clip]{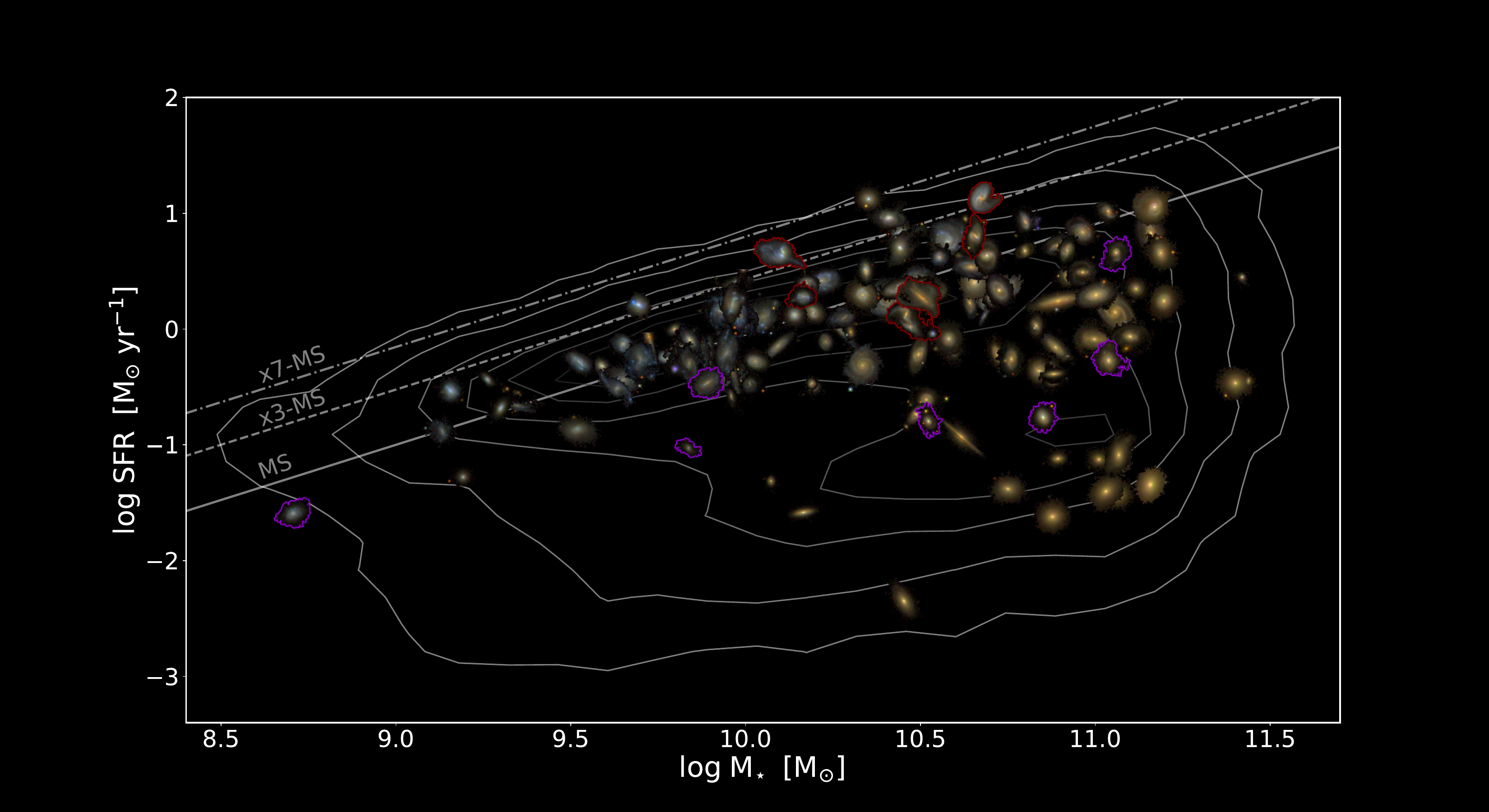}
    \caption[]{Integrated $SFR-M_{\star}$ diagram for the 137 galaxies in the sample as in Fig.~\ref{fig:SFR-M-psb}, with the point values as the respective SDSS colour image of each galaxy.}
    \label{fig:SFR-M-img}
\end{figure*}

As complement of the figures~\ref{fig:SFR-M-whan} and \ref{fig:SFR-M-D4000} presented in Sect.~\ref{sec:mer_res_int}, for reference, we present here as well a version of the SFR-$M_{\star}$ diagram where each data point for each galaxy is replaced by its SDSS three-colour image (see Fig.~\ref{fig:SFR-M-img}).

\section{Data catalogue}
\label{sec:cat}

For the sake of transparency and reproducibility, in Table~\ref{tab:cat} we compile the information related to the sample that we have used for the analysis presented in this work. 

\onecolumn
\begin{longtable}{ccccccccc}
\caption{Physical properties and derived quantities of the galaxies used in this work. } \\
\hline
\hline
(1) & (2) & (3) & (4) & (5) & (6) & (7) & (8) & (9) \\
 Plate-IFU & Catalogue & Member & Morphology & Merger stage & log(SFR) & log(SFR)$_{err}$ & log($M_{\star}$) & log($M_{\star}$)$_{err}$ \\
  & & & & & log [M$_{\odot}$\,yr$^{-1}$] & log [M$_{\odot}$\,yr$^{-1}$] & log[M$_\star$] & log[M$_\star$] \\
\hline
 \endfirsthead
\hline
(1) & (2) & (3) & (4) & (5) & (6) & (7) & (8) & (9) \\
 Plate-IFU & Catalogue & Member & Morphology & Merger stage & log(SFR) & log(SFR)$_{err}$ & log($M_{\star}$) & log($M_{\star}$)$_{err}$ \\
   & & & & & log [M$_{\odot}$\,yr$^{-1}$] & log [M$_{\odot}$\,yr$^{-1}$] & log[M$_\star$] & log[M$_\star$] \\
 \hline
 \endhead
 \hline
 \endfoot 
10839-6101 & SIT & s & ETG & CP & 0.21 & 0.15 & 10.41 & 0.10 \\
12087-3701 & SIT & s & ETG & CP & -0.22 & 0.23 & 10.72 & 0.08 \\
8937-9102 & SIT & s & LTG & CP & -0.11 & 0.20 & 10.23 & 0.11 \\
8986-3703 & SIT & s & ETG & CP & -1.32 & 0.44 & 10.07 & 0.10 \\
8456-12705 & SIT & c & LTG & CP & 0.06 & 0.17 & 10.25 & 0.13 \\
11962-1901 & SIT & c & LTG & CP & -1.14 & 0.46 & 11.06 & 0.06 \\
10221-1902 & SIT & s & ETG & CP & -0.29 & 0.18 & 9.68 & 0.08 \\
8720-1902 & SIT & c & ETG & CP & -0.32 & 0.19 & 9.59 & 0.08 \\
10493-6102 & SIT & c & ETG & CP & -1.13 & 0.37 & 11.01 & 0.08 \\
7968-9102 & SIT & c & ETG & CP & -1.38 & 0.43 & 10.75 & 0.05 \\
8140-3702 & SIP & c & ETG & CP & -1.62 & 0.70 & 10.88 & 0.05 \\
8721-3703 & SIP & c & ETG & CP & -2.35 & 0.61 & 10.45 & 0.06 \\
7972-3703 & SIP & c & ETG & CP & -1.12 & 0.48 & 10.89 & 0.08 \\
8611-12703 & SIP & c & LTG & CP & 0.24 & 0.17 & 10.89 & 0.09 \\
8548-3704 & SIP & c & LTG & CP & -0.24 & 0.22 & 9.77 & 0.06 \\
10842-3702 & SIP & c & LTG & CP & -0.32 & 0.21 & 10.33 & 0.08 \\
8452-12705 & SIP & c & LTG & CP & 0.25 & 0.20 & 10.41 & 0.08 \\
10840-12702 & SIP & c & LTG & CP & -0.26 & 0.20 & 10.76 & 0.07 \\
8459-12703 & SIP & c & LTG & CP & -0.15 & 0.21 & 10.09 & 0.10 \\
8462-3701 & SIP & c & LTG & CP & -0.76 & 0.19 & 10.86 & 0.06 \\
9092-3703 & SIP & c & ETG & CP & -1.43 & 0.82 & 11.06 & 0.06 \\
8330-3702 & SIP & c & ETG & CP & 0.24 & 0.19 & 11.20 & 0.06 \\
8455-6101 & SIP & s & LTG & CP & -0.28 & 0.21 & 10.03 & 0.08 \\
11013-6104 & SIP & c & ETG & CP & 0.35 & 0.21 & 11.12 & 0.11 \\
9496-6103 & SIP & s & LTG & CP & -0.50 & 0.19 & 10.19 & 0.10 \\
7968-3704 & SIP & c & LTG & CP & 0.07 & 0.15 & 10.07 & 0.12 \\
8616-12705 & SIP & s & LTG & CP & -0.12 & 0.19 & 9.81 & 0.12 \\
8981-9101 & SIP & c & LTG & CP & -0.16 & 0.21 & 10.89 & 0.07 \\
11977-6103 & SIP & c & ETG & CP & -1.40 & 0.29 & 11.03 & 0.04 \\
8984-3703 & SIP & c & ETG & CP & -0.37 & 0.35 & 10.86 & 0.06 \\
8441-9101 & SIP & s & LTG & CP & -0.42 & 0.22 & 10.00 & 0.17 \\
8980-3702 & SIP & s & ETG & CP & -0.36 & 0.19 & 10.84 & 0.06 \\
12082-6104 & SIP & c & ETG & CP & -0.43 & 0.24 & 10.88 & 0.08 \\
9878-1902 & SIP & c & ETG & CP & -1.59 & 0.56 & 10.17 & 0.06 \\
8566-12705 & SIP & c & LTG & CP & -0.40 & 0.19 & 9.95 & 0.09 \\
8713-12703 & SIP & s & LTG & CP & -0.29 & 0.20 & 9.89 & 0.10 \\
9040-3704 & SIP & s & LTG & CP & -0.47 & 0.21 & 10.19 & 0.10 \\
8137-6102 & SIP & c & LTG & CP & -1.35 & 0.88 & 11.16 & 0.06 \\
8245-3701 & SIP & s & ETG & CP & -0.00 & 0.17 & 9.80 & 0.12 \\
8311-12702 & SIP & c & ETG & CP & -0.39 & 0.27 & 10.88 & 0.09 \\
8093-12703 & SIT & c & LTG & PrM & -0.09 & 0.22 & 11.00 & 0.07 \\
8570-12704 & SIT & s & LTG & PrM & -0.27 & 0.17 & 9.85 & 0.15 \\
8327-12703 & SIT & s & LTG & PrM & -0.47 & 0.20 & 10.01 & 0.16 \\
8602-12704 & SIT & c & LTG & PrM & -0.47 & 0.55 & 11.40 & 0.08 \\
9509-12701 & SIT & c & LTG & PrM & -0.30 & 0.16 & 9.96 & 0.15 \\
10221-6101 & SIT & s & LTG & PrM & -0.34 & 0.18 & 9.72 & 0.10 \\
8322-12702 & SIT & s & ETG & PrM & 0.02 & 0.19 & 11.07 & 0.08 \\
8081-9102 & SIT & c & LTG & PrM & 0.25 & 0.15 & 10.71 & 0.12 \\
8244-6101 & SIP & c & LTG & PrM & 0.83 & 0.21 & 10.96 & 0.08 \\
11745-1902 & SIP & c & LTG & PrM & 0.22 & 0.19 & 10.64 & 0.09 \\
8551-12705 & SIP & c & LTG & PrM & 0.76 & 0.23 & 10.58 & 0.07 \\
9865-12704 & SIP & c & LTG & PrM & -0.93 & 0.28 & 10.62 & 0.08 \\
11020-3702 & SIP & c & LTG & PrM & -0.13 & 0.15 & 10.58 & 0.11 \\
12069-6102 & SIP & c & LTG & PrM & 0.69 & 0.21 & 10.65 & 0.09 \\
9089-6104 & SIP & c & LTG & PrM & 0.83 & 0.19 & 11.16 & 0.09 \\
11017-6103 & SIP & c & LTG & PrM & 0.82 & 0.22 & 10.44 & 0.10 \\
8318-9101 & SIP & c & LTG & PrM & 0.42 & 0.20 & 10.22 & 0.08 \\
12067-12705 & SIP & c & LTG & PrM & 0.14 & 0.20 & 11.06 & 0.06 \\
12067-9101 & SIP & s & LTG & PrM & -0.58 & 0.22 & 9.96 & 0.20 \\
11980-1901 & SIP & s & ETG & PrM & 0.45 & 0.23 & 11.42 & 1.17 \\
9046-12705 & SIP & s & LTG & PrM & 0.10 & 0.18 & 10.28 & 0.12 \\
8092-12702 & SIP & c & LTG & M & 1.13 & 0.16 & 10.67 & 0.10 \\
10215-6102 & SIP & s & LTG & M & 0.66 & 0.16 & 10.11 & 0.09 \\
8554-6101 & SIP & s & LTG & M & 0.28 & 0.19 & 10.16 & 0.08 \\
10496-12704 & SIP & s & LTG & M & 0.80 & 0.18 & 10.66 & 0.12 \\
8241-12705 & SIP & c & LTG & M & 0.13 & 0.19 & 10.46 & 0.09 \\
12483-12702 & SIP & c & LTG & M & 0.27 & 0.13 & 10.50 & 0.13 \\
9514-9101 & SIT & c & LTG & PsM & 0.45 & 0.20 & 10.92 & 0.09 \\
11828-12705 & SIP & s & LTG & PsM & -0.86 & 0.16 & 9.52 & 0.08 \\
8713-1902 & SIP & s & LTG & PsM & -0.44 & 0.15 & 9.26 & 0.09 \\
11753-12705 & SIP & s & LTG & PsM & -0.66 & 0.16 & 9.35 & 0.14 \\
8250-12703 & SIP & c & LTG & PsM & -0.89 & 0.15 & 9.13 & 0.12 \\
8439-1901 & SIP & s & LTG & PsM & -1.28 & 0.18 & 9.19 & 0.07 \\
8245-6102 & SIP & c & LTG & PsM & 0.38 & 0.20 & 10.50 & 0.08 \\
11953-3702 & SIP & c & ETG & PsM & -0.07 & 0.14 & 11.10 & 0.06 \\
9868-3703 & SIP & c & LTG & PsM & 0.14 & 0.23 & 9.92 & 0.07 \\
8310-3704 & SIP & c & ETG & PsM & -0.68 & 0.16 & 9.30 & 0.07 \\
8442-6103 & SIP & c & LTG & PsM & 0.03 & 0.16 & 10.83 & 0.09 \\
10514-12701 & SIG & - & LTG & PsM & -0.44 & 0.16 & 9.65 & 0.09 \\
8138-9102 & SIG & - & LTG & PsM & 0.49 & 0.19 & 10.96 & 0.09 \\
8244-3702 & SIG & - & LTG & PsM & 0.14 & 0.22 & 10.19 & 0.08 \\
8482-3704 & SIG & - & LTG & PsM & 0.29 & 0.22 & 11.00 & 0.06 \\
7961-12703 & SIG & - & LTG & PsM & 0.35 & 0.21 & 10.44 & 0.10 \\
8095-12703 & SIG & - & LTG & PsM & 0.10 & 0.15 & 10.05 & 0.10 \\
10496-6102 & SIG & - & LTG & PsM & 0.18 & 0.17 & 9.95 & 0.08 \\
10514-12702 & SIG & - & LTG & PsM & 0.06 & 0.17 & 9.99 & 0.09 \\
8553-1901 & SIG & - & LTG & PsM & 1.12 & 0.10 & 10.35 & 0.09 \\
10518-1902 & SIG & - & LTG & PsM & 0.96 & 0.20 & 10.41 & 0.10 \\
12491-6101 & SIG & - & ETG & PsM & 0.65 & 0.20 & 11.19 & 0.08 \\
7958-12705 & SIG & - & LTG & PsM & -0.21 & 0.18 & 9.95 & 0.11 \\
8075-12705 & SIG & - & LTG & PsM & -0.23 & 0.17 & 9.86 & 0.11 \\
8937-6104 & SIG & - & ETG & PsM & -1.09 & 0.65 & 11.07 & 0.07 \\
8144-6101 & SIG & - & LTG & PsM & 0.29 & 0.21 & 10.34 & 0.08 \\
8626-12703 & SIG & - & LTG & PsM & 0.72 & 0.21 & 10.89 & 0.10 \\
8940-6102 & SIG & - & ETG & PsM & 0.70 & 0.23 & 10.44 & 0.08 \\
8606-12705 & SIG & - & LTG & PsM & 0.68 & 0.20 & 10.92 & 0.13 \\
7960-3702 & SIG & - & LTG & PsM & -0.61 & 0.22 & 10.52 & 0.06 \\
9504-3703 & SIG & - & LTG & PsM & -0.13 & 0.16 & 9.65 & 0.07 \\
8440-12701 & SIG & - & LTG & PsM & 0.31 & 0.17 & 10.72 & 0.09 \\
8146-9102 & SIG & - & LTG & PsM & 0.29 & 0.20 & 10.70 & 0.11 \\
8565-6104 & SIG & - & LTG & PsM & -0.53 & 0.15 & 9.16 & 0.08 \\
10509-6101 & SIG & - & LTG & PsM & 0.62 & 0.22 & 10.55 & 0.08 \\
8313-1901 & SIG & - & LTG & PsM & 0.21 & 0.13 & 9.69 & 0.08 \\
8488-3702 & SIG & - & LTG & PsM & -0.32 & 0.20 & 9.81 & 0.06 \\
9036-9102 & SIG & - & LTG & PsM & 0.40 & 0.21 & 10.65 & 0.08 \\
8455-3704 & SIG & - & LTG & PsM & -0.17 & 0.18 & 9.82 & 0.07 \\
11013-6102 & SIG & - & LTG & PsM & -0.29 & 0.16 & 9.53 & 0.07 \\
8313-6103 & SIG & - & ETG & PsM & 0.12 & 0.19 & 10.14 & 0.07 \\
8985-6104 & SIG & - & LTG & PsM & 0.34 & 0.15 & 10.53 & 0.10 \\
9867-9101 & SIG & - & LTG & PsM & 0.54 & 0.19 & 10.64 & 0.09 \\
11760-1901 & SIG & - & LTG & PsM & -0.10 & 0.21 & 9.84 & 0.08 \\
11018-3703 & SIG & - & ETG & PsM & 0.51 & 0.22 & 10.34 & 0.09 \\
10221-6103 & SIG & - & LTG & PsM & 0.33 & 0.21 & 10.73 & 0.08 \\
11944-9101 & SIG & - & LTG & PsM & 1.02 & 0.17 & 11.04 & 0.11 \\
10500-12704 & SIG & - & LTG & PsM & 0.92 & 0.21 & 10.80 & 0.11 \\
8092-3701 & SIG & - & LTG & PsM & -0.74 & 0.23 & 10.49 & 0.07 \\
8652-6103 & SIG & - & LTG & PsM & 0.63 & 0.20 & 10.69 & 0.08 \\
11021-6103 & SIG & - & LTG & PsM & -0.51 & 0.19 & 9.63 & 0.08 \\
8080-3703 & SIG & - & LTG & PsM & 0.39 & 0.18 & 9.97 & 0.09 \\
8259-3702 & SIG & - & LTG & PsM & -0.08 & 0.19 & 10.58 & 0.07 \\
8323-12701 & SIG & - & LTG & PsM & 0.14 & 0.16 & 9.98 & 0.09 \\
8134-6103 & SIG & - & LTG & PsM & 0.40 & 0.18 & 10.17 & 0.08 \\
8140-12704 & SIG & - & LTG & PsM & -0.30 & 0.19 & 9.84 & 0.11 \\
10498-12702 & SIG & - & LTG & PsM & 0.16 & 0.16 & 9.93 & 0.10 \\
8310-3701 & SIG & - & LTG & PsM & 0.67 & 0.15 & 10.80 & 0.08 \\
12483-12704 & SIG & - & LTG & PsM & 0.27 & 0.17 & 10.50 & 0.09 \\
9490-6102 & SIG & - & LTG & PsM & -0.22 & 0.20 & 10.49 & 0.08 \\
9504-12702 & SIM & - & LTG & PsM & -0.26 & 0.14 & 9.70 & 0.09 \\
9484-12705 & SIM & - & LTG & PsM & 1.05 & 0.20 & 11.17 & 0.09 \\
8719-12702 & SIM & - & LTG & PsM & 0.21 & 0.15 & 9.96 & 0.10 \\
9194-3702 & SIT & c & LTG & PSB & 0.66 & 0.10 & 11.06 & 0.08 \\
8483-12702 & SIP & s & LTG & PSB & -1.03 & 0.20 & 9.84 & 0.16 \\
8555-3701 & SIP & c & ETG & PSB & -0.27 & 0.18 & 11.04 & 0.05 \\
11955-6103 & SIP & c & ETG & PSB & -0.76 & 0.45 & 10.85 & 0.06 \\
8088-3704 & SIG & - & LTG & PSB & -0.47 & 0.18 & 9.89 & 0.08 \\
12067-3701 & SIG & - & ETG & PSB & -0.80 & 0.37 & 10.52 & 0.07 \\
8981-12705 & SIG & - & LTG & PSB & -1.59 & 0.14 & 8.71 & 0.12 \\
\end{longtable} \label{tab:cat}
\tablefoot{The columns correspond to: (1) MaNGA Plate-IFU identifier; (2) Isolated system where the galaxy belons (SIG, SIP, SIT, or SIM); (3) Member galaxy in the SIP and SIT, c: central galaxy, s: satellite; (4) Galaxy morphology according to DS18, ETG: early-type galaxy, LTG: late-type galaxy; (5) Merger stage, CP: close pairs, PrM: pre-mergers, M: mergers, PsM: post-mergers, PSB: post-mergers with post-starburst spectral features; (6) and (7) integrated SFR and corresponding error, in log[M$_{\odot}$\,yr$^{-1}$]; (8) and (9) integrated stellar-mass, in  log[M$_\star$].}

\end{document}